\documentclass[11pt,twoside]{article}
\usepackage[makeroom]{cancel}
\usepackage{latexsym}
\usepackage{longtable}
\usepackage{epsfig}
\usepackage{graphicx,bbm,psfrag}
\usepackage{dsfont}
\usepackage{amssymb,amsmath,amsthm,booktabs,mathtools}
\usepackage{bm}
\usepackage{color}
\usepackage{dutchcal}
\usepackage{hyperref}
\hypersetup{
    colorlinks,
    citecolor=red,
    linkcolor=blue,
}
\usepackage{xcolor}
\usepackage{alphabeta}
\usepackage[utf8]{inputenc}
\usepackage{amsfonts}
\usepackage{nameref}
\usepackage{cleveref}
\usepackage[title]{appendix}
\usepackage{array}
\usepackage{setspace}
\RequirePackage{fix-cm}
\newcommand\norm[1]{\lVert#1\rVert}
\newcommand{\floor}[1]{\lfloor #1 \rfloor}

\newtheorem{theorem}{Theorem}[section]
\newtheorem{rema}{Remark}[section]
\newtheorem{propo}[theorem]{Theorem}
\newtheorem{defi}[rema]{Definition}
\newtheorem{lemma}[theorem]{Lemma}
\newtheorem{corol}[theorem]{Corollary}

\newtheorem{property}[theorem]{Property}

\newcommand{\bc}{\begin{center}}
\newcommand{\ec}{\end{center}}
\def\ba#1{\begin{array}{#1}\displaystyle}
\newcommand{\ea}{\end{array}}

\newcommand{\beq}{\begin{equation}}
\newcommand{\eeq}{\end{equation}}
\newcommand{\beqa}{\begin{eqnarray}}
\newcommand{\eeqa}{\end{eqnarray}}
\newcommand{\no}{\nonumber}
\newcommand{\n}{\nonumber\\}
\newcommand{\bi}{\begin{itemize}}
\newcommand{\ei}{\end{itemize}}

\def\lt#1{\left#1}
\def\rt#1{\right#1}
\def\t#1{\tilde{#1}}
\def\h#1{\hat{#1}}
\def\b#1{\bar{#1}}
\def\frc#1#2{\frac{#1}{#2}}

\newcommand{\p}{\partial}

\newcommand{\bra}{\langle}
\newcommand{\ket}{\rangle}
\newcommand{\Z}{{\mathbb{Z}}}
\newcommand{\N}{{\mathbb{N}}}
\newcommand{\R}{{\mathbb{R}}}
\newcommand{\C}{{\mathbb{C}}}
\DeclareMathOperator{\Tr}{{\rm Tr}}

\newcommand{\ep}{\epsilon}
\newcommand{\varep}{\varepsilon}

\DeclareMathOperator{\End}{{\rm End}}
\DeclareMathOperator{\Aut}{{\rm Aut}}

\newcommand{\ri}{{\rm i}}
\newcommand{\re}{{e}}
\newcommand{\dd}{{\rm d}}
\newcommand{\1}{{\bf 1}}

\DeclareMathOperator{\dist}{{\rm dist}}
\DeclareMathOperator{\diam}{{\rm diam}}
\DeclareMathOperator{\supp}{{\rm supp}}

\DeclareMathOperator{\im}{im}

\def\bs#1{\boldsymbol{#1}}

\newcommand{\halmos}{\rule{1ex}{1.4ex}}
\newcommand{\eproof}{\hspace*{\fill}\mbox{$\halmos$}}

\newcommand{\Srat}{\mathbb{S}_{\mathbb Q}}

\begin{document}

\begin{center}
{\Large {\sc Long-time dynamics in quantum spin lattices:\\[0.1cm] ergodicity and hydrodynamic projections\\[0.1cm] at all frequencies and wavelengths}}

\vspace{1cm}

{\em Dedicated to the memory of Krzysztof Gawedski}

\vspace{1cm}

{\large Dimitrios Ampelogiannis and Benjamin Doyon}

\vspace{0.2cm}
Department of Mathematics, King's College London, Strand WC2R 2LS, UK
\ec

Obtaining rigorous and general results about the non-equilibrium dynamics of extended many-body systems is a difficult task. In quantum lattice models with short-range interactions, the Lieb-Robinson bound tells us that the spatial extent of operators grows at most linearly in time. But what happens within this light-cone? We discuss rigorous results on ergodicity and the emergence of the hydrodynamic scale, which establish fundamental principles at the root of non-equilibrium physics.
One key idea of the present work is that general structures of hydrodynamics at the Euler scale follow {\em independently from the details of the microscopic dynamics}, and in particular do not necessitate chaos; they are consequences of ``extensivity". Another crucial observation is that these apply at {\em arbitrary frequencies and wavelengths}. That is, long-time oscillatory behaviours can be reproduced from a natural extension of standard hydrodynamic notions, thus enlarging the hydrodynamic paradigm beyond the zero-frequency / infinite-wavelength point that it traditionally addresses.

\tableofcontents

\section{Introduction}

The dynamics of many-body, extended, isolated, interacting quantum systems has been the focus of much recent theoretical and experimental research \cite{eisert_quantum_2015,gogolin_equilibration_2016,dalessio_quantum_2016}. Of particular interest is their universal, large-scale behaviours away from equilibrium. Given the panoply of phenomena observed, it is crucial to rigorously establish the physical principles that guide them, such as ergodicity and hydrodynamics. Can we prove that certain forms of ergodicity occur, and that hydrodynamics emerges whereby a large amount of information is lost? Can we characterise the remaining degrees of freedom? For reversible, Hamiltonian dynamics in extended quantum systems, these are some of the deepest questions and remain largely open, forming an important part of Hilbert's sixth problem \cite{hilbert_mathematical_1902}. 

On the one hand, ergodicity is the idea that, over long times, the system covers uniformly enough the manifold of states, or at least the part of it that is dynamically accessible, such as the energy shell. In the language of Gibbs' statistical mechanics, this translates into the equivalence of time averages with ensemble averages. In what situations does this happen in extended quantum systems? Conventional wisdom says that this requires the presence of chaos: the exponential separation of initially nearby trajectories, or, in quantum mechanics, certain conditions on the energies and matrix elements of macroscopic observables \cite{neumann_beweis_1929,goldstein_long-time_2010,goldstein_normal_2009}, or on out-of-time-ordered correlators (OTOC) \cite{shenker_black_2014, hosur_chaos_2016, maldacena_bound_2016,hashimoto_out--time-order_2017}. Is there such a link between short-time (chaotic) and long-time (ergodic) behaviours?

On the other hand, hydrodynamics purports that the dynamics of the system reduces to that of a few emergent, slowly-decaying degrees of freedom. One stark realisation of this is the Boltzmann-Gibbs principle \cite{spohn_large_1991,demasi_mathematical_2006,kipnis_scaling_1998}: in hydrodynamic linear response, this is the idea that correlations due to local perturbations are carried by long-lived modes over which local perturbations ``project", and that propagate at hydrodynamic velocities. Indeed, the dominant correlations between a person's vocal chords and another's eardrum are carried by sound waves. The principle holds quite generally, including in extended Hamiltonian quantum models, and including with integrability \cite{doyon_drude_2017,doyoncorrelations,nardis_correlation_2021}. What are the conditions for the emergence of such hydrodynamic projections, and what are the hydrodynamic modes?

In this paper, we report on progress in these directions. First, after reviewing basic aspects of ergodicity including von Neumann's ergodic theorem, we discuss notions of many-body ergodicity. We show that a “non-localisation condition” (there is a $t$ such that no local observable returns to itself under evolution by time $t$) implies time-ergodicity for local observables. We also show that it implies a stability theorem of Kubo-Martin-Schwinger (KMS) states with correlations that vanish quickly enough in space: after local perturbations, over long periods of time, the state returns 100\% of the time arbitrarily close, in a metric based on local observables, to the KMS state. We then discuss the authors’ recent ``almost-everywhere ergodicity" result \cite{ampelogiannis_almost_2021}, which show that for displacements along space-time rays at almost every velocity, these conclusions in fact hold in all generality in short-range hypercubic quantum lattice models (it is always true that, for almost every speed $v$, there is a $t$ such that no local observable returns to itself under evolution by time $t$ and displacement by distance $vt$). In particular, infinite-time averages of local observables along rays in space-time of almost every velocity are non-fluctuating, and, as operators, are ``thin", i.e.~approach the identity (thus become ``classical"). 

Next, we discuss the phenomenon of hydrodynamic projections. Using a form of almost-everywhere ergodicity, we prove hydrodynamic projection theorems in arbitrary dimension, and explain how such results immediately extend to arbitrary frequencies $f$ and wavenumbers $\bs k$. We give a precise definition of the space of hydrodynamic modes, as the kernel of a the (in general $f$-oscillatory) unitary evolution operator on the Hilbert space of (in general $\bs k$-)extensive quantities. In general, this space depends on the state and on the chosen frequency and wavenumber. Again, results apply to all short-range hypercubic quantum lattices, and a large family of states.

Finally, oscillatory hydrodynamic projections give rise to the ``oscillatory linearised Euler equation", and we illustrate in the free-fermionic chain how this reproduces the oscillatory behaviours of two-point functions at large space-time separations. This gives a ``proof of principle" for the use of hydrodynamics to address such oscillatory behaviours at long times. 

\medskip

From the physics perspective, our main messages are:
\begin{enumerate}
\item In contrast to the few-body case, the ``right" notion of ergodicity in many-body systems {\em does not rely on the conventional idea of covering the energy shell}. Rather, in the thermodynamic limit, the fact that timelike averages of local observables become non-fluctuating on large times follows from the interplay between extensivity of the system and locality of the observables. It happens on timescales which are much smaller than the Poincaré recurrence scale, or that for covering the energy shell. This is many-body ergodicity (related to what is sometimes called ``typicality"). Interestingly, despite these physical differences, it is nevertheless the same mathematical principles from the basic ergodicity theory that underpins many-body and few-body ergodicity, including von Neumann's ergodic theorem.
\item Euler-scale hydrodynamics hold for a wide family of systems, {\em independently from their specific dynamics} (integrable or chaotic, constrained or not, etc.). General structures, such as hydrodynamic projections, always emerge at large scales, and only specific properties that control the phenomenology are model dependent. The idea that hydrodynamics is a useful concept beyond its conventional field of applications, not requiring chaos, has come to the fore with the recent development of generalised hydrodynamics for integrable systems \cite{castro-alvaredo_emergent_2016,bertini_transport_2016,bulchandani_bethe-boltzmann_2018,doyon_lecture_2020}, and ties in with the idea that thermodynamic concepts should not rely on short-time behaviours, as emphasised recently in \cite{chakraborti_entropy_2021}. However, {\em rigorous} results about many-body dynamics, especially of such generality, are notoriously difficult to obtain.
\item Hydrodynamics at the Euler scale can be generalised to {\em arbitrary frequencies and wavelengths}, going beyond the conventional zero-frequency and large-wavelength region. Thus, in $D$ dimensions, there is a $(f,\bs k)$-hydrodynamic theory for every frequency $f\in\R$ and wavenumber $\bs k\in\R^D$. The $(f,\bs k)$-hydrodynamic modes control the $(f,\bs k)$-oscillatory behaviours of correlation functions at large separations in space-time (for any given $(f,\bs k)$, if the set of modes is non-empty, then such $(f,\bs k)$-oscillatory behaviours are observed). This shows that recent ideas on dynamical symmetries \cite{buca2019nonstationary,buca2020quantum,medenjak2020rigo,buca2021local,gunawardana2021dynamical} can be extended to the hydrodynamic regime of correlation functions. In fact, the proofs we provide make it clear that the extension goes beyond simple phase oscillations, and include oscillations involving nontrivial internal transformations, such as spin rotations. This in principle allows hydrodynamics to be extended to describe large-scale behaviours involving nontrivial microscopic internal structures, such as helicoidal spin structures.
\end{enumerate}

From a more philosophical perspective, an important message is that {\em it is extremely fruitful to extract the relevant many-body physics out of equilibrium by analysing the operator algebra of local observables already in the thermodynamic limit}. This is to be contrasted with methods based, in quantum systems, on how the spectrum of large Hamiltonians behave as the thermodynamic limit is taken, or, in classical models, on kinetic equations. The operator algebra framework allows one to directly concentrate on the small part of the Hilbert space (or phase space in classical mechanics) that is significant in the thermodynamic limit. In particular, we emphasise how kernels of naturally defined unitary operators on Hilbert spaces of observables encode the relevant aspects of large-scale physics.

We believe most of these notions are valid in a large class of many-body systems, quantum or classical. In order to be precise, however, we will concentrate on translation-invariant, infinitely-extended $D$-dimensional hyper-cubic quantum lattices with finite local spaces and short-range interactions (interaction strength decaying at least exponentially with distance).

The paper is organised as follows. First, in Section \ref{sectergo}, we discuss notions of ergodicity, and in particular many-body ergodicity, and ``almost-everywhere ergodicity" established recently \cite{ampelogiannis_almost_2021} within the context of the $C^*$ algebraic formulation of quantum statistical mechanics. We contrast this with von Neumann's ergodic theorem, and discuss application to the problem of return to equilibrium after local quenches. We then prove rigorously, in Section \ref{sectproj}, the principle of hydrodynamic projection onto conserved modes in the Euler scaling limit, within the same context. This generalises the result of \cite{doyon_hydrodynamic_2022} to any dimension, although the proof requires the introduction of quite a few new ideas. Crucially, mathematical formalism make it clear that the general principle can take into account oscillatory behaviours at any frequency and wavelength. We give an accurate definition of the full, complete space of hydrodynamic modes, which in general depends on the state, and on the frequency and wavelength. Finally, in Section \ref{sectosci} we review the linearised (oscillatory) Euler hydrodynamic equations established in \cite{doyon_hydrodynamic_2022} in one dimension, and discuss how they recover oscillatory behaviours in free fermion models. The structures uncovered are pictorially illustrated in Figure \ref{fig}.


\begin{figure}
    \centering
    \includegraphics[width=6cm]{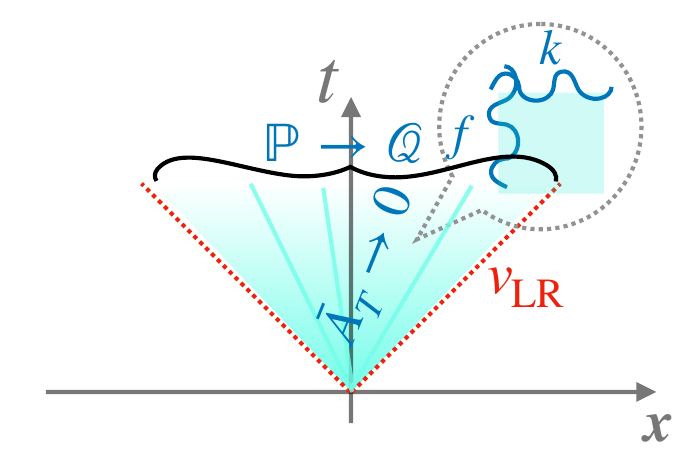}
    \caption{A pictorial representation of the mathematical results of \cite{ampelogiannis_almost_2021}, and of the hydrodynamic projection results of section \ref{sectproj}. Within the Lieb-Robinson cone $|\boldsymbol x|< v_{LR} t$, very little was known until now. We find that ray-averages $\b A_T$ of observables up to time $T$, with respect to every frequency $\omega$ and wavelength $\boldsymbol k$ and at every velocity $\boldsymbol v$, vanish at large times $T\to\infty$ (represented by the colour-gradient shaded region), except perhaps for a set of $\boldsymbol v$ of measure zero (represented by solid ray segments). Observables become thin as time evolve. The large-scale information that remains, after oscillatory fluid-cell means, is their projection onto a smaller space of degrees of freedom $\mathcal Q_{\omega,\boldsymbol k}$, the slowly decaying hydrodynamic modes.}
    \label{fig}
\end{figure}

\section{Ergodicity, many-body ergodicity, almost-everywhere ergodicity} \label{sectergo}

Ergodicity plays a fundamental role in physics. In this section, we pedagogically review some concepts of ergodicity in many-body systems, one of the goals being to put in context some recent results by the authors. We first review the basic rigorous ergodicity result in classical mechanics from Birkhoff's ergodic theorem. We then review von Neumann's fundamental work on the quantum ergodic theorem, recasting the nice discussion in \cite{goldstein_long-time_2010} into the language of local observables in quantum spin lattices. We observe that the eigenstate thermalisation hypothesis (ETH) is in fact closely related to the conditions of von Neumann's quantum ergodic theorem. Both the von Neumann quantum ergodic theorem and the ETH rely heavily on the structure of the spectrum, and how it behaves in the thermodynamic limit. Lastly, we avoid such difficulties by passing to the $C^*$-algebra formalism, in which context we discuss what many-body ergodicity may look like, and review the recent results \cite{ampelogiannis_almost_2021}, where almost-everywhere ergodicity is proven for every quantum spin model on (hyper-)cubic lattices with short-range interactions; we discuss some applications and relations with the problem of return to equilibrium.

We denote throughout this section the time average as
\beq\label{timeaverage}
	\overline{f(t)} := \lim_{T\to\infty} \frc1T \int_0^T\dd t\,f(t).
\eeq

\subsection{Ergodicity in classical mechanics}\label{ssectergoclassical}

In the classical mechanics of interacting particles, the basic notion of ergodicity is the equivalence between time average and ``ensemble average", or average over the uniform measure on the energy shell. 

Consider a system of $N$ particles, with phase-space coordinates $(\bs x_n,\bs p_n) \in \R^{2D}:n=1,\ldots,N$ and time evolution generated by a Hamiltonian $H(\boldsymbol x,\boldsymbol p)$ (we denote $(\boldsymbol x,\boldsymbol p) = ((\bs x_n,\bs p_n) :n=1,\ldots,N)\in\R^{2DN}$). The microcanonical ensemble average of an observable $A(\boldsymbol x, \boldsymbol p)$ on the shell $\Omega_E=\{(\bs x,\bs p): H( \boldsymbol x,\boldsymbol p) = E\}$ of energy $E$ can be defined by a limit on the thickened shell $S_{E_-,E_+} = \{(\bs x,\bs p):E_-<H(\boldsymbol x,\boldsymbol p)<E_+\}$ as
\beq
	\bra A\ket_{E} :=
	\lim_{E_-,E_+\to E} \frc1{{\rm vol}\,S_{E_-,E_+}} \int_{S_{E_-,E_+}}
	\dd^{DN}  x\, \dd^{DN} p\, 
	A(\boldsymbol x,\boldsymbol p)
\eeq
where
\beq
	{\rm vol}\,S_{E_-,E_+} = \int_{S_{E_-,E_+}}
	\dd^{DN}  x\, \dd^{DN} p.
\eeq
Note that this naturally induces a measure $\mu_E$ on $\Omega_E$.

Now consider points $( {\boldsymbol x},{\boldsymbol p})\in\Omega_E$.  The ergodicity statement is that, if there is no nontrivial constant of the motion\footnote{A nontrivial constant of the motion is a measurable function $Q$ on phase space (an observable) that is invariant under the dynamics (constant of the motion), and that breaks the energy shells  (nontrivial) in that for at least one $E$, there exist two non-intersecting subsets $\Gamma,\Gamma'\subset\Omega_E$ of nonzero $\mu_E$-measure such that $Q(\Gamma)\cap Q(\Gamma') = \emptyset$.}, then, for every observable $A$, every $E\in \im H$, and almost every $( {\boldsymbol x}, {\boldsymbol p})\in\Omega_E$,
\beq
	\overline{A( {\boldsymbol x}(t), {\boldsymbol p}(t))}
	= \bra A\ket_E
\eeq
where $( {\boldsymbol x}(t), {\boldsymbol p}(t))$ is the point $( {\boldsymbol x}, {\boldsymbol p})$ time-evolved by time $t$. This is equivalent to a more intuitive statement: the fraction of time that a trajectory $t\mapsto ( {\boldsymbol x}(t), {\boldsymbol p}(t))$ spends in a neighbourhood $\Gamma\subset\Omega_E$ of a point $(\bs x',\bs p')\in\Gamma$ is, over long times, proportional to its ``surface area" $\mu_E(\Gamma)$ on the energy shell, for almost every initial point $( {\boldsymbol x}, {\boldsymbol p})$ and for every $(\bs x',\bs p')$. In particular, over long times, the trajectory comes as close as desired to the point $(\bs x',\bs p')$.

Ergodicity in classical mechanics, as expressed above, follows immediately from Birkhoff's ergodic theorem \cite{birkhoff1931ergodic} . 

\subsection{Von Neumann's quantum ergodic theorem}

It is natural to ask about the equivalent ergodicity notion for quantum systems. A problem arises: in quantum mechanics, phase space is essentially ``discretised" as $\dd x\dd p \simeq \hbar/2$. Now we can not ask a more precise question than if a trajectory will cross a neighbourhood of volume $\hbar/2$ of a point. Requiring that this happens for every point (in some precise way that can be worked out, see below) is a much weaker condition, that of ``quantum recurrence". Of course, at large values of the classical action $S\gg \hbar$, this quantisation is not seen, and one recovers classical mechanics. The above classical ergodicity notion, with the more stringent condition on the measure covered by trajectories, is then a good approximation. But how is this recovered directly from quantum mechanics? Large classical actions naturally occur in macroscopic systems. Can we formulate ergodicity for macroscopic quantum systems, and find the quantum properties that guarantee it? In particular, timescales for quantum recurrence are prohibitively large, while those for ergodicity of the corresponding classical system at $S\gg \hbar$ are short; how to account for this?

Von Neumann's idea is to look at ``macroscopic observables". Intuitively, such observables should behave like classical ones, but one asks what the microscopic quantum evolution says about these observables, and in particular their ergodicity properties. Perhaps the original idea was to consider a classical object (such as a tennis ball) as a macroscopic combination of quantum particles, and macroscopic observables for such objects. But in fact, it turns out that the abstract formulation applies quite generally. Making connections with modern developments, we will recast von Neumann's line of thought to the context of many-body quantum systems with short-range interactions: what about ergodicity in large (macroscopic) quantum spin lattices?

Take a finite spin lattice: a quantum system with $N$-dimensional local spaces lying on $\Lambda_L := [-L,L]^D\cap \Z^D$, and Hilbert space $\mathcal H = \big(\C^{N}\big)^{\otimes \Lambda_L}$. Consider the Hamiltonian
\beq\label{localhamiltonian}
	H = \sum_{\bs x\in\Lambda_L} h_{\bs x}
\eeq
where $h_{\bs x}$ is an operator supported on a finite, $\bs x$-independent number of sites near to $\bs x$. For simplicity we take the system periodic and translation invariant: there is a representation $\bs y\mapsto \iota_{\bs y}$ of the group $(\Z/(2L+1)\Z)^D$ on the group of automorphisms of the operator algebra $\Aut(\End \mathcal H)$, such that $\iota_{\bs y} h_{\bs x} = h_{\bs x+\bs y}$ (and $\iota_{\bs y} A$ is supported on the $\bs y$-translate of the support of $A$). The Hamiltonian has spectrum of eigenvalues $\{E_n\}$ and normalised eigenvectors $|n\ket$.

Consider a microcanonical shell $S = \{n:|\Lambda_L|e_-<E_n<|\Lambda_L|e_+\}$, keeping the energy densities $e_-,e_+$ implicit for readability. The microcanonical ensemble is represented by the density matrix
\beq\label{rhomc}
	\rho_{\rm mc} = \frc1{|S|} \sum_{n\in S} |n\ket\bra n|.
\eeq
It can be shown that \cite{brandao2015equivalence}, in the limit of large volume $L\to\infty$, followed by the limit of small shell $e_-\to e_+\to e$, averages of local observables $A$ (operators supported on finite numbers of sites) in this ensemble tend to averages in the Gibbs ensemble,
\beq\label{microgibbs}
	\lim_{e_-,e_+\to e}
	\lim_{L\to\infty}
	\Tr\rho_{\rm mc} A = \lim_{L\to\infty}\Tr \rho A,\quad \rho = \frc{\re^{-\beta H}}{\Tr \re^{-\beta H}}
\eeq
where $\beta$ is chosen such that $\Tr\rho h_{\bs 0} = e$.

A number of ergodicity-like results can be obtained depending on the strength of the assumptions on the underlying system.

\subsubsection{Assumption of non-degeneracy and quantum recurrence}\label{sssectrecurrence}

Before discussing von Neumann's QET, let us first consider the simplest ideas of ergodicity, keeping $L$ finite and the shell of finite thickness. Let us assume that $E_n\neq E_m$ if $n\neq m$. Take a state in the microcanonical shell,
\beq
	|\psi\ket = \sum_{n\in S}
	c_n |n\ket.
\eeq
Then clearly
\beq\label{psiaverage}
	\overline{\bra \psi(t)|A|\psi(t)\ket}
	=\sum_{m,n\in S} \b c_m c_n \overline{e^{\ri (E_m-E_n)t}}
	\bra m|A|n\ket
	= \sum_{n\in S} |c_n|^2\bra n|A|n\ket.
\eeq
The result is within the so-called ``diagonal ensemble": the ensemble of density matrices that are diagonal in the energy eigenbasis. This is certainly not ergodicity, even in the limit of large $L$ and thin shell! The equivalent of the classical uniform shell-averaging is the uniform sum over $n$'s ($\Tr\rho_{\rm mc}A$), but here, the $c_n$'s may behave wildely as functions of $n$.

One can nevertheless go further, and get a weak statement of ergodicity. Beyond the trivial stationarity statement $\overline{\bra n|A(t)|n\ket} = \bra n|A|n\ket$, in any state $|n\ket$, the observable $A$ is non-fluctuating:
\beq
	\bra n|\overline{A(t)}^{\,2}|n\ket
	=\sum_m \Big|\overline{e^{\ri (E_n-E_m)t}}\Big|^2 \bra n|A|m\ket \bra m|A|n\ket
	= \big(\bra n|A|n\ket\big)^2.
\eeq
This is ``mean-square ergodicity", and in fact implies that the distribution of the random variable $\overline{A(t)}$ in the state $|n\ket$ is supported on $\overline{A(t)} = \bra n|A|n\ket$.

Mean-square ergodicity has a clear technical parallel with the classical ergodicity notion. Indeed, the condition of non-degeneracy is equivalent to the condition that the space of all operators invariant under time evolution is spanned by the projections $P_m = |m\ket\bra m|$, and this parallels the classical condition that there be no nontrivial constants of the motion\footnote{In this respect, it is sometimes stated in the literature that integrability -- in the sense of the existence of large enough amount of nontrivial constants of the motion -- for finite quantum spin lattices does not make sense, because there always are as many projections $P_m$, which commute with $H$, as the dimension of the Hilbert space $\dim \mathcal H$. This is misleading, as a similar statement could be made for classical systems: functions of the Hamiltonian are constants of the motion. But they are not ``nontrivial": they do not divide the energy shells. Similarly, $P_m$ are not nontrivial constants of the motion. Only with degeneracies can new, nontrivial constants of the motion appear: the space of operators is of dimension $(\dim \mathcal H)^2$, and integrability typically leads to enough degeneracies so that there are $\propto \log (\dim \mathcal H) \dim \mathcal H$ linearly independent projections commuting with $H$; this is associated to the presence of a large enough amount of extensive conserved quantities.}. In evaluating time averages, it is von Neumann's ergodic theorem from functional analysis \cite[Theorem II.11]{Simon_Reed_Functional} (not to be confused with von Neumann's quantum ergodic theorem!) that replaces Birkhoff's theorem from measure theory. This theorem implies that, if $U_t$ is a unitary one-parameter group on a Hilbert space, then
\beq
	\overline {U_t} = \mathbb P_1
\eeq
where $\mathbb P_1$ the projector onto its unit-eigenvalue subspace. In order to use this, we must consider time evolution as a unitary operator acting on matrices (say under the Hilbert-Schmidt inner product), and then its long-time average projects onto ${\rm span}(P_m)$. Of course, in the finite-dimensional context considered here, such more involved mathematics is not necessary: it is elementary to see that the long-time average of an observable reduces it to its diagonal, thus to a linear combination of projections $P_m$. Then, because we look at a single state $|n\ket$, this implies that observables are non-fluctuating. Thus, within a fix state $|n\ket$, ergodicity holds, paralleling what happens within a fix classical energy shell (we fix the state $|n\ket$, because the limit of an infinitesimally thin shell, for fixed $L$, is just a single state). As $|n\ket$ is stationary, the statement that $\overline{A(t)}$ be non-fluctuating and supported on $\bra n|A|n\ket$ is the natural ergodicity statement.

But this parallel is essentially technical, as the above simple statements only relate to quantum recurrence, not many-body ergodicity. Indeed, the order of limits is simply incorrect for many-body systems: the limit in time $T\to\infty$, in the time averaging \eqref{timeaverage}, is taken before the limit of large volumes $L\to\infty$ (which is in fact never taken in the mean-square ergodicity statement). This is thus valid, in realistic many-body systems, only at astronomically large times! As we will see in subsection \ref{ssectalmost}, however, with the right space of operators, such a technical parallel becomes fully meaningful.

\subsubsection{Assumptions on maximal deviations  and von Neumann's QET}

Assume that ${\rm max}_{n\in S} \Big(\bra n|A|n\ket - \Tr\rho_{\rm mc}A\Big)^2$ is ``small".

A natural requirement would be that it is uniformly bounded in $L$ (thus stays finite in the large volume limit), and this bound tends to zero in the limit of thin microcanonical shell. Of course, this cannot be expected to be valid for all observables $A$. But, at least for {\em local observables}, this should hold (certainly uniform boundedness in $L$ is immediate by finiteness of the operator norm). A clear example is the energy density $h_{\bs x}$. The states $|n\ket$ may be assumed to diagonalise the translation operator, and therefore
\beq
	\bra n|h_{\bs x}|n\ket
	= \frc1{|\Lambda_L|}
	\sum_{\bs x' \in \Lambda_L}
	\bra n|h_{\bs x + \bs x'}|n\ket
	= \frc1{|\Lambda_L|}
	\bra n|H|n\ket
	= \frc{E_n}{|\Lambda_L|}.
\eeq
Thus
\beq
	e_-\leq\Tr \rho_{\rm mc} h_{\bs x}\leq e_+
\eeq
and ${\rm max}_{n\in S} \Big(\bra n|h_{\bs x}|n\ket - \Tr\rho_{\rm mc}h_{\bs x}\Big)^2 \leq (e_+-e_-)^2$. The requirement is that such a smallness bound holds not just for the energy density, but for all (or at least for a large family of) local observables.

Then, as $\sum_n |c_n|^2 = 1$, we get from \eqref{psiaverage} that
\beq\label{vn_ergodicity}
	\overline{\bra \psi(t)|A|\psi(t)\ket}\
	\mbox{ is near to }\Tr\rho_{\rm mc}A.
\eeq
This looks much more like an ergodicity statement.

Making connection with more modern ideas, we note that the above assumption is in fact equivalent to a part of the eigenstate thermalisation hypothesis (ETH) \cite{gogolin_equilibration_2016}: that in the thermodynamic limit $L\to\infty$, averages of local observables in energy eigenstates vary continuously with the energy density.

One can get further with one more assumption:

Assume that ${\rm max}_{n,m} |\bra n|A|m\ket|^2$ is small, and that the energy differences $E_m-E_n$ are non-degenerate (this implies, but is stronger than, non-degeneracy of energies).

Then, von Neumann proved that $\bra \psi(t)|A|\psi(t)\ket$ has small time variance,
\beq
	\overline{\Big(\bra \psi(t)|A|\psi(t)\ket - \Tr\rho_{\rm mc}A\Big)^2}\ \mbox{ is small.}
\eeq
This is (a loose formulation of) {\em von Neumann's quantum ergodic theorem}.

The assumption that ${\rm max}_{n,m} |\bra n|A|m\ket|^2$ be small is implied by the full formulation of the ETH. Indeed, in the ETH, one equates off-diagonal matrix elements of local operators with an entropy factor that vanishes exponentially with the system size, times matrix elements from random matrix theory. Importantly, the assumption that ${\rm max}_{n,m} |\bra n|A|m\ket|^2$ be small is in fact weaker. The energy-difference non-degeneracy condition would be paralleled by the assumed random-matrix distribution of eigenvalues in quantum chaos; although the full connection is not obvious.

One sees immediately two potential drawbacks from von Neumann's ergodic theorem:
\bi
\item The assumptions on maximal deviations from the microcanonical averages are somewhat contrived; it is not clear, given a short-range interaction on a quantum lattice, how to show their validity. (The same can be said {\em a fortiori} of the ETH.) Is there something more universal that can be said? Are these assumptions really necessary to understand the structure of long-time dynamics in many-body quantum systems?
\item Non-degeneracy assumptions are natural, but they are seemingly unrelated to the thermodynamic limit, where a continuum of states exists. Can something be said when there are degeneracies, most interestingly in the presence of a large number of nontrivial conserved quantities? Can this be formulated directly in the thermodynamic limit?
\ei

\begin{rema}
In von Neumann's work, instead of local observables, it is ``macroscopic observables" that are the main focus.  One should consider a family of macroscopic observables $M_i$, with the property that they almost mutually commute, $[M_i,M_j]\approx 0$, and thus can be almost simultaneously diagonalised. The associated subspaces correspond to different ``macrostates", intuitively different cells in the classical phase space. Von Neumann's theory can be formulated in terms of projections on these subspaces, and conditions on these. One may take spatial averages of local observables, $\h A := |\Lambda_L|^{-1} \sum_{\bs x\in\Lambda_L} A(\bs x)$, which indeed have vanishing commutators, $|| [\h A,\h B]||\sim |\Lambda_L|^{-1}$. We find that it is more transparent to forgo the abstract notion of macroscopic observables and their associated eigenspaces, and instead concentrate on local observables and their matrix elements.
\end{rema}

\subsection{Many-body and almost-everywhere ergodicity theorems}\label{ssectalmost}

Recall that in von Neumann's quantum setup, we considered the ``macroscopic", or many-body, limit $L\to\infty$ in order to say something about ergodicity.

It is indeed commonly believed that, in typical interacting many-body systems and states $\bra\cdots\ket$ of physical importance, auto-correlation functions  of local observables in the thermodynamic limit vanish at long times, $\bra A(\boldsymbol 0,t)B(\boldsymbol 0,0)\ket - \bra A\ket\bra B\ket\to 0$ ($t\to\infty$). This is the notion of mixing \footnote{In quantum statistical mechanics, ``mixing'' is sometimes used to describe what we will refer to as spatial mixing, Eq. (\ref{clustering}) below, and likewise ``ergodicity'' is used to describe a property related to spatial integrals. Here we reserve these words to the more physical notions involving time. }: the effect of the small localised perturbation $B(\boldsymbol0 ,0)$ mixes with its surrounding and, locally, decays in time. A weaker notion, implied by mixing,  is that of ergodicity: $T^{-1}\int_0^T \dd t\,\bra A(\boldsymbol 0,t) B(\boldsymbol0,0)\ket \to \bra A\ket\bra B\ket$ ($T\to\infty$). This is equivalent to mean-square ergodicity: the vanishing of the variance of the observable
\beq\label{timeaverageT}
	\b A_T:= T^{-1}\int_0^T \dd t\, A(0,t)
\eeq
at long times, $\bra \b A_T^2\ket - \bra A\ket^2\to 0$. As explained above, mean-square ergodicity says that infinite-time averages of observables, in the state $\bra\cdots\ket$, do not fluctuate, and are simply equal to ensemble averages.

But an important observation, which we believe was mostly overlooked in the literature, is that the resulting question of ergodicity, in fact, has less to do with quantum mechanics, and much more with the {\bf emergence of large-scale behaviours in many-body systems}. This is about the passage from microscopic laws to macroscopic dynamics. As we mentioned, in many-body quantum systems, {\bf quantum recurrence}, as discussed in paragraph \ref{sssectrecurrence}, {\bf is irrelevant to this question}. But also, in many-body classical systems, {\bf the uniform covering of the energy shell}, as discussed in subsection \ref{ssectergoclassical}, {\bf is also irrelevant to this question}. In both cases, timescales are too large. The question of emergent ergodicity in many-body systems is of a different nature, and it brings out similar concepts both in quantum and classical many-body systems; ``quantumness" does not play a fundamental role.

In this subsection, we take the $C^*$ algebra  viewpoint on statistical mechanics. This, we argue, is much better adapted to the thermodynamic limit and the question of many-body ergodicity. We discuss this question, and in particular review results on almost-everywhere ergodicity established recently by the authors, which are extremely general and free from hard-to-check conditions. For clarity we still concentrate on quantum spin lattices, however this viewpoint is much more easily adapted to other many-body systems, with only small differences between the quantum and classical cases.

Interestingly, as we will see, the basic mathematical ideas behind quantum recurrence (especially as ``mean-square ergodicity") and classical ergodicity are still the right ones for many-body ergodicity, however only after {\bf the space of observables has been reduced to that relevant for local physics in the thermodynamic limit}. It is the reduction to this space that allows these ideas to extract the correct form of ergodicity seen in many-body systems, associated to times that do not grow with the system size.

We note that the notions of ergodicity discussed in this subsection are not to be confused with what is sometimes called the ``ergodic principle" (see e.g.~\cite{ollahydro}). This principle stipulates that relaxation must lead to a state of the Gibbs form. What we discuss here is about correlation functions of local observables, and is associated to conserved quantities that are ``localised": time-invariant local observables. By contrast, thermalisation, the ergodic principle, and the Gibbs form, are concepts related to extensive conserved quantities. Extensive conserved quantities are involved in the hydrodynamic projection theorem proved in Section \ref{sectproj}, and in the thermalisation theorems shown in \cite{doyon_thermalization_2017}, where it is proposed that the Hilbert space of extensive conserved quantities is in fact the tangent space to the manifold of maximal entropy states. Localised and extensive conserved quantities span the kernels of different evolution operators, acting on different spaces of observables (local observable, and extensive observables, repsectively).

We believe that analysing the structure of local observables in thermodynamic state is much more directly relevant to local physics than analysing the structure of spectra and how they behave as $L\to\infty$. But, perhaps most importantly, in the thermodynamic limit, more can be said, giving almost-everywhere ergodicity, which ultimately is at the basis of the emergence of hydrodynamic structures.

\subsubsection{$C^*$ algebra formulation and basic results}

The strength of the $C^*$ algebra formulation is that we can directly study dynamics in the thermodynamic limit: the limit $L\to\infty$ is already taken, and we concentrate on the remaining, relevant degrees of freedom.

We thus consider a quantum spin model on an infinite lattice, with local spaces $\simeq\C^N$ on $\Z^D$ ($N,D\geq 1$). The Hamiltonian is homogeneous (space-translation invariant) and with short-range interaction (exponentially decaying or faster, see Appendix \ref{setup} for the precise formulation); this generalises slightly the form \eqref{localhamiltonian}, which was of finite-range interaction. Observables form a $C^*$ algebra $\mathfrak U$, the norm-completion of the algebra  $\mathfrak U_{\rm loc}$ of local observables (supported on finite numbers of sites). As above, we denote $A(\boldsymbol x,t),\; B(\boldsymbol x,t),\ldots$ observables $A\in\mathfrak U$ translated to space-time points $\boldsymbol x,t$, and sometimes omit the time argument if $t=0$. A state\footnote{Here we use the $C^*$ algebra notation $\omega(\cdots)$ for states, instead of the notation $\bra\cdots\ket$ from statistical physics or quantum mechanics.} $\omega$ is a continuous, positive linear functional on $\mathfrak U$, interpreted as giving ensemble averages, normalised to $\omega( \1) = 1$. It is assumed homogeneous and stationary, $\omega (A(\boldsymbol x,t)) = \omega( A)\ \forall\ \boldsymbol x\in\Z^D,\,t\in\R$. 

Most importantly, the state is assumed to satisfy a property which we will simply refer to as ``{\bf spatial mixing}". We consider two different implementation: ``factoriality", and a uniform-enough spatial mixing, see Appendix \ref{setup}. Both include, in particular, the clustering of correlations at large separations,
\beq\label{clustering}
    \lim_{\boldsymbol x\to\infty} \omega(  A(\boldsymbol x) B)
    = \omega( A)\omega( B), \quad A,B\in\mathfrak U.
\eeq
Results reviewed below on almost-everywhere ergodicity apply either if the state is factor (\cite[p. 81, Def. 2.4.8]{bratteli_operator_1987}), or uniformly-enough spatially mixing (or both).

Any thermal state that is in a single thermodynamic phase (i.e.~Kubo-Martin-Schwinger (KMS) state satisfying \eqref{clustering}, see below) is an example of a state with all above properties: for $D=1$ this is every thermal state at nonzero temperature as there are no thermal phase transitions, for $D>1$ this includes every thermal state above a certain model-dependent temperature \cite[Section 5.3]{bratteli_operator_1997}.

Fundamental theorems have been established that guarantee that the infinite-volume limit of finite-volume thermal states exist and gives a state $\omega$ on $\mathfrak U$; this is so at least at large enough temperatures if $D>1$ (see e.g.~\cite[Section 5.3]{bratteli_operator_1997}), and at all temperatures if $D=1$ \cite{Araki1975uniqueness}. For instance, for $H$ of the form \eqref{localhamiltonian}, there exists $\beta_c$, with $\beta_c=\infty$ if $D=1$, such that for all $\beta<\beta_c$ and for all local observable $A\in\mathfrak U_{\rm loc}$, the limit
\beq\label{gibbs}
	\lim_{L\to\infty} \frc{\Tr \Big(e^{-\beta H} A\Big)}{\Tr \Big(e^{-\beta H} \Big)} = \omega(A)
\eeq
exists and gives rise to a positive, continuous (with respect to the topology induced by the operator norm), normalised linear functional, that is invariant under space-time translations,
\beq\label{basicomega}
	\omega(A^\dag A)\geq 0,\quad
	|\omega(A)|\leq ||A||,\quad
	\omega(\1) = 1,\quad
	\omega(A(\bs x,t)) = \omega(A).
\eeq
The result is extended by continuity to $\mathfrak U$.

The thermodynamic limit point in \Cref{gibbs} can be shown to be what's called a KMS state \cite[ Proposition 6.2.15]{bratteli_operator_1997}, which is defined as follows:
\begin{defi}[KMS state] \label{defn:KMS}
Consider a dynamical system $(\mathfrak{U},ι,τ)$. A state $ω$ is called a $(τ,β)$-KMS state, at inverse temperature $β$, if 
\begin{equation}
    ω(A τ_{iβ}B) = ω(BA) \label{eq:KMS_state}
\end{equation}
for all $A,B$ in a dense (with respect to the norm topology) $τ$-invariant $^*$-subalgebra of $\mathfrak{U}_τ$, where $\mathfrak{U}_τ$ is the set of entire analytic elements for $τ$.
\end{defi}  
At $D>1$, there is a unique KMS state at high enough temperatures \cite[Proposition 6.2.45]{bratteli_operator_1997}, while at D=1 the KMS state is unique at all temperatures \cite{Araki1975uniqueness}. This unique state is factor, as an immediate consequence of \cite[ Theorem 5.3.30]{bratteli_operator_1997}.  By the result of existence of space invariant KMS states \cite[pg. 296]{bratteli_operator_1997}, this unique state will also be space invariant.  Note that time invariance is an immediate consequence of the KMS condition, \Cref{eq:KMS_state}.

Further, it is guaranteed, again for $\beta<\beta_c$, that the state $\omega$ from Eq.~\eqref{gibbs} is exponentially clustering: there exists $\gamma>0$ such that for every local $A,B\in\mathfrak U_{\rm loc}$, there is $c>0$ such that
\beq\label{LRnonorm}
	|\omega(AB)-\omega(A)\omega(B)|\leq
	c e^{-\gamma \dist(A,B)}
\eeq
(the distance can be taken as the $L_1$ or $L_2$ distance -- as chosen for convenience -- on the square lattice between the supports of $A$ and $B$). In particular, $\omega$ satisfies the spatial mixing condition. More generally, any Kubo-Martin-Schwinger (KMS) state satisfying \eqref{clustering} satisfies also \eqref{basicomega} and the spatial mixing condition.

Finally, by the Lieb-Robinson bound \cite{Lieb:1972wy}, this exponential decay can be extended in space-time \cite{ampelogiannis_almost_2021}, to the full region outside of a Lieb-Robinson ``light-cone": there exists $v_{\rm LR}>0$ and $\gamma>0$ such that for every local $A,B$ and $v>v_{\rm LR}$, there is $c>0$ such that
\beq
	|\omega(A(t)B)-\omega(A)\omega(B)|\leq
	c e^{-\gamma (\dist(A,B)-vt)}.
\eeq
The structure in space-time is depicted in Figure \ref{fig2}.
\begin{figure}
    \centering
    \includegraphics[width=10cm]{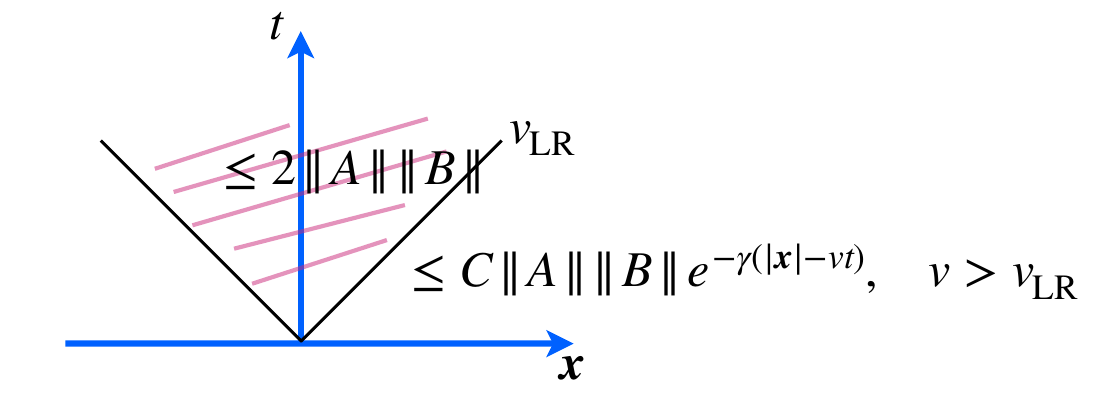}
    \caption{A pictorial representation of the basic bounds on correlation functions in space-time. The connected correlation function $\omega(A(\bs x,t)B)-\omega(A)\omega(B)$ is bounded by the operator norm in general, and by the sharper Lieb-Robinson bound outside of the Lieb-Robinson light-cone (here the latter is expressed more precisely than \eqref{LRnonorm} by including the operator norm).}
    \label{fig2}
\end{figure}

We note that in the $C^*$ algebra formulation, the concept of microcanonical shell, Eq.~\eqref{rhomc}, does not make immediate sense, especially in view of the ensemble equivalence \eqref{microgibbs}. Nevertheless, for the purposes of interpretation of the results, it will be convenient to consider any state $\psi$ that is ``near to" $\omega$ as being part of the microcanonical shell in the thermodynamic limit, $\psi\in\mathfrak S_{\rm micro}$. Here, ``near to" simply means ``absolutely continuous" with respect to $\omega$, that is, of the form
\beq\label{cstarmicro}
	\psi \in \mathfrak S_{\rm micro} \quad :\quad \psi(A) = \frc{\omega(B^\dag A B)}{\omega (B^\dag B)} \;\forall \;A\in\mathfrak U\quad
	\mbox{for some $B\in\mathfrak U$.}
\eeq
These states are still positive, continuous, normalised linear functionals, but not space-time translation invariant (which is of course as expected for states in the microcanonical shell). Further, by spatial mixing, the density of any extensive observable in $\psi$ is the same as it is in $\omega$:
\beq\label{microdensity}
	\lim_{L\to\infty}\frc1{|\Lambda_L|}\sum_{{\bs x}\in\Lambda_L} \psi(A(\bs x)) = \omega(A).
\eeq
In particular, the energy density is the same in $\psi$ as in $\omega$.

Finally, in the theory of $C^*$ algebras, a very important role is played by the Gelfand-Naimark-Segal (GNS) construction. In this construction, one associates to any state $\omega$ a Hilbert space $\mathcal H_{\omega}$, constructed from the observables in $\mathfrak U$, with in particular the vector $|\Omega\ket$ associated with the identity operator $\1$. This Hilbert space is such that $\omega(A) = \bra\Omega|A|\Omega\ket$, and more generally $\omega (B^*AC) = \bra B|A|C\ket$ \cite[Chapter 2.3]{bratteli_operator_1987} (here we use the same notation for $A\in\mathfrak U$ and its representation as a bounded operator on $\mathcal H_{\omega}$). More precisely, the Hilbert space is that arising from the completion of $\mathfrak U/\mathfrak N$ with respect to the inner product induced by the sesquilinear form $(A,B)\mapsto \omega(A^* B)$ on the equivalence classes $A + \mathfrak N\in \mathfrak U/\mathfrak N$, with null space $\mathfrak N$ (satisfying $\omega(B^* \mathfrak N)=0\;\forall\; B\in\mathfrak U$). Operators acting on $\mathcal H_{\omega}$ are sometimes called ``superoperators" in the condensed matter literature. For $\omega$ being the infinite-temperature state or trace state (see above), the resulting norm is in fact (the infinite-dimensional completion of) the  Hilbert–Schmidt norm, but the GNS construction is more general. Physically, in thermal states, $\mathcal H_{\omega}$ is the space of particle and hole excitations above the thermal ``vacuum".

Space and time translations can be extended to unitary group actions on the GNS representation, see also \cite[Theorem 2.3.16, Corollary 2.3.17]{bratteli_operator_1987}. Here we express the related theorem; below we will use the same notation $\tau_t$ and $\iota_{\bs x}$ for time and space translation unitaries on $\mathcal H_\omega$.
\begin{propo}[GNS representation] \label{prop:gns}
Given a state $ω \in E_{\mathfrak{U}}$ of a unital C$^*$ algebra $\mathfrak{U}$ there exists  a (unique, up to unitary equivalence)  triple $(H_ω,π_ω,Ω_ω)$ where $H_ω$ is a Hilbert space with inner product $\langle \cdot, \cdot \rangle$, $π_ω$ is a representation of the C$^*$ algebra by bounded operators acting on $H_ω$ and $Ω_ω$ is a cyclic vector for $π_ω$, i.e.\ the $\operatorname{span} \{ π_ω(A)Ω_ω : A \in \mathfrak{U}\}$ is dense in $H_ω$, such that
\begin{equation}
    ω(A) = \langle Ω_ω , π_ω(A) Ω_ω \rangle ,  \ \ A \in \mathfrak{U}.
\end{equation}
If additionally we have a group $G$ of automorphisms $\{ τ_g \}_{g \in G}$ of $\mathfrak{U}$ and $ω$ is $τ$-invariant, then there exists a representation of $G$ by unitary operators $U_ω(g)$ acting on $H_ω$. This representation is uniquely determined by 
\begin{equation}
    U_ω(g) π_ω(A) U_ω(g)^* = π_ω (τ_g (A)) \ , \ \ A \in \mathfrak{U}, g \in G
\end{equation}
and invariance of the cyclic vector
\begin{equation}
    U_ω(g)Ω_ω = Ω_ω \ , \forall g \in G.
\end{equation}
\end{propo}
See Appendix \ref{setup} for the full mathematical setup.

\subsubsection{Many-body ergodicity}\label{sssectmanybodyergo}

In order to look at ergodicity, in the classical realm we needed to consider the energy shell to which the initial condition belonged, and in the von Neumann quantum setup, we needed a small quantum energy shell. So, in the $C^*$ algebra context, it is natural to take a state $\omega$ assumed spacetime stationary and spatially mixing as described above, and states $\psi$ in its microcanonical shell, as in \eqref{cstarmicro}.

In the $C^*$-operator algebra context, one can still consider time-averaging \eqref{timeaverage} of operators (either as elements of the $C^*$ algebra, or bounded operators on $\mathcal H_\omega$), where time integration is now to be understood as a Bochner integral \cite{hille_functional_1996}. For clarity, we will write explicitly the long-time limits, hence we will use \eqref{timeaverageT}.

The basic formulation of many-body ergodicity is conceptually similar to that of subsection \ref{ssectergoclassical} (classical case), and paragraph \ref{sssectrecurrence} (quantum case):
\begin{theorem}\label{theomeansquare} {\em (Mean-square ergodicity.)}
Let $\omega$ be spacetime stationary and spatially mixing. Suppose that the unitary time evolution operator $\tau_t$ on the GNS space $\mathcal H_\omega$ has trivial kernel, $ \ker (\tau_s-1) = \C|\Omega\ket$ for some $s\in(0,\infty)$. Then for all integers $n\geq 1$,
\beq\label{manybodyergo}
	\lim_{T\to\infty} \omega\big(\,\big(\b{A}_T\big)^{n}\,\big)
	=\omega(A)^n.
\eeq
That is, we have mean-square ergodicity, and in fact the ``random variable" $\overline{A}$ is non-fluctuating within the state $\omega$.
\end{theorem}
The theorem follows from von Neumann's ergodic theorem for unitary operators on Hilbert spaces, as in paragraph \ref{sssectrecurrence}, but now with the GNS space $\mathcal H_\omega$ as the Hilbert space. Using von Neumann's ergodic theorem the time average gives a projection onto the subspce of $\mathcal H_\omega$ that is invariant under $τ_s$, for some $s \in (0,\infty)$, see also \cite[Theorem 3.1]{BergelsonDiscreteToContinuous}. If then  $ \ker (\tau_s-1) = \C|\Omega\ket$, this projection is the rank one projection onto $|Ω\rangle$, and the result is straightforward, as a direct application of the techniques used in \cite{ampelogiannis_almost_2021} (also discussed in \cite[Section 2]{fidaleo2014Nonconventional}). Relevant results exist in various forms, including \cite[Proposition 6.3.5]{ruelle_statistical_1974}, \cite{duvenhage2009Bergelson}, \cite{Fidaleo2009Ergodic}.

Clearly, as we have taken already the limit $L\to\infty$, the timescales for convergence are not controlled by the volume of the system, contrary to the phenomenon of quantum recurrence as discussed in paragraph \ref{sssectrecurrence}; in \eqref{manybodyergo} convergence of long-time averages happen much before the recurrence time. Yet, the techniques are essentially the same. What has happened? The most important aspect, as mentioned, is that in the $C^*$ algebra formulation, {\bf we have reduced the set of degrees of freedom to the relevant ones for local physics}. Thus, in contrast to the condition of non-degeneracy, {\bf the condition of triviality of the kernel of the time evolution properly encodes many-body ergodicity.}

In a formal sense, this kernel triviality condition is akin to asking that there be no nontrivial constants of the motion in the classical case, or to asking that there be no degeneracies in the quantum case. However, in a more physical sense, it is starkly different. In the classical context, for instance, integrability implies that there are nontrivial constants of the motion (as many as degrees of freedom). But in extensive systems, integrability is not related to the presence of nontrivial elements in $\ker (\tau_t-1)$. Instead, it is related to the presence of extensive conserved quantities in addition to the Hamiltonian. These span the kernel of a different evolution operator, that which acts not on the GNS space built from the local observables, but on a Hilbert space of extensive observables, see section \ref{sectproj}. Extensive conserved quantities may exist, the system may be integrable, and yet many-body ergodicity may still hold. Thus, {\bf we have separated the condition for many-body ergodicity to hold, from the requirement on nontrivial extensive conserved quantities, such as those found in integrable systems}.

The ``nontrivial constants of the motion" that should not be present for many-body ergodicity to hold, are the {\bf time-invariant local observables} (locality is here in the weak sense of elements of the $C^*$ algebra). Typically, except in special systems such as those with dynamical constraints, there are no such observables beyond (the trivial) multiples of $\1$. In fact, we note that time-invariant local observables play a r\^ole in the phenomenon of localisation -- thus many-body ergodicity occurs if there is no ``localisation" (in the sense that there is no time-invariant local observables). As mentioned, extensive conserved quantities, instead, play a r\^ole in integrability. And in fact, most importantly for us, they are at the basis of the emergent hydrodynamic structures, which we explain in Section \ref{sectproj}.

It turns out that one can go further if the state $\omega$ is also a KMS state, Eq.~\ref{eq:KMS_state}. In fact, it is not necessary for $H$ in \eqref{gibbs} to be the evolution Hamiltonian, or equivalently for $\tau_t$ in \eqref{eq:KMS_state} to be the dynamics we take for time averaging. Any other $H'$, any other dynamics $\tau_t'$, that keeps the state spacetime stationary will do -- for instance, any other extensive conserved quantity, if available. Thus, in integrable model, any generalised Gibbs ensemble \cite{rigol_relaxation_2007,ilievski_complete_2015,essler_quench_2016}.
\begin{theorem}\label{theoergo} {\em (Ergoditicy.)} Under the same conditions as those of Theorem \ref{theomeansquare}, if further $\omega$ is a $\tau'$-KMS state, then, for all $A\in\mathfrak U$,
\beq\begin{aligned}
	&\lim_{T\to\infty} \psi(\b{A}_T) = \omega(A)\ \forall \ \psi \in \mathfrak S_{\rm micro},\quad\mbox{and}\\
	&\lim_{T\to\infty} \b A_T = A\1\ 
	\mbox{as operator on $\mathcal H_\omega$, in the strong operator topology.}
	\end{aligned}
\eeq
\end{theorem}
Again, the proof is omitted, but follows quite directly from the techniques developed in \cite{ampelogiannis_almost_2021}.  Recall that $C_T\to 0$ in the strong operator topology means $|| C_T |\psi\ket||\to0$ for all $|\psi\ket\in\mathcal{H}_\omega$.

Now the result is much stronger: this is truly an ergodicity property, stating that the time average of any observable, in any state within a microcanonical shell surrounding $\omega$, gives the average in $\omega$. Time averaging gives back ensemble averaging. The operatorial equation, for $A$ as an operator on the GNS space $\mathcal H_\omega$, is in fact an equivalent statement.

Finally, an additional insight may be gained into many-body ergodicity by the following construction. Recall that in the classical case, ergodicity implies that every point in the energy shell is approached arbitrarily closely by the trajectory as time passes. Here, instead, with states playing the role of ``points in phase space", a trajectory starting from $\psi$ approaches arbitrarily closely the final state $\omega$: {\bf for every $\ep>0$, there is a finite time $t$ such that the time-evolved state $\psi\circ\tau_t$ is a distance less than $\ep$ from $\omega$, and in fact, this is true for a fraction of time that tends to 100\% over long periods.}

This construction starts with the choice of a sequence of unit-norm, strictly local observables $A_n \in \mathfrak U_{\rm loc},\,||A_n||=1,\,n=1,2,3,\ldots$ which spans $\mathfrak U_{\rm loc}$ (every element of $\mathfrak U_{\rm loc}$ may be written as a finite linear combination of these). Such a sequence certainly exists. Then, we define the following distance function between states:
\beq \label{eq:states_metric}
    d(\psi,\psi') = \sqrt{
    \sum_{n=1}^\infty 2^{-n}|\psi(A_n)-\psi'(A_n)|^2
    }\ \leq\  2. 
\eeq
We note that there is a lot of freedom in defining the distance function; in particular, any finite set of $C^*$-algebra elements $A_n'\in\mathfrak U,\,n=1,2,\ldots,N$ may be adjoined to the sequence $A_n$. Thus, the general result on the distance function can be used to deduce results for any finite-dimensional subspace of such elements.

Eq.~\eqref{eq:states_metric} indeed defines a distance function. It is simple to see that $d(\psi,\psi'') \leq d(\psi,\psi')+d(\psi',\psi'')$, and that if $\psi=\psi'$ then $d(\psi,\psi')=0$. Further, if $d(\psi,\psi')=0$, then $\psi=\psi'$. This is shown as follows. Suppose $d(\psi,\psi')=0$, and consider some $B\in\mathfrak U$. For every $\ep>0$ we can find $A\in\mathfrak U_{\rm loc}$ such that $||B-A||\leq \ep$. Therefore $|\psi(B)-\psi'(B) - (\psi(A)-\psi'(A))| \leq 2\ep$. As $d(\psi,\psi')=0$, then $\psi(A_n)=\psi'(A_n)$ for all $n$. As $A_n$'s span $\mathfrak U_{\rm loc}$, then also $\psi(A)=\psi'(A)$. Therefore by the triangle inequality, $|\psi(B)-\psi'(B)|\leq 2\ep$. As this holds for every $\ep>0$, we conclude that $|\psi(B)-\psi'(B)|=0$.

Now consider $d(\psi_t,\omega)^2$ where $\psi_t = \psi\circ \tau_t$ is the time-evolved state. We have
\beqa
    d(\psi_t,\omega)^2
    &=&
    \sum_{n=1}^\infty 2^{-n}|\psi(\tau_t A_n)-\omega(A_n)|^2 \\
    &=& \sum_{n=1}^{\infty} 2^{-n} \big( ψ^2(τ_tA_n) - 2 ψ(τ_tA_n) ω(A_n) + ω^2 (A_n) \big)
\eeqa
and we are interested in the long time limit of the time average:
\begin{equation} \label{eq:distance_average}
 \begin{array}{*3{>{\displaystyle}lc}p{5cm}}
 \lim_{T \to \infty}\lefteqn{\frc1T \int_0^T \dd t\,d(\psi_t,\omega)^2} && \\
   &=&\lim_{T \to \infty}\sum_{n=1}^\infty 2^{-n} \Big(\frc1T\int_0^T ψ^2(τ_t A_n) \, \dd t -2 ω(A_n)\frc1T\int_0^T ψ(τ_t A_n) \, \dd t+ ω^2(A_n) \Big).
\end{array}
\end{equation}
It is easy to see, by the dominated convergence theorem, that we can move the $\lim_{T \to \infty}$ past the summation $\sum_{n=1}^{\infty}$. We then want to apply \Cref{theoergo}. We can immediately do this for the second term: $ \lim_{T \to \infty} 2 ω(A_n)\frc1T\int_0^T ψ(τ_t A_n) \, \dd t = 2 ω^2(A_n)$.

For the first term, $\frc1T\int_0^T ψ^2(τ_t A_n)\, \dd t$ , we notice that its limit coincides with the limit of $ω(A)\frc1T\int_0^T ψ(τ_t A_n)\, \dd t$, by the following argument:
\beqa
 \big| \frc1T\int_0^T ψ^2(τ_t A_n)\, \dd t - ω(A)\frc1T\int_0^T ψ(τ_t A_n)\, \dd t\big| &= &
  \big| \frc1T\int_0^T ψ(τ_t A_n)( ψ(τ_t A_n)  - ω(A)\, \dd t\big|\\
  &\leq& \frc1T\int_0^T |ψ(τ_t A_n)| |ψ(τ_t A_n)  - ω(A)| \, \dd t 
\eeqa   
and $|ψ(τ_t A_n)| \leq \norm{A}$, while $\lim_{T \to \infty} \frc1T\int_0^T  |ψ(τ_t A_n)  - ω(A)| \, \dd t =0 $, by \Cref{theoergo}. Thus, 
\begin{equation}
\lim_{T \to \infty} \frc1T\int_0^T ψ^2(τ_t A_n)\, \dd t = \lim_{T \to \infty} ω(A)\frc1T\int_0^T ψ(τ_t A_n)\, \dd t = ω^2 (A_n) \ , \ \forall n
\end{equation}
 again by applying  \Cref{theoergo}. Finally, applying the limits in \Cref{eq:distance_average}, we show:
\beq\label{ergostate0}
    \lim_{T\to\infty} \frc1T \int_0^T \dd t\,d(\psi_t,\omega)^2 = 0.
\eeq
This implies the following.
\begin{theorem}
    Let $\psi\in\mathfrak S_{\rm micro}$ be a state in the microcanonical shell of $\omega$. Under the conditions of Theorem \ref{theoergo}, for every $\ep>0$, there exists $t>0$ such that $d(\psi_t,\omega)<\ep$. Moreover, $d(\psi_{t},\omega)<\ep$ for 100\% of the times $t$ (that is, the ratio of the Lebesgue measure for the set $\{t\in[0,T]:d(\psi_{t},\omega)\geq\ep\}$, to that for the set $\{t\in[0,T]:d(\psi_{t},\omega)<\ep\}$, tends to zero as $T\to\infty$).
\end{theorem}
\proof
We prove the last statement as it implies for first. Let us denote $\mu_T^+$ and $\mu_T^-$ the measures for the sets $\{t\in[0,T]:d(\psi_{t},\omega)\geq\ep\}$ and $\{t\in[0,T]:d(\psi_{t},\omega)<\ep\}$, respectively. Note that $\mu_T^+ + \mu_T^- = T$. Then $\frc1T \int_0^T \dd t\,d(\psi_t,\omega)^2\geq \ep^2\frc{\mu_T^+}T = \ep^2\Big(1+\frc{\mu_T^-}{\mu_T^+}\Big)^{-1}$, hence by \eqref{ergostate0} we must have $\lim_{T\to\infty} \Big(1+\frc{\mu_T^-}{\mu_T^+}\Big)^{-1} = 0$ and therefore $\lim_{T\to\infty}\frc{\mu_T^+}{\mu_T^-} = 0$.
\eproof

In particular, from the above theorem we conclude that, for any $A\in\mathfrak U$, we have $|\psi(\tau_t A) - \omega(A)|<\epsilon$ for 100\% of the times: the average in a state in a microcanonical shell, is almost always as near as desired to the average in the stationary state $\omega$.

Some remarks are in order:
\begin{enumerate}
\item Because the results apply to the $C^*$ algebra obtained from local observables, the results address {\em local relaxation}, a form of typicality (see e.g.~\cite{goldstein_normal_2009}) which expresses {\em many-body ergodicity}.
\item Note how both Theorems \ref{theomeansquare} and \ref{theoergo} apply not just to thermal states, but to more general spacetime stationary states, such as generalised Gibbs ensembles. This therefore takes into account the possibility that the model possesses a large number of {\em extensive conserved quantities} similar to the Hamiltonian; these would lead, in the finite-volume setup, to a large number of degeneracies. Here, no condition arises on the presence or not of extensive conserved quantities.
\item The microcanonical shell represented by the states $\psi$ should be understood as a being a shell around the microcanonical state {\em with respect to all extensive conserved quantities}, not just the energy: indeed all extensive variables have densities that are the same as in $\omega$, Eq.~\eqref{microdensity}. This is why ergodicity can hold in $\psi$ without conditions on the presence or not of extensive conserved quantities.
\item Requiring the KMS condition is not too strong a requirement. It is expected that most, or maybe all, spacetime stationary spatially mixing states are in fact KMS states. Perhaps the Tomita–Takesaki theory \cite{bratteli_operator_1997} could be used.
\item A similar $C^*$ algebra construction can be made for classical systems, such as a classical gas, where the limit of infinitely many particles is taken. We expect that similar ergodicity statements can be obtained. In the classical case, the GNS space can also be constructed, and time evolution is again -- at least in well-behaved systems -- a one-parameter unitary group. Thus, von Neumann's ergodic theorem for unitary operators applies. Again, as we have taken the limit of infinitely-many particles, timescales are not those that would be necessary to cover the full energy shell -- {\bf it is not true that over long times, the many-particle trajectory comes as close (in the conventional many-particle metric) as desired to a given point in the many-particle phase space.} Particle numbers are so large that the phase space is immense, and ``long times" are never enough.
\end{enumerate}

Thus we have solved the problem of obtaining ergodicity in many-body systems with timescales below the quantum recurrence timescale (in a different way than in von Neumann's QET, with different-looking conditions), or below the classical shell-covering timescales. We have also solved the problem of generalising to cases where nontrivial conservation laws exist (the extensive conserved quantities), and of having a formulation that is directly in the thermodynamic limit.

There remains the problem that the condition of triviality of the kernel of the time-evolution unitary one-parameter group on $\mathcal H_{\omega}$ is rather difficult to verify. It is also perhaps not too clear if the resulting ergodicity statement is the one that is most physically relevant. These two problems are solved by {\em almost-everywhere ergodicity}, to which we now turn.

\subsubsection{Almost-everywhere ergodicity}

We now obtain ergodicity results {\em which are valid in every quantum spin lattices} as described above, and {\em in every spacetime stationary spatially mixing states}. No special, hard-to-check condition is required.

Such general results however are not valid for the standard notion of ergodicity, with integration purely in the time direction. Instead, we must {\em look in spacetime}. On the lattice, it is natural to restrict to the set of rational spatial directions $\Srat^{D-1} = \{\boldsymbol x / |\boldsymbol x|:\boldsymbol x\in\Z^D\}$ (this is dense on the sphere). Given $T>0$, $v\in\R$ and $\boldsymbol n \in \Srat^{D-1}$, we consider the average of $A\in\mathfrak U$ {\em along the ray} with velocity vector $\boldsymbol v = v\boldsymbol n$,
\beq\label{average}
	\b A_T^{\boldsymbol v} = \frc1T\int_0^T \dd t\,A(\lfloor\boldsymbol vt\rfloor, t)
\eeq
where $\lfloor \boldsymbol a \rfloor = (\lfloor a_i\rfloor)_i$ is the vector of the integer parts. 

In fact, the results apply more generally. That is, instead of considering the flat averages, we {\em may modulate the average with oscillating factors}. So, we consider instead
\beq\label{average2}
	\b A_T^{\boldsymbol v} = \frc1T\int_0^T \dd t\, 
	e^{\ri f t - \ri \bs k \cdot \bs v t}
	A(\lfloor\boldsymbol vt\rfloor, t)
\eeq
(keeping the wavenumber $\bs k$ and frequency $f$ implicit in the notation for the average, for lightness of notation). Let us also denote
\beq
	E_A
	:= \lim_{T\to\infty}
	\omega(\b A_T^{\boldsymbol v})
	= \lt\{\ba{ll}
	\omega(A) & (f-\bs k\cdot \bs v = 0) \\
	0 & (\mbox{otherwise})
	\ea\rt.
\eeq
See Fig.~\ref{fig} for a pictorial representation of what is going on. We show in \cite{ampelogiannis_almost_2021}:
\begin{theorem}\label{t1}
For every rational direction $\boldsymbol n\in\Srat^{D-1}$ and almost every $v\in\R$ with respect to the Lebesgue measure: {\em (Mean-square ergodicity)}
\beq\label{ergo1}
	\lim_{T\to\infty} \omega( \big(\b A_T^{\boldsymbol v}\big)^m) = 
	E_A^m
\eeq
for every $A\in\mathfrak U$, $m\in\N$. If, further, $\omega$ is a KMS state, then: {\em (Ergodicity)}
\beq\label{ergo2}
    \lim_{T\to\infty} \psi(\b A_T^{\boldsymbol v}) = E_A ,\quad\lim_{T\to\infty} \b A_T^{\boldsymbol v} = E_A\mathds{1},\quad \lim_{T\to\infty} [\b A_T^{\boldsymbol v},B] = 0
\eeq
for every $\psi\in\mathfrak S_{\rm micro}$, and for every $A,B\in\mathfrak U$, in the strong operator topology.
\end{theorem}

Additionally, under the same assumptions of the Theorem above, and similarly to the derivation around \Cref{eq:distance_average}, one can show that for any $\psi\in\mathfrak S_{\rm micro}$ in the microcanonical shell of $\omega$, the ray-averaged distance (with respect to the metric of \Cref{eq:states_metric}) between $ψ$ and $ω$ tends to zero, for almost every ray. Denoting $ψ_{υt} \coloneqq ψ \circ ι_{\floor{υt}} τ_t$, we have:
\begin{equation}
   \lim_{T \to\infty} \frac1T \int_0^T \dd t \,d( ψ_{υt}, ω)^2  =0
\end{equation}

Taking the language of probability, Eq.~\eqref{ergo1} means ``$\b A_T^{\boldsymbol v}\to E_A$ in law". This is {\em almost-everywhere ergodicity}: along almost every velocity, $\b A_T^{\boldsymbol v}$ tends to a non-fluctuating quantity. As a consequence, by the Cauchy-Schwartz inequality, $\omega(\, \b A_T^{\boldsymbol v}B\,) \to E_A\bra B\ket$ and $\omega(\, [\b A_T^{\boldsymbol v},B]\,) \to 0$ for every $A,B\in\mathfrak U$.

Eqs.~\eqref{ergo2} are starker statements. They hold in spatially mixing, space-time translation invariant KMS states, including Gibbs and generalised Gibbs states. They imply that the limits in \eqref{ergo2} hold within any expectation value, multiplied by any other observables in any order. We thus obtain an extension of (a weak version of) the Lieb-Robinson bound {\em to within the Lieb-Robinson cone}: {\bf the ray-averaged operator $A_T^{\boldsymbol v}$ becomes ``thin" as $T\to\infty$, being un-observable by any $B$, at every frequency and wavelength, and at almost every velocity $v$}

Further, Theorem \ref{t1} means that {\bf time-averages, at every frequency and wavelength and almost every velocity $v$, converge to non-fluctuating (cluster out), classical (commutators vanish) variables.}

In particular, we have $\bra [\b A_T^{\boldsymbol v},B]^2\ket\to0$. At short times, the OTOC $\bra [A(\lfloor \boldsymbol v t\rfloor ,t), B]^2\ket$ is expected to grow exponentially in chaotic systems \cite{maldacena_bound_2016, khemani_velocity-dependent_2018} (eventually reaching $O(1)$ values by boundedness of the state). The velocities where this happens are bounded by a state-dependent ``butterfly velocity" $v_{\rm B}$, $|\boldsymbol v|<v_{\rm B}<v_{\rm LR}$. We find that the {\em time-averaged} version of this quantity, $\bra [\b A_T^{\boldsymbol v},B]^2\ket = T^{-2}\int_0^T \dd t\dd t'\,\bra [A(\lfloor \boldsymbol v t\rfloor ,t), B]\,[A(\lfloor \boldsymbol v t'\rfloor ,t'), B]\ket$, in fact decays for almost every velocity.

The results do not say anything about what happens in the pure time direction, $\boldsymbol v = \boldsymbol0$. For instance, there are non-interacting Hamiltonians for which the velocity $\boldsymbol v=\boldsymbol 0$ is not ergodic; as mentioned, this is in fact related to localisation. However, all results hold for velocities as near as desired to $\boldsymbol 0$.

Again, Theorem \ref{t1} is based on the von Neumann ergodic theorem \cite[Theorem II.11]{Simon_Reed_Functional}, which relates time averages to projections onto invariant spaces. As far as we know, before our work there were no nontrivial results concerning the invariant subspace of $\mathcal H_\omega$ for unitary time evolution in quantum lattice models. We show it to be spanned by $|\Omega\ket$ at almost every velocity. This happens because finite-dimensional local spaces (in fact, we believe that countable-dimensionality is sufficient) are too small to allow for operators to extend thickly on large distances over time: the set of speeds where the operator remains supported under averaging should be ``small enough".

Remarks 1 - 5 of paragraph \ref{sssectmanybodyergo} apply to almost-everywhere ergodicity results as well. We emphasise again that the results of Theorem \ref{t1} do not require any nontrivial condition on the interaction or the dynamics (such as chaos, non-degenerate eigenvalues, absence of nontrivial constants of the motion, non-localisation condition, etc.). The results are {\bf purely consequences of extensivity of the system} -- that is, the fact that the interaction range is short enough, and that the thermodynamic, large-volume limit has been taken.

Thus we have now addressed the question of the conditions under which notions of ergodicity may hold: almost-everywhere ergodicity only requires extensivity. However, we have not yet addressed the question of the physical relevance of this notion. Surprisingly, it turns out that almost-everywhere ergodicity is in fact closely related to the hydrodynamic structure emerging on large scales. We now turn to this important result.

\section{Hydrodynamic projections and spaces of extensive charges} \label{sectproj}
The hydrodynamic projection theorem says that the Euler-scale connected correlations decompose into  the conserved charges of the system. For its formulation, we will construct Hilbert spaces of extensive quantities: intuitively to every local observable $A\in \mathfrak{U}_{\rm loc}$ (density) we will associate an element, formally, $ΣA= \sum_{\bm x \in \Z^D}  A(\bm x)$, defined as an appropriate equivalence class\footnote{Note that this series does not converge in the $C^*$ algebra, but it converges weakly in the GNS space, with respect to the dense subspace corresponding to $\mathfrak U_{\rm loc}$.}. The Cauchy-completion of the space of such elements will give a Hilbert space of extensive quantities $\mathcal{H}_0$. Under appropriate clustering assumptions, we can define the time evolution as a unitary action on the Hilbert spaces of extensive quantities, and thus have an unambiguous definition of the conserved extensive charges $\mathcal{Q}_0$. This will be a closed subspace of $\mathcal{H}_0$, thus allowing us to define the orthogonal projection $\mathbb{P} : \mathcal{H}_0 \to \mathcal{Q}_0$.

The objects of interest are the time-averaged, long-wavelength Fourier transform of two-point connected correlations:
\begin{equation} \label{eq:spacetime}
    S_{ΣA,ΣB} (\bm{\kappa}) \coloneqq \widetilde{\lim_{T \to \infty}} \frac{1}{T} \int_0^T \dd t \, \sum_{\bm x \in \Z^D}e^{\ri \boldsymbol\kappa\cdot \boldsymbol x/t} (A (\bm x, t),B)
\end{equation}
where $(A,B) = \omega(A^\dag B) - \omega(A^\dag)\omega(B)$ is the sesquilinear connected correlation. We do not know how to show the existence of the limit $\lim_{T \to \infty}$, but we can show that $\sum_{\boldsymbol x\in\Z^D}\,e^{\ri \boldsymbol\kappa\cdot \boldsymbol x/t} (A(\boldsymbol x,t), B)$ is uniformly bounded. This allows us to use the notion of a Banach limit $\widetilde{\lim_{T \to \infty}}$, see \cite[Chapter III.7]{conwayFunctionalAnalysis2007} and \cite[Appendix A]{doyon_hydrodynamic_2022}. We can also show that the result is indeed a function not merely of $A,\,B$, but in fact of the equivalence classes $\Sigma A,\,\Sigma B$, as the notation suggests; and that it is continuous, on both variables, with respect to the norm $||\cdot||_0$ on $\mathcal H_0$. The essence of the hydrodynamic projection is that inside correlation functions every extensive observable will project to a conserved charge, irrespectively of the choice of a Banach limit:
\begin{equation}\label{hydroprojstandard}
    S_{\Sigma A,\Sigma B}(\boldsymbol\kappa) = S_{\mathbb{P}(\Sigma A),\mathbb{P}(\Sigma B)}(\boldsymbol\kappa).
\end{equation}
This is an expression of the Boltzmann-Gibbs principle \cite{doyon_drude_2017,doyoncorrelations,nardis_correlation_2021}: the reduction of the number of degrees of freedom at large space-time separations by projections over hydrodynamic modes. The physical idea is that the initial, dynamically complicated disturbance quickly relaxes and projects, at the Euler scale of long time and large distances, onto the extensive conserved quantities, that then carry correlations. This occurs thanks almost-everywhere ergodicity, as is made apparent in the proof of Theorem \ref{hydroprojection} in \Cref{section:proof} (see \cite{doyon_hydrodynamic_2022} for proof in D=1).

We believe this is the first general, rigorous result concerning the Boltzmann-Gibbs principle in deterministic interacting systems of arbitrary dimensions. The principle is applicable to the large class of quantum lattices, going beyond interacting particle systems conventionally studied in statistical physics and hydrodynamics. It provides further support to the idea that the basic principles of hydrodynamics hold independently from the details of the microscopic dynamics.

Surprisingly, the hydrodynamic projection can be generalised to describe oscillatory behaviours: to any frequency $f \in \R$, and wavenumber $\bm k \in \R^D$.  We show that the  projection still takes place for $\bm k$-extensive quantities, formally $\Sigma^{\boldsymbol k} A = \sum e^{-\ri {\boldsymbol k}\cdot {\boldsymbol x}} A(\bs x)$, which will project to $(f,\bm k)$-conserved charges, defined as those $\Sigma^{\boldsymbol k} A$ such that their time evolution take the form $ \Sigma^{\boldsymbol k} A (t) =e^{\ri ft} \Sigma^{\boldsymbol k} A$.

Oscillatory hydrodynamic projection describes oscillatory behaviours that emerge at large space and time separations in correlation functions. One in general expects that $(f,\bm k)$-conserved charges only exist for certain frequency-wavenumber pairs $(f,\bm k)$, depending on the specific model.  A simple example is the free fermionic lattice, with some dispersion relation $E(\boldsymbol k)$. In this system, creation and annihilation operators at momentum $\boldsymbol k$ are $(E(\boldsymbol k),\boldsymbol k)$-extensive conserved quantities. In Section \ref{sectosci}, in order to illustrate the phenomemon we will explain (in a non-rigorous fashion) how oscillatory hydrodynamic projection recovers the oscillatory algebraic decay of fermion two-point functions in a free fermionic quantum chain, in agreement with a saddle-point analysis.

In the present section, we concentrate on the rigorous and general results of hydrodynamic projections in quantum lattice models.


\subsection{Clustering and basic assumptions} \label{section:clustering}
Throughout all of \Cref{sectproj} we consider a dynamical system $(\mathfrak{U},ι,τ)$, see \Cref{defn:dynamicalsystem} and Appendix \ref{setup} for more details. We use the notation $A(\bm x,t)$ for space-time translated observables $ι_{\bm x}τ_t A$. We also denote $A(\bm x)=A(\bm x,0)$, while we keep the notation $τ_tA$ for pure time translations. 
The system will be in a space-time invariant state $ω$, \Cref{defn:state-invariance}, which will be assumed to have clustering properties with respect to space translations. The almost-everywhere ergodicity theorems shown in \cite{ampelogiannis_almost_2021}, and discussed in \Cref{ssectalmost}, require the state $ω$ to be clustering in space, in particular $\lim_{\boldsymbol x\to\infty}\big( ω(  A(\bm x) B)  - ω( A)ω( B) \big)=0$ for any $A,B\in\mathfrak U$. Almost-everywhere ergodicity holds no matter how fast the connected correlations of observables decay at large spatial separations.  For the hydrodynamic projection to occur stronger clustering assumptions need to be imposed. In particular, clustering of correlations at large space separations will need to happen faster than $|\bm x|^p$, for $p$ large enough:
\begin{equation}
    \lim_{\boldsymbol x\to\infty}|\bm x|^p\big( ω(  A(\bm x) B)  - ω( A)ω( B) \big)=0.
\end{equation}
Additionally we will need to control the dependence on the size of the supports of the observables in the clustering bound; the support of a local observable $A$ is defined as 
$Λ_{A} \coloneqq \bigcap \{ X \subset \Z^D : A \in \mathfrak{U}_X\} $. That is because time evolving observables ``spreads out'' their support. A state satisfying this form of clustering will be  called $r$-sizeably $p$-clustering:
\begin{defi}[p-clustering] \label{defn:mixing}
A state $ω$ of a dynamical system  $(\mathfrak{U},ι,τ)$  is called $r$-sizeably $p$-clustering for $p>r>0$ if there exists a constant $u$ such that  for any  $A,B \in \mathfrak{U}_{\rm loc}$ 
\begin{equation}
    |ω(A(\bm x)B) - ω(A)ω(B) | \leq C_{A,B} (1+ |\boldsymbol x|)^{-p} , \text{ for all $\bm x \in \Z^D$}
\end{equation}
and $C_{A,B}$ is bounded as:
\begin{equation}
   C_{A,B} \leq u \norm{A} \norm{B} |Λ_A|^r |Λ_B|^r.
\end{equation}

A subset $\mathfrak{C} \subset \mathfrak{U}$ is called uniformly $p$-clustering for $p>D$ if there exist a $C>0$ s.t. for all $A,B \in \mathfrak{C}$:
\begin{equation} \label{eq:uniformclustering}
    | ω( A(\boldsymbol x) B )- ω(A) ω(B)| \leq C (1 + |\boldsymbol x|)^{-p} , \ \ \forall \boldsymbol{x} \in \Z^D.
\end{equation}
\end{defi}
Of course $C_{A,B}$ will, in general, depend on the pair of observables, i.e.\ on their norm and the size of their support.   In  $D=1$ quantum spin chains, \cite{doyon_hydrodynamic_2022}, the state is assumed to be $r$-sizeably $p$-clustering; this notion originates from \cite[Definition 4.2]{doyon_thermalization_2017}. 
The definition of $p$-clustering, for every $p>0$, includes high temperature KMS states, which are exponentially clustering in space, as shown in \cite{frohlich_properties_2015, Eisert_locality}.

\begin{defi} \label{defn:expclustering}
A state $ω$ of a dynamical system  $(\mathfrak{U},ι,τ)$  is called exponentially clustering  if there exist constants $u,r,λ>0$ such that for any  $A,B \in \mathfrak{U}_{\rm loc}$ 
\begin{equation}
    |ω(A(\bm x)B) - ω(A)ω(B) | \leq k_{A,B} e^{-λ |\bm x|} \text{, for all $\bm x \in \Z^D$.}
\end{equation}
and
\begin{equation}
    k_{A,B}< u \norm{A} \norm{B} |Λ_A|^r |Λ_B|^r
\end{equation}
Uniform (exponential) clustering of a subset $\mathfrak{C} \subset \mathfrak{U}$ is defined similarly.
\end{defi}

In order for the time evolution to be well defined as a unitary group action on the Hilbert spaces of extensive quantities, we need to
assume that the time translations of local observables  cluster in a uniform enough manner. First, we require that any element $A$ in ${\rm span} \{ τ_t A : A \in  \mathfrak{U}_{\rm loc}, t \in \mathbb{R} \}$ is approximated by a sequence $σ_nA$ of local elements. We will show that this follows from the Lieb-Robinson bound. Additionally, we require that for any local $A,B$ the set of pairs $\{ (\sigma_n A, \sigma_n B ) \}$ is uniformly clustering in space, as per \Cref{eq:uniformclustering}.
Combining spatial clustering  of the state $ω$ with the Lieb-Robinson bound, one can obtain space-like $p_c$ clustering, defined as follows:
\begin{defi} \label{defn:spacelikeclustering}
The dynamical system $(\mathfrak{U},ι,τ)$ in the state $ω$ is called space-like $p_c$-clustering  with velocity $υ_c$, if
\begin{enumerate}
    \item $\forall A \in \hat{ \mathfrak{U}}_{\rm loc} = {\rm span} \{ τ_t A : A \in  \mathfrak{U}_{\rm loc}, t \in \mathbb{R} \}$ there exists a sequence $ \sigma_n A \in \mathfrak{U}_{\rm loc}$, $n \in \mathbb{N}$, such that 
$\lim σ_n A = A$. For any $A \in  \mathfrak{U}_{\rm loc}$ we define $\sigma_n A = A$, $\forall n$. \label{spacelike1}
    \item  $\forall A,B \in \hat{ \mathfrak{U}}_{\rm loc}$ the set of pairs $\{ (\sigma_n A, \sigma_n B ) \}$ is uniformly $p$-clustering for some $p>p_c$. \label{spacelike2}
    \item $\forall A,B \in \mathfrak{U}_{\rm loc}$ there exist $p> p_c$, $0<V<υ_c$ and $C_{A,B}>0$ such that
\begin{equation}
\big| ω( A(\bm x,t)  B ) - ω(A)ω(B)  \big|\leq  \frac{C_{A,B}}{ ( |\boldsymbol x| +1)^p} \label{eq:spacelikeclustering}
\end{equation}
for all $\boldsymbol x\in\Z^D$, $|\boldsymbol x|\geq V|t|$. \label{spacelike3}
\end{enumerate}
Similarly, we define exponential space-like clustering by the same conditions, with Condition 2 replaced by uniform exponential clustering and \Cref{eq:spacelikeclustering} replaced by  $\big| ω( A(\bm x,t) , B) - ω(A)ω(B)  \big|\leq C_{A,B} {\rm e}^{-λ |\bm x|}$, for some $λ>0$. 
\end{defi}


\textbf{Space-like $p_c$-clustering for $p_c>D$}, where $D$ is the lattice dimension, \textbf{is a sufficient condition for showing the hydrodynamic projection theorem}.   
In turn, for space-like $p_c$-clustering to hold, it is sufficient to have the following two conditions:
\begin{enumerate}
    \item fast decaying interactions, satisfying \Cref{eq:interaction}, so that the Lieb-Robinson bound holds
    \item  a $p$-clustering (spatially mixing) state $ω$, \Cref{defn:mixing}, with appropriately controlled growth of $C_{A,B}$ with respect to $|Λ_A|$, $|Λ_B|$.
\end{enumerate}
This is realised, for example,  in any quantum spin lattice with nearest-neighbor, finite range, or exponentially decaying two-body interactions that is in a high temperature thermal state. This is because high temperature thermal states are exponentially clustering \footnote{Note that the constant $C_{A,B}$ grows linearly with $|Λ_{A}|$, $|Λ_{B}|$ in the case of exponential clustering of KMS states}, \cite[Theorem 3.2]{frohlich_properties_2015} \cite[Theorem 2]{Eisert_locality} . In the case of $r$-sizeably $p$-clustering states \cite[Definition 4.2]{doyon_thermalization_2017}, we can  extend the arguments in \cite[Section 8]{doyon_hydrodynamic_2022} to arbitrary lattice dimension $D$, to show that whenever $p>D(r+1)$, a $r$-sizeably $p$-clustering state will be space-like $p_c$-clustering for $p_c>D$, in any system with interaction that satisfies \Cref{eq:interaction}. This is slightly more technical in $D>1$, but it largely follows the same ideas as in the $D=1$ proof in \cite{doyon_hydrodynamic_2022}, hence we omit it. Instead, in Appendix \ref{appC} we show how exponential space clustering and the Lieb-Robinson bound lead to  exponential space-like clustering.

\subsection{Spaces of extensive charges} \label{section:charges}

We start by constructing Hilbert spaces of extensive quantities from the dynamical system $(\mathfrak{U},ι,τ)$. For each wavenumber $\bm k\in \mathbb{R}^D$ we define the positive-semidefinite sesquilinear form 
\beq\label{inner}
    \langle A,B\rangle_{\boldsymbol k} = \sum_{\boldsymbol x\in \Z^D}\,
    \re^{\ri \boldsymbol k \cdot \boldsymbol x} ( A(\boldsymbol x), B)
        \quad A,B\in\mathfrak U_{\rm loc},
\eeq
 where we recall that $(A,B) = \omega(A^\dag B) - \omega(A^\dag)\omega(B)$ is the sesquilinear connected correlation. We define the equivalence relation $A \sim^{\bm k} A^{\prime}$ on $\mathfrak{U}_{\rm loc}$  by $\langle A'-A,A'-A\rangle_{\boldsymbol k}=0$. The Hilbert spaces of extensive quantities are the norm-completion of the quotient spaces  formed of the set of equivalence classes of $\sim^{\bm k}$. The equivalence class of $A \in \mathfrak{U_{\rm loc}}$ is $Σ^{\bm k} A \coloneqq \{ A^{\prime}\in \mathfrak{U_{\rm loc}} : A^{\prime} \sim^{\bm k}A \}$ and it is to be understood as the $\bm k$-extensive observable associated to the density $A$. Formally, we can associate it with the infinite series $Σ^{\bm k} A = \sum_{\bm x \in \Z^D} e^{-i \bm k \cdot \bm x} A(\bm x)$, the ``total'' $A$ of the full quantum lattice. Of course this series does not converge within $\mathfrak{U}$, but it does converge weakly in the GNS representation associated to the state: this weak limit coincides with the above definition of $Σ^{\bm k} A$. The set of equivalence classes is $\mathcal{V}_{\bm k} \coloneqq \mathfrak{U_{\rm loc}} ~/ \sim^{\bm k} = \{ Σ^{\bm k}A : A \in \mathfrak{U_{\rm loc}}\}$ and its Cauchy completion gives the Hilbert space of $\bm k$-extensive observables  $\mathcal{H}_{\boldsymbol k} = \overline{\mathcal{V}}_{\bm k}$.
 
 In order to examine the long time dynamics of these extensive observables we have to extend the action of time evolution $τ_t,\,t\in\R$ to the Hilbert spaces $\mathcal H_{\boldsymbol k}$. This is done rigorously in \cite[Section 5.3]{doyon_hydrodynamic_2022} for $D=1$ (spin chains) and can be immediately extended to arbitrary dimension $D$. The key in order to be able to do this is the uniform clustering condition 2 in \Cref{defn:spacelikeclustering}. Thus, we can  show that time evolution acts as a unitary operator $τ_t^{\bm k}$ on $\mathcal H_{\boldsymbol k}$. For lightness of notation, we omit the superscript and simply write $τ_t$ for the unitary action of time evolution on these spaces.
 
 With this construction, we can define the subspace of conserved extensive charges as those elements of $\mathcal H_{\boldsymbol k}$
 that are invariant under $τ_t$:
 \begin{equation}
     \mathcal{Q}_{\bm k} = \{ q \in \mathcal{H}_{\bm k} : τ_t q = q, \forall t \in \R \}.
 \end{equation}
 In fact, we can go one step further and define the subspace of $(f,\boldsymbol k)$-extensive conserved quantities (or $f$-oscillatory $\bs k$-extensive charges) as:
\begin{equation}\label{Qfk}
\mathcal Q_{(f,\boldsymbol k)} = \{\mathcal a\in \mathcal H_{\boldsymbol k}: τ_t \mathcal a = e^{-\ri f t} \mathcal a ,\forall t\in\R\}.
\end{equation}
See Section \ref{sectosci} for a discussion of these charges and examples in a free fermion chain.
It is immediate that $\mathcal Q_{(f,\boldsymbol k)}$ is a closed subspace of $\mathcal{H}_{\bm k}$, hence there is an orthogonal projection
\begin{equation}
    \mathbb P_{(f,\boldsymbol k)}: \mathcal{H}_{\boldsymbol k} \to \mathcal Q_{(f,\boldsymbol k)}.
\end{equation}
 
\subsection{A hydrodynamic projection theorem}
The correlation functions with $(f,\boldsymbol k)$-fluid-cell averaging are given by
\begin{equation} \label{spacetime}
    S_{\Sigma^{\boldsymbol k} A,\Sigma^{\boldsymbol k} B}^{(f,\boldsymbol k)}(\boldsymbol\kappa) := \widetilde{\lim_{T\to\infty}} \frc1T \int_0^T \dd t\,
    \sum_{\boldsymbol x\in\Z^D}\,\re^{\ri \boldsymbol k \cdot \boldsymbol x - \ri f t}\re^{\ri \boldsymbol\kappa\cdot \boldsymbol x/t} ( A(\boldsymbol x,t), B).
\end{equation}
Note how one extracts, thanks to the time integral and factor $e^{-ift}$, the time-oscillatory behaviour of the correlation function with frequency $f$. Note also how the full wavenumber is $\bs k + \bs\kappa/t$, representing, in the large-time limit, a long-wavelength modulation of a $\bs k$-oscillatory factor; this extracts the space-oscillatory behaviour with wavenumber $\bs k$. 

The limit on $T$ is in general a Banach limit \cite[Chapter III.7]{conwayFunctionalAnalysis2007}, \cite[Appendix A]{doyon_hydrodynamic_2022}. The result will hold irrespectively of the choice of Banach limit, and all the aspects of the proof regarding the Banach limit are the same as in $D=1$, as shown in \cite[Section 6]{doyon_hydrodynamic_2022}. As the notation implies, the result only depends on the equivalence classes $\Sigma^{\boldsymbol k} A,\Sigma^{\boldsymbol k} B$, which we'll denote with the respective lowercase letters $\mathcal{a}, \mathcal{b}$. In fact this defines a continuous sesquilinear form on $\mathcal H_{\bm k}$. The hydrodynamic projection theorem is rigorously stated as follows:  

\begin{theorem}[Hydrodynamic Projection]\label{hydroprojection}
Consider a dynamical system $(\mathfrak{U},ι,τ)$ with interactions satisfying \Cref{eq:interaction}, in an $r$-sizeably $D(r+1)$-clustering  state $ω$, as per \Cref{defn:mixing}.  For every frequency-wavenumber pair $(f,\bm k) \in \R \times \R^D$, rational vector $\boldsymbol\kappa\in \R^D : \exists r\in\R \,|\,r\boldsymbol k \in\Z^D$, and any $\bm k$-extensive elements $\mathcal a,\mathcal{b} \in \mathcal H_{\boldsymbol k}$,
\beq\label{proj}
    S_{\mathcal a,\mathcal b}^{(f,\boldsymbol k)}(\boldsymbol\kappa) =
    S_{\mathbb P_{(f,\boldsymbol k)}\mathcal a,\mathbb P_{(f,\boldsymbol k)}\mathcal b}^{(f,\boldsymbol k)}(\boldsymbol\kappa).
\eeq
Specifically for the simple case $(f,\bm k)=(0, \bm 0)$ we have
\begin{equation}
     S_{\mathcal a,\mathcal b} (\bm \kappa) =
    S_{\mathbb P\mathcal a,\mathbb P \mathcal  b}(\bm \kappa).
\end{equation}
\end{theorem}
The case $\boldsymbol \kappa=0$ is of special interest, and is known at the (oscillatory version of the) Drude weight:
\begin{corol} The oscillatory Drude weight
\begin{equation}
    \mathsf D_{\Sigma^{\boldsymbol k} A,\Sigma^{\boldsymbol k} B}^{(f,\boldsymbol k)} := \lim_{T\to\infty} \frc1T \int_0^T \dd t\,\sum_{\boldsymbol x\in\Z^D} \re^{\ri \boldsymbol k \cdot \boldsymbol x - \ri f t} ( A(\boldsymbol x,t), B)
\end{equation}
satisfies the projection formula
\begin{equation}
    \mathsf D_{\mathcal a,\mathcal b}^{(f,\boldsymbol k)} = \mathsf D_{\mathbb P_{(f,\boldsymbol k)}\mathcal a,\mathbb P_{(f,\boldsymbol k)}\mathcal b}^{(f,\boldsymbol k)} .
\end{equation}
\end{corol}
Here the limit is in fact an ordinary limit. The projection formula, when written explicitly in a basis decomposition, is exactly (an oscillatory version of) the so-called Mazur bound, here shown rigorously to be saturated. This is a very immediate extension to the oscillatory realm, and to arbitrary dimensions, of the result \cite[Thm 6.1]{doyon_hydrodynamic_2022}, and can be proven quite straightforwardly without the extensive machinery we develop below for dealing with the case $\boldsymbol \kappa\neq0$, see the proof of \cite[Thm 6.1]{doyon_hydrodynamic_2022} and the discussion there for more details.

We proceed with the proof of the hydrodynamic projection theorem. We will first prove it for $(f,\bm k)=(0, \bm 0)$, and then easily generalise the proof to arbitrary $(f,\bm k)$. We note again that for $\bs\kappa=\bs 0$, the theorem is a straightforward application of von Neumann’s
ergodic theorem for unitary operators. We thus concentrate on the case $\bs\kappa\neq\bs 0$.

\subsection{Proof of hydrodynamic projection} \label{section:proof}
\textbf{The main idea for the proof of the hydrodynamic projection formula in quantum spin lattices is this}: By using an appropriate geometric construction, we recast the $D$-dimensional problem of hydrodynamic projection into the a 1-dimensional problem. This is done by identifying the summation
over space coordinates in a plane perpendicular to the wavenumber direction, with a sesquilinear form and its new associated Hilbert space, which is to play the role of the sesquilinear correlation
    in an effective one-dimensional problem. Once this is done, the proof follows that of \cite{doyon_hydrodynamic_2022} done for $D=1$. The schematics of our proof is as follows: We assume a dynamical system in a $p$-clustering state $ω$, \Cref{defn:mixing} and with sufficiently fast decaying interactions, \Cref{eq:interaction}, so that the Lieb-Robinson bounds, \Cref{eq:liebrobinsonbound}, holds. We combine $p$-clustering with the Lieb-Robinson bound (proof in Appendix \ref{appC}) in order to obtain space-like $p_c$-clustering, \Cref{defn:spacelikeclustering}. This leads to the special Property \ref{PROPERTY}, below, for the dynamical system. Using this property we can prove Lemma \ref{thelemma}, which forms the basis for the hydrodynamic projection formula, Theorem \ref{hydroprojection}.

The most important difference from the $D=1$ case is the new geometric constructon, which we now explain. We define the rational unit sphere in $D$-dimensional space as $\mathbb S_{\mathbb Q}^{D-1} = \{\boldsymbol \kappa/|\boldsymbol \kappa|:\boldsymbol \kappa\in\Z^D\}$. We also denote the set of vectors in rational directions in $\R^D\setminus \{\boldsymbol 0\}$ by $\R^D_{\mathbb Q} = \{\boldsymbol \kappa \in \R^D\setminus\{\bs 0\} : \exists r\in\R \,|\,r\boldsymbol \kappa \in\Z^D\}$. Clearly, by definition, $\mathbb S_{\mathbb Q}^{D-1}\subset \R^D_{\mathbb Q}$. For any such vector $\boldsymbol \kappa\in \R^D_{\mathbb Q}$, there is a unique unit vector $\widehat{\boldsymbol \kappa}' = \pm\boldsymbol \kappa/|\boldsymbol \kappa|\in \mathbb S_{\mathbb Q}^{D-1}$ that lies on the half-unit sphere: such that $\widehat \kappa_1'> 0$, or $\widehat \kappa_1'= 0,\,\widehat \kappa_2'>0$, or $\cdots$, or $\widehat \kappa_1'=0,\ldots,\widehat \kappa_{D-1}'=0,\, \widehat \kappa_D'>0$. Then, given such a unit vector $\widehat{\boldsymbol \kappa}'$, we denote by $\widehat{\boldsymbol \kappa} = r \widehat{\boldsymbol \kappa}'$ the ``smallest" integer vector associated to it, where $r>0$ is the smallest positive number such that $\widehat{\boldsymbol \kappa} \in \mathbb Z^D$ ($\widehat{\boldsymbol \kappa}$ exists because $\widehat{\boldsymbol \kappa}'\in \R^D_{\mathbb Q}$).  Clearly $\widehat{r\boldsymbol \kappa} = \widehat{\boldsymbol \kappa}$ for every $r\in\R$, and in fact, $\widehat{\boldsymbol \kappa}$ is an element of $\Z^D$ that identifies the ``rational ray" on which $\boldsymbol \kappa$ lies: there is a bijection between the set of $\widehat{\boldsymbol \kappa}$'s and the set of $\{r\boldsymbol \kappa:r\in\R\}$'s for $\boldsymbol \kappa\in\R^D_{\mathbb Q}$. Below we will refer to $\widehat{\boldsymbol \kappa}$ as a rational ray, and it will be important that it lies in $\Z^D$.
For any rational ray $\widehat{\boldsymbol \kappa}$, we may build its perpendicular plane in $\Z^D$: a sublattice of $\Z^D$ that is isomorphic to $\Z^{D-1}$, and whose vectors have vanishing vector dot product with $\widehat{\boldsymbol \kappa}$. Given $\widehat{\boldsymbol \kappa}$, there is a set (possibly empty) of zero components $\zeta = \{i: \widehat \kappa_i = 0\}$ and its complement (always non-empty) $\bar\zeta = \{1,\ldots,D\}\setminus \zeta = \{j_1,\ldots,j_d\}$ for unique $1\leq d = |\b\zeta|\leq D$ and $j_1<\ldots<j_d$. We construct the vectors
\beqa
    \boldsymbol h^{(i)}\in\Z^D&:&
    h^{(i)}_m = \delta_{i,m} \quad (m\in\{1,\ldots,D\},\,i\in\zeta)\n
    && h^{(j_l)}_{j_l} = \widehat \kappa_{j_{l+1}},\ 
    h^{(j_l)}_{j_{l+1}} = -\widehat \kappa_{j_l},\ 
    h^{(j_l)}_{m} = 0 \quad (l\in \{1,\ldots,d-1\}, m\in \{1,\ldots,D\}\setminus\{j_l,j_{l+1}\}).\n
\eeqa
This defines $D-1$ vectors: all $\boldsymbol h^{(i)}, i=1,\ldots,D$, except $\boldsymbol h^{(j_d)}$ which has not been defined. If $j_d \neq D$, then $D\in\zeta$,     and for convenience, in this case, we simply define $\boldsymbol h^{(j_d)} = \boldsymbol h^{(D)}$ as well as $\t\zeta = \zeta \setminus \{D\}\cup \{j_d\}$ (otherwise $\t\zeta = \zeta$). Thus we may always concentrate on $\boldsymbol h^{(i)}:i=1,\ldots,D-1$.

By construction, the set $\{\boldsymbol h^{(i)}:i=1,\ldots,D-1\}$ is linearly independent, and satisfy $\boldsymbol h^{(i)} \cdot \widehat{\boldsymbol \kappa} = \boldsymbol h^{(i)}  \cdot \boldsymbol \kappa = 0$. The perpendicular plane is $\Z^D\supset \mathbb H_{\widehat{\boldsymbol \kappa}}^{D-1} \coloneqq {\rm span}_{\Z}(\boldsymbol h^{(1)},\ldots,\boldsymbol h^{(D-1)})\simeq \Z^{D-1}$, i.e.\ $\mathbb H_{\widehat{\boldsymbol \kappa}}^{D-1}$ is the $\mathbb{Z}$-module freely generated by the $\bm h^{(i)}$. The isomorphism between $\mathbb{Z}^{D-1}$ and $\mathbb H_{\widehat{\boldsymbol \kappa}}^{D-1} $, as $\Z$-modules, is clear by  construction, in particular
\beq
     \mathbb H_{\widehat{\boldsymbol \kappa}}^{D-1} = \{ x_i\boldsymbol h^{(i)}:{\boldsymbol x}\in\Z^{D-1}\}
\eeq
(with implied summation over repeated indices). Of course, we could have taken away the greatest common divisor of the components of $\boldsymbol h^{(i)}$ in order to have ``denser" planes, but this is not necessary in the following construction.

By stacking the perpendicular planes parallely, we obtain a sublattice of $\Z^D$ which is itself isomorphic to $\Z^D$, defined as the $\Z$-module freely generated by the set $\{ \bm h^{(i)},\widehat{\bm \kappa} \}$:
\beq
    \Z_{\widehat{\boldsymbol \kappa}}^D \coloneqq \{x_i\boldsymbol h^{(i)}+z\widehat{\boldsymbol \kappa}:(\boldsymbol x,z)\in\Z^D\}.
\eeq
In-between the points of this lattice lie ``foundamental cells". That is, consider the region $R_{\widehat{\boldsymbol \kappa}} =\{x_i\boldsymbol h^{(i)}+z\widehat{\boldsymbol \kappa} : x_i\in[0,1)\forall i,\,z\in[0,1)\}\subset\R^D$ and the cell in $\Z^D$ given by
\beq
    \Lambda_{\widehat{\boldsymbol \kappa}} = \Z^D\cap R_{\widehat{\boldsymbol \kappa}} = 
    \Z^D \cap \{x_i\boldsymbol h^{(i)}+z\widehat{\boldsymbol \kappa}
    : x_i\in[0,1)\forall i,\,z\in[0,1)\}.
\eeq
Then the union of all unit cells shifted by the sublattice gives back $\Z^D$:
\beq\label{ZDdecomp}
    \Z^D = \bigcup_{\boldsymbol r\in \Z_{\widehat{\boldsymbol \kappa}}^D} \Big(\Lambda_{\widehat{\boldsymbol \kappa}} + \boldsymbol r\Big).
\eeq
Indeed, by linear independence and the fact that the Jacobian $J$ is finite and nonzero, every $\boldsymbol u\in\R^D$ may be written in a unique way as $\boldsymbol u = x_i\boldsymbol h^{(i)}+z\widehat{\boldsymbol \kappa}$ for some $(\boldsymbol x,z)\in\R^D$, and thus $\boldsymbol u = \floor{x_i}\boldsymbol h^{(i)}+\floor{z}\widehat{\boldsymbol \kappa} + \boldsymbol\lambda$ where $\boldsymbol\lambda \in R_{\widehat{\boldsymbol \kappa}}$. This holds in particular for $\bm u\in\Z^D$, and since $\floor{x_i}\boldsymbol h^{(i)}+\floor{z}\widehat{\boldsymbol \kappa}\in\Z^D_{\widehat{\boldsymbol \kappa}}$ it gives \eqref{ZDdecomp}.

We define for any wavenumber $\boldsymbol \kappa\in\R^D_{\mathbb Q}$ and $\boldsymbol \kappa'\in\R^D$, some positive-semidefinite sesquiliniear forms, and associated Hilbert spaces, in a way similar to that done above: first by summation over the perpendicular planes $\mathbb H^{D-1}_{\widehat{\boldsymbol \kappa}}$, and second by summation over the full sublattices $\Z^D_{\widehat{\boldsymbol \kappa}}:= \{x_i\bs h^{(i)} + z\widehat{\bs \kappa}:(\bs x,z)\in\Z^D\}$. Let
\begin{equation} \label{lemma_sesquiliniar}
( A , B )^{\widehat {\boldsymbol \kappa},\perp} \coloneqq \sum_{\boldsymbol x \in \mathbb H_{\widehat{\boldsymbol \kappa}}^{D-1}} ( A(\bm x) , B),\quad
( A , B )^{\widehat {\boldsymbol \kappa}}_{\boldsymbol \kappa'} \coloneqq \sum_{\boldsymbol x \in \Z_{\widehat{\boldsymbol \kappa}}^{D}} 
e^{\ri \boldsymbol \kappa'\boldsymbol x}
(  A( \bm x) , B), \quad A,B \in \mathfrak U_{\rm loc}.
\end{equation}
Note that these are well defined by clustering of the connected correlator $(A,B)=ω(A^\dagger B)-ω(A^\dagger )ω(B)$. Similarly to the construction of the spaces $\mathcal{H}_{\boldsymbol \kappa}$, we define the Hilbert spaces $\mathcal{H}^{\widehat {\boldsymbol \kappa},\perp}$ and $\mathcal H^{\widehat {\boldsymbol \kappa}}_{\boldsymbol \kappa'}$ as the norm completions of $\mathcal{V}^{\widehat {\boldsymbol \kappa},\perp}\coloneqq \mathfrak U_{\rm loc}/ \sim^{\widehat {\boldsymbol \kappa},\perp}$ and $\mathcal V^{\widehat {\boldsymbol \kappa}}_{\boldsymbol \kappa'}\coloneqq \mathfrak U_{\rm loc}/ \sim^{\widehat {\boldsymbol \kappa}}_{\boldsymbol \kappa'}$, respectively, where $A \sim^{\widehat {\boldsymbol \kappa},\perp} A^{\prime} \Leftrightarrow ( A - A^{\prime}, A-A^{\prime} )^{\widehat {\boldsymbol \kappa},\perp}=0$ and $A\sim^{\widehat {\boldsymbol \kappa}}_{\boldsymbol \kappa'} A' \Leftrightarrow ( A - A^{\prime}, A-A^{\prime} )^{\widehat {\boldsymbol \kappa}}_{\boldsymbol \kappa'} = 0$. The equivalence class of $A \in \mathfrak{U}_{\rm loc}$ is denoted by the respective lowercase $a$ and it is immediate that $( \cdot , \cdot )^{\widehat {\boldsymbol \kappa},\perp}$ and $(\cdot,\cdot)^{\widehat {\boldsymbol \kappa}}_{\boldsymbol \kappa'}$ are inner products on their respective Hilbert spaces, and thus satisfies the Cauchy-Schwarz inequality. See Appendix \ref{forms} for proofs of basic properties. It can be established that $ι_x \coloneqq \iota_{x \widehat{\boldsymbol \kappa}}$ ($x\in\Z$) and $τ_t$ ($t\in\R$) act unitarily on $\mathcal{H}^{\widehat {\boldsymbol \kappa},\perp}$ as representations of the groups $\Z$ and $\R$ respectively, and that $τ_t$ ($t\in\R$) act unitarily on $\mathcal H^{\widehat {\boldsymbol \kappa}}_{\boldsymbol \kappa'}$ as a representation of the group $\R$. Details and other basic results are discussed in Appendix \ref{appendix:sesquilinear}.  We denote space-time translations of elements $a \in \mathcal{H}^{\hat{\kappa},\perp}$ by $ι_z τ_t a$, time translations of $a \in \mathcal{H}^{\bm \hat \kappa}_{\bm \kappa^{\prime}}$ by $τ_t a$, while keeping the notation $A(\bm x,t)$ for $A \in \mathfrak{U}$.
\begin{property}\label{PROPERTY}
Consider a dynamical system $(\mathfrak{U},ι,τ)$ (\Cref{defn:dynamicalsystem}) that is space-like $D+1$-clustering (\Cref{defn:spacelikeclustering}). For every $\boldsymbol \kappa \in \R^{D}_{\mathbb Q}$, it follows that:
\begin{enumerate}
\item $\forall\,a,b \in \mathcal{H}^{\widehat {\boldsymbol \kappa},\perp}$, for almost all $v \in \mathbb{R}$,
\begin{equation}
\lim_{T \to \infty} \frac{1}{T} \int_0^T  ( ι_{\lfloor vt \rfloor}τ_t a, b )^{\widehat {\boldsymbol \kappa},\perp} \dd t=0.
\end{equation} \label{property1}
\item $\forall\, a,b \in  \mathcal{V}^{\widehat {\boldsymbol \kappa},\perp}$, there exist $T>0$ (independent of $\widehat{\boldsymbol \kappa}$) and a Lebesgue measurable function $f : \mathbb{R} \mapsto \mathbb{R}_+$ such that:
\begin{equation}
  | ( ι_{  \floor{vt}} τ_{t} a,b )^{\widehat {\boldsymbol \kappa},\perp}| \leq f(v) \ \forall \  v \in \mathbb{R}, \ t\geq T \text{ satisfying } \int_{-\infty}^{\infty} \dd v \,( |v| +1) f(v) < \infty.
\end{equation}\label{property2}
\end{enumerate} \label{property}
\end{property} 
\begin{proof}
Property \ref{PROPERTY}.\ref{property1} is a consequence of the almost-everywhere ergodicity theorem, \cite[Theorem 3.2]{ampelogiannis_almost_2021}.  One can see this by proving that $( ι_{\lfloor vt \rfloor}τ_t a, b )^{\widehat {\boldsymbol \kappa},\perp}$ is space-like ergodic, \cite[Definition 3.1]{ampelogiannis_almost_2021}, see Appendix \ref{appendixB} for the general idea. That is, first, note that our assumptions immediately imply that the state $ω$ is space-like ergodic:
\begin{equation}
  \lim_N  \frac{1}{N} \sum_{m=0}^{N-1} \bigg( A\big( m \bm n , m υ^{-1} |\bm n|\big) , B\bigg) = 0  \ , υ>υ_{LR} \ , A,B \in \mathfrak{U}_{\rm loc}.
\end{equation}
Consider $a,b \in \mathcal{V}^{\bm{\hat{\kappa}},\perp}$ and class representatives $A,B \in \mathfrak{U}_{\rm loc}$ respectively. Then,
\begin{equation}
\label{eq63}
  \lim_N  \frac{1}{N} \sum_{m=0}^{N-1} \big( ι_{\floor{m z}} τ_{mυ^{-1}z}a,b \big)^{\hat{\bm \kappa}, \perp} =
  \lim_N  \frac{1}{N} \sum_{m=0}^{N-1}  \sum_{\bm x \in \mathbb{H}_{\widehat{\boldsymbol \kappa}}^{D-1}} \bigg(A \big( \floor{m z}\bm{\hat{\kappa}} + \bm x, mυ^{-1}z\big),B \bigg).
\end{equation}
Using space-like clustering we can see for any $z\in \R$, $υ>υ_c$ and  $A,B \in \mathfrak{U}_{\rm loc}$
\begin{equation}
    \Big| \frac{1}{N} \sum_{m=0}^{N}
    \bigg( A \big(\floor{m z}\bm{\hat{\kappa}} + \bm x, mυ^{-1}z\big)
    ,B \bigg) \Big| \leq \frac{1}{N} \sum_{m=0}^{N} \frac{c}{ (\sqrt{\floor{mz}^2|\bm{\hat{\kappa}}|^2 + |\bm x|^2 } +1)^p}
\end{equation}
where the summand is $N$-independent, $\bs x$-summable, and uniformly bounded by a $m$-summable function, allowing us to apply the dominated convergence theorem in order to move $\lim_N$ inside the sum $\sum_{\bm x \in \mathbb{H}_{\widehat{\boldsymbol \kappa}}^{D-1}}$ in Eq.~\eqref{eq63}. Thus, 
\begin{equation}
     \lim_N  \frac{1}{N} \sum_{m=0}^{N-1} \big( ι_{\floor{m z}} τ_{mυ^{-1}z}a,b
     \big)^{\hat{\bm \kappa}, \perp} =0 , \text{ for $υ>υ_{LR}$ }.
\end{equation}
Applying the almost-everywhere ergodicity theorem for the system $(\mathcal{H}^{\widehat {\boldsymbol \kappa},\perp},ι,τ)$ (see Appendix \ref{appendixB}) we get Property \ref{property1}.

For Property \ref{property}.\ref{property2}: Choose a $V$ as per the definition of space-like clustering, Def.~\ref{defn:spacelikeclustering}, choose $T>0$, and consider all $t\geq T$. Note that the choice of $T$ and $V$ does not depend on $\widehat{\boldsymbol \kappa}$.

First consider the case $|v|\geq V/|\widehat{\boldsymbol \kappa}|+T^{-1}$. Note that $(|v|t-1)|\widehat{\boldsymbol \kappa}| \geq Vt$ and $(|v|t-1)|\widehat{\boldsymbol \kappa}|\geq (|v|T-1)|\widehat{\boldsymbol \kappa}|\geq VT$.  We have, for some $p>D+1$ (each step is explained below):
\beqa
    | ( ι_{  \floor{vt}} τ_{t} a,b )^{\widehat {\boldsymbol \kappa},\perp} |
    &\leq &
    \sum_{{\boldsymbol x}\in\Z^{D-1}}
    \Big|
    \big(  A(\floor{vt}{\widehat{\boldsymbol \kappa}}
    + x_i\boldsymbol h^{(i)}, t), B \big)
    \Big|
    \n
    &\leq &
    \sum_{{\boldsymbol x}\in\Z^{D-1}}
    \frc c{\lt(\sqrt{(|v|t-1)^2|\widehat{\boldsymbol \kappa}|^2
    +| x_i\boldsymbol h^{(i)}|^2}+1\rt)^p}
    \n
    &\leq& \sum_{{\boldsymbol x}\in\Z^{D-1}}
    \frc c{\lt((|v|t-1)^2|\widehat{\boldsymbol \kappa}|^2
    +| x_i\boldsymbol h^{(i)}|^2\rt)^{p/2}}.
\eeqa
 Note that the right-hand side never gets infinite, since $\hat{ \bm \kappa}$ is fixed and $(|v|t-1)|\widehat{\boldsymbol \kappa}|\geq VT>0$. In the second line we used the fact that
\beqa
|\floor{vt}\widehat{\boldsymbol \kappa}+ x_i\boldsymbol h^{(i)}|^2 
    &=&
    |\floor{vt}|^2\,|\widehat{\boldsymbol \kappa}|^2  +|x_i\boldsymbol h^{(i)}|^2 \quad \mbox{(by orthogonality and the $L^2$ norm)}\n
    &\geq& (|v|t-1)^2|\widehat{\boldsymbol \kappa}|^2 +| x_i\boldsymbol h^{(i)}|^2
    \quad \mbox{(as $\floor{vt}> vt-1$ and $|vt-1|\geq ||v|t-1|$)}\n
    &\geq & (Vt)^2 \quad \mbox{(as $(|v|t-1)|\widehat{\boldsymbol \kappa}| \geq Vt$)}
\eeqa
which allowed us to use space-like $p$-clustering. We now use the fact that $\sum_{\boldsymbol x\in\Z^{D-1}} F(\boldsymbol x) = \int \dd^{D-1}x\,F(\floor{\boldsymbol x})$, hence we'll have to deal with $\big|\,\floor{ x_i} \bm h^{(i)} \, \big|$. The idea is to do the following transformation:
\beqa \label{eq:chgy}
    |x_i\boldsymbol h^{(i)}|^2 &=& \sum_{m\in\t\zeta}
    |x_m|^2 + |x_{j_1} \widehat \kappa_{j_2}|^2
    + |x_{j_2}\widehat \kappa_{j_3}- x_{j_1}\widehat \kappa_{j_1}|^2
    + \ldots
    + |x_{j_{d-1}}\widehat k_{j_d}- x_{j_{d-2}}\widehat \kappa_{j_{d-2}}|^2\n
    &=& \sum_{m=1}^{D-1} |y_m|^2 = |\boldsymbol y|^2,\quad
    y_m = \lt\{\ba{ll}
    x_m & (m\in\t\zeta)\\
    x_{j_1} \widehat \kappa_{j_2} & (m = j_1)\\
    x_{j_l}\widehat \kappa_{j_{l+1}} - x_{j_{l-1}}\widehat \kappa_{j_{l-1}} & (m = j_l \in\{j_2,\ldots,j_{d-1}\}).
    \ea\rt.
    \label{chgy}
\eeqa

The  finite Jacobian of the change of variable above is (nonzero)
\beq
    J = \Big|\frc{\p \boldsymbol y}{\p\boldsymbol x}\Big|
    = \prod_{l=1}^{d-1} |\widehat \kappa_{j_{l+1}}|,\quad 0<J<\infty.
\eeq

We have by the triangle inequality:
\beqa
    |\floor{x_i}\boldsymbol h^{(i)}| = |(x_i-\varep_i)\boldsymbol h^{(i)}|
    \geq |x_i\boldsymbol h^{(i)}| - |\varep_i\boldsymbol h^{(i)}|
    \geq |x_i\boldsymbol h^{(i)}| - h
\eeqa
where we used $\floor{x_i}=x_i -ε_i$ for some $ε_i\in[0,1)$, and where $h = \sum_i|\boldsymbol h^{(i)}|$. Hence, doing the transformation $\bm x \mapsto \bm y$, defined in Eq. \eqref{chgy}:
\beqa
     | ( ι_{  \floor{vt}} τ_{t} a,b )^{\widehat {\boldsymbol \kappa},\perp} | &\leq& \int_{\R^{D-1}} \dd^{D-1} x\,
    \frc c{\big((|v|T-1)^2|\widehat{\boldsymbol \kappa}|^2+(  | x_i\boldsymbol h^{(i)}| - h)^2\big)^{p/2}}
    \n
    &= &
   J^{-1} \int_{\R^{D-1}} \dd^{D-1} y\,
    \frc c{((|v|T-1)^2|\widehat{\boldsymbol \kappa}|^2+(|\boldsymbol y|-h)^2)^{p/2}}
    \n
    &= &
    J^{-1}((|v|T-1)|\widehat{\boldsymbol \kappa}|)^{D-1-p} \int_{\R^{D-1}} \dd^{D-1} y\,
    \frc c{\lt(1+\lt(|\boldsymbol y|-\frc h{(|v|T-1)|\widehat{\boldsymbol \kappa}|}\rt)^2\rt)^{p/2}}
    \n
    &\leq &
   cIJ^{-1}((|v|T-1)|\widehat{\boldsymbol \kappa}|)^{D-1-p}\quad
    (\mbox{$I$ is defined in \eqref{defI}}).
\eeqa
In the third line we used the fact that $(|v|T-1)|\widehat{\boldsymbol \kappa}|>0$. In the final line, we used $(|v|T-1)|\widehat{\boldsymbol \kappa}|\geq VT$ and, for every $\ell\in(0, h/VT]$,
\beqa
    \lefteqn{\int_{\R^{D-1}} \dd^{D-1} y\,
    \frc 1{\lt(1+\lt(|\boldsymbol y|-\ell\rt)^2\rt)^{p/2}}} \n
    &\leq&
    \int_{\R^{D-1},|\boldsymbol y|\geq\frc{h}{VT}} \dd^{D-1} y\,
    \frc 1{\lt(1+\lt(|\boldsymbol y|-\frc{h}{VT}\rt)^2\rt)^{p/2}}
    +
    \int_{\R^{D-1},|\boldsymbol y|<\frc{h}{VT}} \dd^{D-1} y\,1
   \  =:\  I\label{defI}
\eeqa
where $I$, as defined by the right-hand side of the inequality, is finite, as $p>D+1>D-1$, and only depends on $h/(VT)$ and $D$.

For $|v|< V/|\widehat{\boldsymbol \kappa}|+T^{-1}$, we instead use the fact that 
\beq
    |( ι_{  \floor{vt}} τ_{t} a,b )^{\widehat {\boldsymbol \kappa},\perp} |
    \leq ||a||^{\widehat {\boldsymbol \kappa},\perp}\,
    ||b||^{\widehat {\boldsymbol \kappa},\perp}
\eeq
by the Cauchy-Schwartz inequality and space-time translation invariance.

Thus we set
\beq
    f(v) = \lt\{\ba{ll}
    c IJ^{-1}
    ((|v|T-1)|\widehat{\boldsymbol \kappa}|)^{D-1-p}
    & (|v|\geq V/|\widehat{\boldsymbol \kappa}|+T^{-1}) \\
    ||a||^{\widehat {\boldsymbol \kappa},\perp}\,
    ||b||^{\widehat {\boldsymbol \kappa},\perp}
    & (|v|< V/|\widehat{\boldsymbol \kappa}|+T^{-1}).
    \ea\rt.
\eeq
We see that, for $p>D+1$, this indeed satisfies the right properties, and in particular this lower bound on $p$ is necessary for the integral of $(|v|+1)f(v)$ to exists on $\R$.
\end{proof}

From this, we obtain the following crucial lemma:
\begin{lemma}\label{thelemma}
If the dynamical system satisfies Property \ref{property}, then $\forall A,B \in \mathfrak{U}_{\rm loc}$, $\forall s \in \mathbb{R}$, there exists $T_0>0$ such that for every ${\boldsymbol \kappa}\in \mathbb{R}^D_{\mathbb Q}$, the following holds:
\begin{equation}\label{basiclem}
\lim_{T \to \infty} \frac{1}{T} \int_{T_0}^T \dd t \,g(t) = 0 \ , \ \ g(t)= \sum_{{\boldsymbol x} \in \mathbb{Z}^D} \big( e^{\ri{\boldsymbol \kappa}{\boldsymbol x}/t} -e^{\ri{\boldsymbol \kappa}{\boldsymbol x}/(t+s)} \big) \big(A(\bm x,t), B \big).
\end{equation} 
\end{lemma}

\begin{proof}
First we write
\beq
    g(t) = G_0(t) - G_s(t),\quad
    G_s(t) = \sum_{{\boldsymbol x} \in \mathbb{Z}^D} e^{\ri{\boldsymbol \kappa}{\boldsymbol x}/(t+s)} \big(A(\bm x,t), B \big).
\eeq
Let us first simplify $G_s(t)$ for arbitrary $s$. Writing (in a unique fashion) $\boldsymbol x = \boldsymbol r + \boldsymbol \lambda$ where $\boldsymbol r \in \Z^D_{\widehat{\boldsymbol \kappa}}$ and $\bm \lambda\in\Lambda_{\widehat{\boldsymbol \kappa}}$, we get
\beq
    G_s(t) = \sum_{\boldsymbol r\in \Z^{D}_{\widehat{\boldsymbol \kappa}}}
    e^{\ri {\boldsymbol \kappa}\boldsymbol r/(t+s) } \big( \t A_{t+s}(\bm r, t), B \big)
    = (\tau_t \t a_{t+s},b)^{\widehat{\boldsymbol \kappa}}_{\boldsymbol \kappa/(t+s)}
\eeq
where (recall that $\Lambda_{\widehat{\boldsymbol \kappa}}$ is finite)
\beq
    \t A_{u} = \sum_{\boldsymbol\lambda\in\Lambda_{\widehat{\boldsymbol \kappa}}}
    e^{\ri  \boldsymbol \kappa
    \boldsymbol\lambda /u}\iota_{\boldsymbol \lambda}A
\eeq
and the lowercase $\t a_u,b\in \mathcal V^{\widehat {\boldsymbol k}}_{\boldsymbol \kappa/(t+s)}$ are the respective equivalence classes of $\t A_u,B$ (see \Cref{lemma_sesquiliniar} and discussion below). We note that
\beqa
    \lim_{t\to\infty} |G_s(t) - (τ_t \t a_{\infty},b)^{\widehat{\boldsymbol \kappa}}_{\boldsymbol \kappa/(t+s)}|
    &\leq& \lim_{t\to\infty}
    ||τ_t(\t a_{t+s} - \t a_{\infty})||^{\widehat{\boldsymbol \kappa}}_{\boldsymbol \kappa/(t+s)}
    ||b||^{\widehat{\boldsymbol \kappa}}_{\boldsymbol \kappa/(t+s)}\n
    &\leq& \lim_{t\to\infty}
    {\rm max}\{
    |\sin(\boldsymbol \kappa
    \boldsymbol\lambda /(2(t+s)))|
    :{\boldsymbol\lambda \in \Lambda_{\widehat{\boldsymbol \kappa}}}\}
    \, ||\t a_{\infty}||^{\widehat{\boldsymbol \kappa}}_{\boldsymbol \kappa/(t+s)}
    ||b||^{\widehat{\boldsymbol \kappa}}_{\boldsymbol \kappa/(t+s)}\n
    &=& 0
\eeqa
where we used the fact that $\lim_{\boldsymbol \kappa'\to\boldsymbol 0} ||c||_{\boldsymbol \kappa'}^{\widehat{\boldsymbol \kappa}} = ||c||_{\boldsymbol 0}^{\widehat{\boldsymbol \kappa}}$ exists for any $c$. Therefore, using the form $\boldsymbol r = x_i\boldsymbol h^{(i)}+z\widehat{\boldsymbol \kappa} \in \Z^D_{\widehat{\boldsymbol \kappa}}$,
\beqa
    \lim_{T \to \infty} \frac{1}{T} \int_{T_0}^T \dd t \,g(t) &=&
    \lim_{T \to \infty} \frac{1}{T} \int_{T_0}^T \dd t \,\Big(
    (τ_t \t a_{\infty},b)^{\widehat{\boldsymbol \kappa}}_{\boldsymbol \kappa/t}
    -
    (τ_t \t a_{\infty},b)^{\widehat{\boldsymbol \kappa}}_{\boldsymbol \kappa/(t+s)}\Big)\n
    &=&
    \lim_{T \to \infty} \frac{1}{T} \int_{T_0}^T \dd t \,
    \sum_{z\in\Z}
    \Big(
    e^{\ri |{\boldsymbol \kappa}| |\widehat{\boldsymbol \kappa}|z/t }- 
    e^{\ri |{\boldsymbol \kappa}| |\widehat{\boldsymbol \kappa}|z/(t+s) }\Big) ( ι_{z} τ_t  \t a_{\infty}, b )^{\widehat {\boldsymbol \kappa},\perp}.
\eeqa
At this point, we have managed to recast the problem into an effectively one-dimensional problem. The proof now broadly follows that in $D=1$, \cite{doyon_hydrodynamic_2022}. In order to show the Lemma, i.e\ show $\lim_{T \to \infty} \frac{1}{T}\int_{T_0}^{T} \dd t g(t)=0$, we want to commute the limit and the time integral past the summation $\sum_{z \in \Z}$ in $g(t)$. The result would then follow by applying Property \ref{PROPERTY}.\ref{property1}. We proceed to uniformly (in $t$) bound the summand in $g(t)$. As in the proof of Lemma 6.6 in \cite{doyon_hydrodynamic_2022}, we write (denoting $κ=|\bm \kappa| | \bm{ \hat \kappa|}$):
\begin{equation}
 \begin{array}{*3{>{\displaystyle}lc}p{5cm}}
    g(t) &=& \sum_{z \in \Z} 2i \exp{\frac{iκz}{2} \big( \frac{1}{t} +\frac{1}{t+s}\big)} \operatorname{sin}\big[ \frac{κz}{2} \big( \frac{1}{t} - \frac{1}{t+s} \big) \big] ( ι_z τ_t \t a_{\infty},b)^{\bm{\hat \kappa},\perp} \\
    &=& \sum_{υ \in t^{-1} \Z} \frac{i κυ}{2} \exp{\frac{iκυ}{2} \big( 1+\frac{t}{t+s}\big)} \frac{2t}{κυ}\operatorname{sin}\big[ \frac{κυ}{2t} \big( t - \frac{t^2}{t+s} \big) \big] ( ι_{υt} τ_t \t a_{\infty},b)^{\bm{\hat \kappa},\perp}\\
    &= &\int_{\R} i κ υ_t \exp{\frac{ i κ υ_t}{2}\big(1 + \frac{t}{t+s})} 
     \frac{2t}{ κ υ_t}  \operatorname{sin}\big(\frac{κ υ_t}{2t}  \big(t - \frac{t^2}{t+s}\big)\big) ( ι_{\floor{υt}} τ_t \t a_{\infty} ,b)^{\bm{\hat \kappa}, \perp}\, \dd υ
\end{array}
\end{equation}
where $υ_t= t^{-1} \floor{ υt}$. Having established Property \ref{PROPERTY}, the rest of the proof follows exactly as in \cite[Lemma 6.6]{doyon_hydrodynamic_2022}.

\end{proof}

Finally, having shown Lemma \ref{basiclem}, we have for all $A,B \in \mathfrak{U}_{\rm loc}$, any rational vector ${\boldsymbol \kappa} \in \mathbb{R}^D$, $s \in \mathbb{R}$ and any choice of Banach Limit $\widetilde{\lim_{t \to \infty}}$:

\begin{equation}
\begin{array}{*3{>{\displaystyle}lc}p{5cm}}
0&=& \widetilde{\lim_{t \to \infty}}\frac{1}{T}\int_0^T \bigg( \langle τ_t a,b \rangle_{{\boldsymbol \kappa}/t} - \langle τ_t a,b \rangle_{{\boldsymbol \kappa}/(t+s)} \bigg) \,\dd t \\
&=& \widetilde{\lim_{t \to \infty}}\frac{1}{T}\int_0^T \langle τ_t a,b \rangle_{{\boldsymbol \kappa}/t} \, \dd t- \widetilde{\lim_{t \to \infty}}\frac{1}{T}\int_0^T \langle τ_t a,b \rangle_{{\boldsymbol \kappa}/(t+s)}  \,\dd t
\end{array}
\end{equation}
where $a,b \in \mathcal{V}_{\bm \kappa/t}$ are the respective equivalence classes (see \Cref{inner} and below).
We then proceed exactly as in the proof  \cite[Theorem 6.7]{doyon_hydrodynamic_2022}, which completes the proof of the Hydrodynamic projection theorem for any dimension $D$ and for zero wavenumber and frequency.

The proofs can easily be generalised to all frequencies and wavelengths, as all the bounds remain essentially unchanged.  We consider different representations of the groups of space and time translations on the C$^*$-algebra $\mathfrak{U}$, for each frequency-wavenumber pair, defined as
\begin{equation}
    \t {ι}_{\bm x} \coloneqq e^{\ri \bm k \bm x} ι_{\bm x} , \ \t {τ_t} \coloneqq e^{\ri f t} τ_t
    \quad \mbox{for}\  \bs k\in\R^D, f\in\R.
\end{equation}
These form representations of the groups $\Z^D$, $\R$ respectively, by bijective linear maps on $\mathfrak{U}$. They are not $^*$-automorphisms like $ι_{x}$ and $τ_{t}$. However, linearity and unitarity on the Hilbert spaces defined from the state are all that is used in our proofs (and of course the group representation properties). Hence we can relax the Definition of a dynamical system, \Cref{defn:dynamicalsystem}, to a triplet $(\mathfrak{U},\t{ι},\t{τ})$ where $\t{τ}$ is a strongly continuous representation of the group $\R$ by bijective linear maps $\{ \t{τ}_t : \mathfrak{U} \xrightarrow{\sim} \mathfrak{U} \}_{t \in \R}$, and $\t{ι}$ is a representation of the translation group $\Z^D$ by bijective linear maps $\{ \t{ι}_{\boldsymbol x}: \mathfrak{U} \xrightarrow{\sim} \mathfrak{U} \}_{\boldsymbol x \in \Z^D}$, and the state is required to have the invariance property $(\t\iota_{\bs x}\t\tau_t(A),\t\iota_{\bs x}\t\tau_t(B)) = (A,B)$.
It is clear that if  uniform $p$-clustering holds for a  subset of elements in $\mathfrak{U}_{\rm loc}$ under
$\iota_{\boldsymbol x}$, then it also holds under $\t\iota_{\boldsymbol x}$. Under this new represenations of space translations we can construct the Hilbert space of extensive quantities as
\beq
    \t{\mathcal H}_{\bs 0} = \mathcal H_{\bs k}.
\eeq
Further, it is immediate that the Lieb-Robinson bound also holds for $\t\tau_t = e^{\ri ft}\tau_t$.  Thus, by the Lieb-Robinson bound, in this case $\t\tau_t$ is also a one-parameter unitary group on $\t{\mathcal H}_{\bs 0}$. 
We then can define the space of $\t\tau_t$-invariants $\t{\mathcal Q}$,
\beq
    \t{\mathcal Q} = \{a\in\mathcal H_{\bs k}:
    \tau_t a = e^{-\ri f t}a\;\forall\;
    t\in\R\},
\eeq
and the orthogonal projection
\beq
    \t{\mathbb P}:\t{\mathcal H}_{\bs 0} \to \t{\mathcal Q}.
\eeq

Finally, it is easily seen (by $p$-clustering and the Lieb-Robison bound) that if the dynamical system $(\mathfrak U,\iota,\tau)$ and state $ω$ is space-like $p_c$ clustering according to \Cref{defn:spacelikeclustering}, then so is $(\mathfrak U,\t\iota,\t\tau)$, $ω$.  From there on, the results we proved above also hold for these new quantities:
\begin{equation}
\t S_{a,b} ({\boldsymbol \kappa}) = \lim_{t\to \infty} \frac1T \int_0^Te^{-\ri f t}\langle  τ_t a,b \rangle_{\boldsymbol k + {\boldsymbol \kappa}/t}	 \, \dd t.
\end{equation}

\section{Linearised oscillatory Euler equation in one dimension} \label{sectosci}

In \cite{doyon_hydrodynamic_2022}, it is shown, in one-dimensional quantum lattices (quantum chains), that {\bf the theorem of hydrodynamic projection can be used to rigorously obtain linearised Euler equations}. These are equations for two-point functions of local conserved densities, in the limit of large times and large wavelengths. In the previous sections, we have seen how both the concepts of ergodiciy and that of hydrodynamic projections hold as well with respect to any wavenumber $\bs k$ and frequency $f$. In \cite{doyon_hydrodynamic_2022}, it was commented that indeed also the linearised Euler equation will hold at arbitrary wavenumber and frequency. Thus, this allows us to {\bf extend the principles of hydrodynamic correlations to predict oscillatory behaviours}. This is perhaps the most important consequence of the observation that large-scale concepts in fact hold with oscillatory factors. In essence, the hydrodynamic expansion becomes an expansion near a nonzero point in the $(f,\bs k)$-plane.

In this section, we illustrate this principle in a simple quantum chain, in a non-rigorous fashion. Recall $\mathcal Q_{f,\bs k}$, Eq.~\eqref{Qfk}: this is the {\em $f$-oscillatory} closed subspace of the Hilbert space $\mathcal H_{\boldsymbol k}$ generated by the {\em $\boldsymbol k$-extensive observables}. One may consider the corresponding $(f,\boldsymbol k)$-conserved densities and currents $A,B\in\mathfrak U_{\rm loc}$: the density gives rise to the $\boldsymbol k$-extensive charge $\Sigma^{\bs k}A = \sum_{\bs x\in\Z^D} e^{-\ri \bs k\cdot\bs x} A(\bs x)\in\mathcal Q_{f,\bs k}$ which is $f$-oscillatory $\tau_t \big(\Sigma^{\bs k}A\big) = e^{-\ri ft}\Sigma^{\bs k}A$. Such densities and currents should satisfy the oscillatory continuity equation
\beq\label{consdensomega}
    0=\frc{\dd}{\dd t}A(\boldsymbol x,t) + \ri f A(\boldsymbol x,t) + \sum_i \Big(B(\boldsymbol x+\boldsymbol e_i,t)-B(\boldsymbol x,t) + (e^{-\ri \boldsymbol k\cdot \boldsymbol e_i}-1)B(\boldsymbol x+\boldsymbol e_i,t)\Big).
\eeq
In terms of the infinitesimal generator $\ri [H,\cdot]$ for $\tau_t$, the $\boldsymbol k$-extensive charge satisfies $[H,\Sigma^{\bs k} A] + f \Sigma^{\bs k} A = 0$. If such a density and current exist, then two-point correlation functions of observables that have nonzero overlap with this density (or more precisely, with  $\Sigma^{\bs k} A$), should present oscillatory behaviours at large scales of space and time. Indeed, under the $(f,\boldsymbol k)$-fluid-cell mean, defined, as on the r.h.s.~of \eqref{spacetime}, with a factor $e^{\ri\boldsymbol k\cdot \boldsymbol x - \ri f t}$, correlation functions at large scales project onto $\mathcal Q_{f,\boldsymbol k}$, as per the projection Theorem \ref{hydroprojection}. As this holds for the oscillatory fluid-cell mean, that is under averaging with oscillatory factors, this extracts the oscillatory behaviour of the correlation function.

For any given model, one expects only certain values of $(f,\boldsymbol k)$ that would give nontrivial ${\mathcal Q}_{f,\boldsymbol k} \neq \{0\}$. That is, only certain type of oscillatory behaviour can be observed (if any). A simple example is the free fermionic lattice, with some dispersion relation $E(\boldsymbol k)$. In this system, creation and annihilation operators at momentum $\boldsymbol k$ are $(E(\boldsymbol k),\boldsymbol k)$-extensive conserved quantities, and oscillatory hydrodynamic projection corresponds to the oscillatory algebraic decay of fermion two-point functions conventionally obtained by a saddle-point analysis. In fact, it is convenient to go beyond just the hydrodynamic projection principle, and get the oscillatory linearised Euler hydrodynamic equations, as in \cite{doyon_hydrodynamic_2022}. This gives the explicit oscillatory behaviours of correlation functions of conserved densities. Below we discuss this, and illustrate the phenomenon in a free fermion model. We concentrate on the one-dimensional case
\beq
    D=1.
\eeq

$(f,\boldsymbol k)$-hydrodynamic projection is a hydrodynamic expansion, of linear-response type, in the neighbourhood of arbitrary $(f,\boldsymbol k)$, which extends the paradigm of hydrodynamics to oscillatory behaviours. It generalises ideas in recent works on time crystals, where dynamical symmetries are used to explain persistent oscillations in Drude weights \cite{buca2019nonstationary,medenjak2020rigo}.

\subsection{Oscillatory linearised Euler hydrodynamics}

As mentioned, in one-dimensional models, the results of \cite{doyon_hydrodynamic_2022} are stronger than those reported in Section \ref{sectproj}, as the linearised Euler equation is also proven. This makes the asymptotic of correlation functions more explicit. At a non-rigorous level, the formulae obtained are reviewed in \cite{nardis_correlation_2021} in the usual (non-oscillatory) case. For our purpose, in the non-oscillatory case, it is sufficient to recall that if $A_i, B_i$ are pairs of conserved densities and currents, such that the extensive charges $\mathcal a_i =\Sigma A_i$ form a basis for the space of conserved quantities $\mathcal Q$, then the connected correlation function behaves as
\beq\label{AAusual}
( \overline{A_i(x,t)},A_j(0,0))
\sim  \ell^{-1}\big[\delta(\b x-\mathsf A \b t)\mathsf C\big]_{ij}\qquad(x = \ell \b x,\;t=\ell \b t,\; \ell\to\infty)
\eeq
where the flux Jacobian is
\beq
\mathsf A_i^{~j}
= \sum_l\bra B_i,A_l\ket_{0}\mathsf C^{lj},
\eeq
the statistic susceptibility matrix is
\beq
\mathsf C_{ij} = \bra A_i,A_j\ket_0\quad (\mbox{with $\mathsf C^{ij}$ the inverse matrix,}\ \sum_j \mathsf C^{ij}\mathsf C_{jl} = \delta^i_l)
\eeq
and it is expected that, for the results to hold, the fluid-cell mean can be taken in the more intuitive form of an average over a cell in space-time:
\beq\label{meanordinary}
    \overline{A(x,t)}
    = \frc1{L^2}\sum_{y= -\frc L2}^{\frc  L2}\int_{-\frc  L2}^{\frc L2} A(x+y,t+s) \,\dd s.
\eeq
Here the mesoscopic length can be taken as $L=L(\ell)$ with $L\to\infty$ fast enough as $\ell\to\infty$, and $L/\ell\to0$ (or as $L=\epsilon \ell$, then taking $\epsilon\to0^+$ on the asymptotic large-$\ell$ result).

Three remarks are in order, see the explanations in \cite{nardis_correlation_2021}:

(i) The right-hand side of eq.~\eqref{AAusual} is obtained from the right-hand side of eq.~\eqref{hydroprojstandard} by (1) expressing the continuity equation relating $( A_i(x,t), A_j(0,0))$ and $(B_i(x,t),A_j(0,0))$; (2) using the projection formula, eq.~\eqref{hydroprojstandard}, expressing $S_{\Sigma B_i,\Sigma A_j}(\kappa)$ in terms of $S_{\Sigma A_l,\Sigma A_j}(\kappa)$, Fourier transforming back to real space to get $( B_i(x,t),A_j(0,0))$ in terms of $( A_l(x,t)A_j(0,0))$; and (3) solving the resulting continuity equation for the matrix of correlators $( A_i(x,t) A_j(0,0))$, with appropriate initial condition. The solution is \eqref{AAusual}.

(ii) The fluid-cell mean \eqref{meanordinary} is different from that in \eqref{eq:spacetime}. The latter is used to express the rigorous projection result, but the result is expected to hold for a variety of possible definitions of fluid-cell means. The fluid-cell mean chosen above is more physically transparent and more convenient for our purposes.

(iii) If there is a finite set of basis charges (a finite set of indices $i$), the right-hand side of \eqref{AAusual} is a ``generalised function". Its meaning is that for $\b x/\b t$ equal to an eigenvalue of $\mathsf A$, the large-$\ell$ asypmtotic decays more slowly than $1/\ell$, while for other velocities, it decays more rapidly. The delta-function comes about from the normalisation condition:
\beq
    \mathsf C_{ij}
    =\sum_{x\in\Z} ( A_i(x,t),A_j(0,0))
    = \ell \int \dd \b x\,
    (\overline{A_i(x,t)},A_j(0,0))
\eeq
where we use $1 = \ell \dd \b x$ and the fact that $\sum_{x\in\Z} A_i(x,t)$ is independent of time. If there is a continuous set of basis charges, as is typical in integrable models, then the large-$\ell$ asymptotics is exactly $1/\ell$, and the right-hand side of \eqref{AAusual} is an ordinary function. See \cite{doyoncorrelations,nardis_correlation_2021}.

The observation made in \cite{doyon_hydrodynamic_2022} is that all formulae apply equally well in the oscillatory case: with the space-time translation group defined as
\beq
    \eta_{x,t}^{\omega,k} : A \mapsto \re^{\ri \omega t - \ri kx}A(x,t).
\eeq
This is because $\eta_{x,t}^{\omega,k}$ still is a unitary operator on the GNS space. Thus we have a space of $k$-extensive observables, the equivalence classes $\Sigma^k A$ built using the positive semidefinite sesquilinear form
\beq
\bra A,B\ket_k =\sum_{k\in\Z}\re^{\ri kx} ( A(x),B(0)),
\eeq
and from it we may construct, formally, the $(f,k)$-conserved charges, $\mathcal Q_{f,k}$.

Thus, for frequency-wavenumber $f,k$, if $A_i, B_i$ are pairs of oscillatory conserved densities and currents, eq.~\eqref{consdensomega}, such that $\mathcal a_i =\Sigma^k A_i$ form a basis for $\mathcal Q_{f,k}$, then
\beq
(\, \overline{A_i(x,t)}^{f,k},\,A_j(0,0)\,)
\sim \re^{\ri f t - \ri kx} \ell^{-1}\big[\delta(\b x-\mathsf A^{f,k}\b t)\mathsf C^{f,k}\big]_{ij}\qquad(x = \ell \b x,\;t=\ell \b t,\; \ell\to\infty)
\eeq
where the oscillatory flux Jacobian is
\beq
\big[\mathsf A^{f,k}\big]_i^{~j}
= \sum_l\bra B_i,A_l\ket_k \big[\mathsf C^{f,k}\big]^{lj},
\eeq
the statistic susceptibility matrix is
\beq
\big[\mathsf C^{f,k}\big]_{ij} = \bra A_i,A_j\ket_k
\eeq
(note that these depend on $f$ because the basis $A_i$'s does), and the fluid-cell mean is
\beq\label{omegakmean}
    \overline{A(x,t)}^{f,k}
    = \frc1{L^2}\sum_{y= -\frc L2}^{\frc  L2}\int_{-\frc  L2}^{\frc L2} \re^{\ri f s - \ri ky} A(x+y,t+s) \,\dd s,
\eeq

\subsection{Oscillatory behaviour from a saddle-point analysis} \label{section:fermionic}

For simplicity, and in order to verify the ideas, we consider a one-dimensional quadratic model, where asymptotics of correlation functions can be obtained by an elementary saddle-point analysis. Such models are integrable, and the usual hydrodynamic projection principle has been studied widely in the context of generalised hydrodynamics \cite{doyon_drude_2017,doyoncorrelations,doyon_lecture_2020,nardis_correlation_2021}, and verified to agree with a saddle-point analysis \cite{delvecchio2021hydro}.

The mathematical results reported in the main text apply to quantum lattice models with finite local spaces, and are based on the bosonic version of the $C^*$ algebra formulation of quantum statistical mechanics, where local operators commute with each other. Hence, they do not cover free bosonic chains (infinite-dimensional local space), nor free fermionic chains (fermionic formulation). Nevertheless, all results are expected to hold in both cases, as they only rely on general properties of correlation functions; in particular, for free fermionic chains, all such properties are well established \cite{bratteli_operator_1987}. In this section, we give the example of the free fermionic chain, with Hamiltonian 
\beq\label{Hfree}
H = -\frc J2\sum_{x\in\Z}\big(c^\dag_{x+1}c_x + c^\dag_x c_{x+1}\big),\qquad
\{c_x^\dag,c_y\} = \delta_{x,y}.
\eeq

We are looking to evaluate the asymptotic form of the thermal correlation function
\beq
(c_x(t),c_0(0)) = \omega( c^\dag_x(t)c_0(0))
= \frc{\Tr \Big(e^{-\beta H} c^\dag_x(t)c_0(0)\Big)}{\Tr \Big(e^{-\beta H}\Big)},\quad
c_x(t) = \re^{\ri H t}c_x \re^{-\ri H t}
\eeq
as $x,t\to\infty$ (as one-point functions of fermions are zero, this is the connected correlator). The Hamiltonian \eqref{Hfree} is diagonalised by the Fourier transform
\beq
    c_x = \int_{-\pi}^{\pi} \frc{\dd k}{\sqrt{2\pi}}\,\re^{\ri kx} a(k),\qquad
    \{a^\dag(k),a(k')\} = \delta(k-k')
\eeq
as
\beq
    H = -\int_{-\pi}^{\pi} \dd k\, J\cos k\, a^\dag(k)a(k)
\eeq
with energy spectrum $E(k) = -J\cos k$ and
\beq
    a(k,t) = \re^{\ri Ht} a(k) \re^{-\ri H t} = \re^{-\ri E(k)t} a(k).
\eeq
Therefore we obtain the usual expression for the correlator
\beq\label{corre}
    ( c_x(t),c_0(0))
    = \int_{-\pi}^\pi \frc{\dd k}{2\pi}
    n(k) \re^{\ri E(k)t-\ri kx},\qquad
    n(k) = \frc1{1+\re^{\beta E(k)}}.
\eeq
A similar analysis as that below can be performed in a generalised Gibbs ensemble (GGE), where $n(k)$ takes an arbitrary form that characterises the GGE.

The asymptotic behaviour is easily obtained by a saddle point analysis. With $k_\pm$ solving $v(k_\pm) = x/t$ where $v(k) = E'(k) = J\sin k$ and $k_+\in[-\pi/2,\pi/2]$, $k_- = {\rm sgn}(k_+)(\pi-k_+)$, we obtain
\beq
    ( c_x(t),c_0(0))
    \sim \sum_{\pm} \frc{n(k_\pm)}{\sqrt{2\pi \ri t E''(k_\pm)}} 
    \re^{\ri E(k_\pm)t - \ri k_\pm x}
\eeq
as $t\to\infty$ with $x/t=\xi$ fixed. The decay in $1/\sqrt{t}$, which is slower than $1/t$, indicates that the hydrodynamic projection formula should give a generalised function. As all factors explicitly written as functions of $k_\pm$ are slowly varying, they may be assumed to be constant within the fluid cell. Then we see that the support of the $(f,k)$-fluid-cell mean \eqref{omegakmean} is on $k=k_\pm$ and $f = E(k_\pm)$, which is
\beq
    \xi=v(k),\quad f = E(k).
\eeq

In fact, this saddle point analysis does not provide the full information that we need about the shape of the correlation function around the velocity $\xi=v(k)$. However, because we know the support, we may immediately write, as a consequence of this saddle-point result,
\beq\label{resfluidcell}
    (\,\overline{ c_x(t)}^{f,k}, \,c_0(0)\,) \sim \re^{\ri f t - \ri kx}
    \ell^{-1} \delta(\b x-v(k)\b t)\,
    R(k).
\eeq
The normalisation $R(k)$ is obtained by evaluating
\beq
    \frc1L \int_{-\frc L2}^{\frc L2} \dd s\,\sum_{x
    \in\Z} \re^{\ri kx-\ri f s}( c_x(t+s), c_0(0))
    \sim \ell \int \dd \b x\,\re^{\ri k x}
    (\,\overline{ c_x(t)}^{f,k},\, c_0(0)\,)
\eeq
where again we use $1= \ell \dd \b x$. The left-hand side is found to be $e^{\ri f t}n(k)$ from \eqref{corre}, while the right-hand side is found to be $\re^{\ri f t} R(k)$ from \eqref{resfluidcell}, giving
\beq\label{Rn}
    R(k) = n(k).
\eeq

\subsection{Oscillatory behaviour from oscillatory linearised Euler hydrodynamics}

We note that
\beq
    \re^{\ri E(k) t} a(k,t)
\eeq
is independent of time. Further,
\beq
    a(k) = \frc1{\sqrt{2\pi}} \sum_{x\in\Z} \re^{-\ri kx} c_x.
\eeq
Therefore, $a(k)$ is a $k$-extensive, $f$-oscillatory charge, $a(k)\in \mathcal Q_{f,k}$, for $f = E(k)$, with local density
\beq
    A = \frc{c_0}{\sqrt{2\pi}}.
\eeq
It is a simple matter to verify that the associated current has the form
\beq
    B = \frc{\ri J}{2\sqrt{2\pi}}
    \Big(c_{-1}-\re^{\ri k}c_0\Big)
\eeq
in that $A(x,t)$ and $B(x,t)$ satisfy eq.~\eqref{consdensomega}. Then we evaluate
\beq
    \mathsf C^{f,k}
    = \frc1{\sqrt{2\pi}}( a(k), c_0)
    = \frc{n(k)}{2\pi},\quad
    \bra B,A\ket_k = ( B,  a(k))
    = \frc{n(k)v(k)}{2\pi}
\eeq
and thus
\beq
    \mathsf A^{\omega,k} = v(k).
\eeq
Hence, the prediction from hydrodynamic projections is
\beq\label{reshp}
    (\,\overline{ c_x(t)}^{f,k},\, c_0(0)\,)\sim \re^{\ri f t - \ri kx}
    \ell^{-1} \delta(\b x-v(k)\b t)\,
    n(k)
\eeq
in agreement with \eqref{resfluidcell} and \eqref{Rn}.

\section{Conclusion}

We have analysed the large-time, long-wavelength behaviours of short-range quantum spin models on $\Z^D$.

We have first considered many-body ergodicity, which is ergodicity as viewed from the viewpoint of local physics in the thermodynamic limit. We contrasted it with the standard notion of ergodicity in systems with finite degrees of freedom, and von Neumann's quantum ergodic theorem. In particular, we have obtained strong results in KMS states. We have shown that, under a condition of non-localisation, time averages of local observables become classical, taking the values of their ensemble averages -- this is a natural notion of ergodicity. We have also shown, under the same condition, that a time-evolved state within a ``microcanonical shell" of the KMS state -- defined as perturbations by elements of the $C^*$ algebra -- is arbitrarily closed to the KMS state 100\% of the times over long periods, under a natural metric based on local observables. These results show that the long-time evolution locally looses a large amount of information, in agreement with expected ergodicity.

The non-localisation condition is that the kernel of the time-evolution operator on the GNS space should be ``trivial" (one-dimensional): there should not be non-trivial observables (elements of the $C^*$ algebra different from $\1$) that come back to themselves after some time. This is in general hard to check, and it would be interesting to study this condition is explicit models.

Only the new aspects of the proofs of the above many-body ergodicity statements were presented, as the main parts follow directly from the techniques presented in \cite{ampelogiannis_almost_2021}.

We have then reviewed the notion of almost-everywhere ergodicity, introduced in \cite{doyon_hydrodynamic_2022,ampelogiannis_almost_2021}. This is ergodicity, as above, for almost every ray in space-time. It is shown in \cite{ampelogiannis_almost_2021} to hold {\em for all short-range quantum lattices on $\Z^D$}. That is, when considering displacements not just in time, but in space-time, {\em there is no need for non-localisation conditions in order for ergodicity to hold} almost everywhere. This includes the parallel result on the distance of states in the microcanonical shell, under space-time displacements. This strongly constrains the structure of time-evolved local observables at large times: essentially, local observables become ``thin" over time. Intuitively, this is because the finite-dimensional local space of quantum spin lattices does not allow to have operators nontrivially supported over growing regions -- only a countable set of space-time rays may remain where the observable is nontrivial.

We have also observed that all these ergodicity results in fact hold {\em with arbitrary oscillatory factors}, going beyond conventional ergodicity to strong constraints on how observables may oscillate over long times.

Almost everywhere ergodicity was then used to obtain new rigorous results on {\em hydrodynamic projections}. We have shown that the principle of hydrodynamic projection -- by which correlation functions, at large wavelengths and long times, project onto the extensive conserved quantities admitted by the model -- hold in every short-range quantum spin model on $\Z^D$, and at every frequency $f$ and wavenumber $\boldsymbol k$. Projection occurs on $f,\bs k$-dependent spaces, which are kernels of the time evolution operator on {\em $f,\bs k$-dependent Hilbert spaces}, different from, but constructed in a similar way to, the GNS space.

The hydrodynamic projection theorem naturally generalises that shown in \cite{doyon_hydrodynamic_2022} to arbitrary dimension $D>1$. However, the proof is a nontrivial extension of that presented there for $D=1$, requiring the construction of new Hilbert spaces and the analysis of their properties. The hydrodynamic projection theorem was presented in its most general form, involving arbitrary frequency $f$ and wavenumber $\bs k$.

The almost-everywhere ergodicity and hydrodynamic projection results underline the large universality of general ergodic and hydrodynamic principles: they emerge solely from {\em the separation of scales, the large gap between extensivity and locality}. Further, the mathematical construction based on Hilbert spaces built from local observables and their correlations shows the inherent flexibility in defining notions of extensivity (or homogeneity) and stationarity, allowing for oscillations.


Finally, using this flexibility, we have considered the oscillatory linearised Euler equation: the linearised Euler equation, but obtained from oscillatory hydrodynamic projections instead of the conventional ones. This was proposed in \cite{doyon_hydrodynamic_2022}, where the linearised Euler equation was shown for finite-range quantum spin chains. The equation describes the oscillatory behaviours of correlation functions at large space-time separations. We have illustrated it, in a non-rigorous fashion, on a free-fermion chain, showing that the oscillatory hydrodynamic principle correctly reproduces the oscillatory behaviour that can be inferred from a simple saddle-point evaluation of two-point fermion correlation functions at large space-time separations. This gives a proof-of-principle for the idea that conventional hydrodynamics can be generalised to describe not only averaged, smoothed-out behaviours at large space-time, but also oscillatory behaviours, with frequencies that are far from 0 and wavelengths that are microscopic.

The results presented open many doors for further studies. Applying the results to more examples, including specific structures inside the LR light-cone, the kernel of the GNS evolution operator and the related many-body ergodicity, and oscillatory hydrodynamic projections in interacting systems generalising the free-fermion example, would be very interesting. It is likely that, in integrable systems, oscillatory hydrodynamic projection would connect with the finite-density form factor expansions of quantum models; this may also provide a way of defining and studying such form factor expansions in classical models.

The results illustrate how the choice of Hilbert space relates to the large-scale physics of interest; the important elements always lie within the kernel of the evolution operator, but the latter can be taken as acting on local observables, or extensive observables. The idea that different Hilbert spaces relate to different scales was introduced in \cite{Doyon:2019oaf}, and it would be interesting to extend the rigorous analysis presented here to the ``diffusive" Hilbert space constructed there.

The $(f,\boldsymbol k)$-hydrodynamic projection result paves the way for a full $(f,\boldsymbol k)$-hydrodynamics, a subject which should help uncover new universal dynamics and which we hope to investigate in the near future. In fact, going beyond simple oscillatory phases, it is also possible to re-define space- and time-translation operators by adjoining internal motions of local observables (such as spin rotations). It is clear that all proofs provided immediately generalise to this case whenever the internal motion is an extensive internal symmetry transformation (it commutes with time evolution and acts tensorially on individual spatial cells). It would be extremely interesting to study the related hydrodynamics.

Finally, it would be interesting to go beyond clean spin lattices, towards quantum and classical gases of particles, and also systems with disorder. We believe that the general methods of operator algebras, and in particular the principles presented here, give an alternative set of tools to the traditional ones (such as kinetic or Boltzmann equations), which have the potential to provide new fruitful ways of understanding the emergence of large-scale behaviours from the microscopic physics.

\medskip

{\bf Acknowledgments}

 We are grateful to Berislav Bu\c{c}a and Olalla Castro Alvaredo for insightful comments and a current collaboration on related aspects. We also thank Abhishek Dhar for sharing a draft of \cite{chakraborti_entropy_2021} and for related discussions on ergodicity. BD was supported by EPSRC under the grant ``Emergence of hydrodynamics in many-body systems: new rigorous avenues from functional analysis", ref.~EP/W000458/1. DA is supported by a studentship from the Engineering and Physical Sciences Research Council.

\begin{appendices}

\section{Mathematical Set-up} \label{setup}
We consider a hypercubic lattice $\Z^D$ and to each site $\boldsymbol x \in \Z^D$ we associate  a quantum spin described by the (matrix) algebra of observables $\mathfrak{U}_{\boldsymbol x} \coloneqq \C^{N_{\bm x}}$, $N_{\bm x} \in \N$,  with the operator norm $||\cdot ||$, and for all lattice points $\boldsymbol x \in \Z^D$ the dimension of the spin matrix algebra is uniformly bounded, $N_{\boldsymbol x} \leq N$ for some $N \in \N$. To each finite $Λ \subset \Z^D$ we have the algebra 
$\mathfrak{U}_Λ\coloneqq \bigotimes_{\boldsymbol x \in Λ} \mathfrak{U}_{\boldsymbol x}$
and the algebra of local observables is defined  as the direct limit of the increasing net of algebras $\{ \mathfrak{U}_Λ \}_{Λ \in P_f(\Z^D)}$, where $P_f(\Z^D)$ denotes the set of finite subsets of $\Z^D$: $\mathfrak{U}_{\rm loc} \coloneqq \displaystyle\lim_{\longrightarrow} \mathfrak{U}_Λ$. The norm completion of $\mathfrak{U}_{\rm loc}$ defines the quasi-local C$^*$-algebra of the quantum spin lattice: $\mathfrak{U} \coloneqq \overline{\mathfrak{U}_{\rm loc}}$. For details on the definition of quasi-local C$^*$-algebras see \cite[Chapter 3.2.3]{naaijkens_quantum_2017}. 

Space translations  are naturally defined on $\mathfrak{U}_{\rm loc}$ and extended (by continuity) to $\mathfrak{U}$, see \cite[Chapter 3.2.6]{naaijkens_quantum_2017}. Time evolution is defined explicitly from the interaction of the quantum spin lattice, see \cite[Chapter 6.2]{bratteli_operator_1997}. An interaction is defined as a map $Φ: P_f(\Z^D) \to \mathfrak{U}_{\rm loc}$, s.t. $Φ(Λ) \in \mathfrak{U}_Λ$ and $Φ(Λ)=Φ^*(Λ)$, $\forall Λ \in P_f(\Z^D)$.  The Hamiltonian associated with any $Λ \in P_f(\Z^D)$ is $H_Λ \coloneqq \sum_{X \subset Λ} Φ(X)$ which defines the local time evolution as
$τ_t^{Λ} (A) \coloneqq e^{itH_Λ}A e^{-itH_Λ},\quad A \in \mathfrak{U}_Λ ,\; t\in \R$. Time evolution of the infinite system is defined when the limit $\lim_{Λ \to \infty}τ_t^{Λ}(A)$ exists in the norm for all $A \in \mathfrak{U}_{\rm loc}$ and can be uniquely extended to a strongly continuous $^*$-automorphism $τ_t$ of $\mathfrak{U}$. This can be proved for a large class of interactions, including finite range and exponentially decaying ones \cite[Theorem 6.2.11]{bratteli_operator_1997}. In particular, this holds for dynamical systems $(\mathfrak{U},ι,τ)$ with interaction that satisfies for some $λ>0$:
\begin{equation} \label{eq:interaction}
    \norm{Φ}_λ \coloneqq \sup_{\boldsymbol n \in \Z^D} \sum_{X \ni \boldsymbol n} \norm{Φ(X)} |X| N^{2|X|} e^{λ \diam(X)} < \infty .
\end{equation}
where $\diam(X) = \max \{ |x-y|: x,y \in X \}$. This is the type of interactions that we consider: $Φ$ can include $m$-body interactions for any $m$, as long as the interaction drops at least exponentially with $m$ and at least exponentially with the distance. 

We thus define a dynamical system of a $D$-dimensional quantum spin lattice model as follows:
\begin{defi}[Dynamical System] \label{defn:dynamicalsystem}
A dynamical system of a  quantum spin lattice is a triple $(\mathfrak{U},ι,τ)$ where $\mathfrak{U}$ is a quasi-local $C^*$-algebra (as constructed above), $τ$ is a strongly continuous representation of the group $\R$ by $^*$-automorphisms $\{ τ_t : \mathfrak{U} \xrightarrow{\sim} \mathfrak{U} \}_{t \in \R}$, and $ι$ is a representation of the translation group $\Z^D$ by $^*$-automorphisms $\{ ι_{\boldsymbol x}: \mathfrak{U} \xrightarrow{\sim} \mathfrak{U} \}_{\boldsymbol x \in \Z^D}$, such that for any $Λ \in P_f(\Z^D)$: $A \in \mathfrak{U}_Λ \implies ι_{\boldsymbol x }(A) \in \mathfrak{U}_{Λ+\boldsymbol x }$ for all $\boldsymbol x \in \Z^D$. We further assume that $τ$ is such that $τ_t ι_{\boldsymbol x} = ι_{\boldsymbol x} τ_t  $, $\forall t \in \R$, $\boldsymbol x \in \Z^D$, i.e.\ time evolution is homogeneous. We use the notation $A(\bm x,t) \coloneqq ι_{\bm x} τ_t A $.
\end{defi}

A state of the dynamical system is defined as: 

\begin{defi}[States, Invariance] \label{defn:state-invariance}
A state  of a dynamical system $(\mathfrak{U},ι,τ)$ is a positive linear functional $ω: \mathfrak{U} \to \C$ such that $\norm{ω}=1$. The set of states is denoted by $E_{\mathfrak{U}}$. A state $ω \in E_{\mathfrak{U}}$ is called space invariant if $ω( ι_{\boldsymbol n}(A) ) = ω(A)$, $\forall A \in \mathfrak{U}$, $\boldsymbol n \in \Z^D$ and time invariant if  $ω( τ_t(A) ) = ω(A)$, $\forall A \in \mathfrak{U}$, $t \in \R$. We will refer to a space and time invariant state simply as invariant.
\end{defi}


Of great importance in all our results is the Lieb-Robinson bound, which we state as in \cite[Corollary 4.3.3]{naaijkens_quantum_2017}:
\begin{propo}[Lieb-Robinson Bound] \label{lem:liebrobinsonbound}
Consider a dynamical system $(\mathfrak{U},ι,τ)$ (\Cref{defn:dynamicalsystem}) with interaction $Φ$ that satisfies Equation \eqref{eq:interaction}, that is
\begin{equation} 
    \norm{Φ}_λ \coloneqq \sup_{\boldsymbol n \in \Z^D} \sum_{X \ni \boldsymbol n} \norm{Φ(X)} |X| N^{2|X|} e^{λ \diam(X)} < \infty 
\end{equation}
for some $λ>0$.
Then, there exists a $υ_{LR}>0$ such that for all local $A,B \in \mathfrak{U}_{\rm loc}$ with supports $\supp(A)=Λ_A \in P_f( \mathbb{Z}^D)$, $\supp(B)=Λ_B \in P_f( \mathbb{Z}^D)$ respectively, and all $t \in \mathbb{R}$ we have the bound
\begin{equation}\label{eq:liebrobinsonbound}
    \norm{[ τ_t (A) , B]} \leq 4 \norm{A}\norm{B}|Λ_A| |Λ_B| N^{2 | Λ_A |} \exp\big( {-λ( \dist ( A,B) - υ_{LR}|t|)} \big)
\end{equation}
 with $υ_{LR}=2 \frac{\norm{Φ}_λ}{λ}$ called the Lieb-Robinson velocity.
\end{propo}

\section{Ergodicity in the various completions of the space of local observables}\label{appendixB}

Consider any  positive-semidefinite sesquiliniear form $\langle \cdot , \cdot \rangle$ defined on elements of $\mathfrak{U}_{\rm loc}$. For example, this can be the connected correlator $(A,B)=ω(A^{\dag}B)-ω(A^{\dag})ω(B)$, or other forms defined in the main text, such as $( A , B )^{\widehat {\boldsymbol k},\perp}$. We define, by the usual construction, a Hilbert space $\mathcal H$ from $\langle\cdot,\cdot\rangle$. On it, we assume that we have space-time translation group actions, $\iota_{\bs x}$ (for $\bs x\in\Z^d$) and $\tau_t$. For the case of $( A , B )^{\widehat {\boldsymbol k},\perp}$, as summation over the perpendicular plane is done, there remain only a one-dimensional translation group, $d=1$. We say that we have the system $(\mathcal H,\iota,\tau)$. 

We can generalise the definitions of clustering of \Cref{section:clustering} for any such forms, so that we define: a subset  $\mathfrak{C} \subset \mathfrak{U}_{\rm loc}$ to be called uniformly $p$-clustering, with respect to $\langle \cdot , \cdot \rangle$, for $p>D$ if there exist a $C>0$ s.t. for all $A,B \in \mathfrak{C}$:
\begin{equation}
    | \langle \iota_{\bs x} A,B\rangle | \leq C (1 + |\boldsymbol x|)^{-p} , \ \ \forall \boldsymbol{x} \in \Z^d.
\end{equation}
Likewise, we can adapt \Cref{defn:spacelikeclustering} of space-like $p_c$ clustering to $\langle \cdot , \cdot \rangle$, and we define, as per \cite[Def 3.1, Theorem 3.2]{ampelogiannis_almost_2021}, $\langle \cdot , \cdot \rangle$ to be space-like ergodic when there exists a $υ_c$ such that for every $A,B \in \mathfrak{U}_{\rm loc}$ and $υ>υ_c$:
\begin{equation}\label{spacelike-ergodic-form}
  \lim_N  \frac{1}{N} \sum_{m=0}^{N-1} \langle ι_{\boldsymbol n}^m τ_{υ^{-1}|\boldsymbol n|}^m (A), B \rangle = 0.
\end{equation}

The proof of the almost everywhere ergodicity theorem \cite[Theorem 3.2]{ampelogiannis_almost_2021} can now be adapted to the system $(\mathcal H,\iota,\tau)$, as long as space-like ergodicity holds.

Further, with space-like $p_c$-clustering for $p_c>d$, we can define an equivalence relation and similarly to the construction of the spaces of extensive quantities, we can construct a Hilbert space by the completion of the space of equivalence classes. We can also extend the action of $ι,τ$ on the Hilbert space and it will be unitary. This is a matter of replicating the methods in \cite[Section 5]{doyon_hydrodynamic_2022}.

\section{Space-like clustering proof}\label{appC}
Throughout this appendix we use the notation $ι_{\bm x}τ_tA$ for space-time translations of observables $A\in \mathfrak{U}$, in order to maintain visual clarity of expressions. We prove the following proposition:
\begin{propo}
    Consider a dynamical system $(\mathfrak{U},ι,τ)$ with interaction satisfying \Cref{eq:interaction}, in an exponentially clustering state (\Cref{defn:expclustering}). It follows that the system is space-like exponentially clustering with respect to $ω$, \Cref{defn:spacelikeclustering}, for the Lieb-Robison velocity $υ_{LR}$.
\end{propo}

 First, we show the first condition of \Cref{defn:spacelikeclustering}. 
The Lieb-Robinson bound allows us to approximate the time evolved observables by local ones, by projecting the time evoluted $τ_t(A)$, $A \in \mathfrak{U}_{loc}$ onto local ones $σ_{Λ}(τ_t (A) )$ supported on finite $Λ \subset \Z^D$. This is done by using the result \cite[Corollary 4.4]{nachtergaele_quasi-locality_2019} and satisfies the first condition of \Cref{defn:spacelikeclustering}:
\begin{lemma} \label{lem:localapprox}
Let $A\in \mathfrak{U}$ and consider a finite $Λ \subset \mathbb{Z}^D$. If there is an $ε>0$ such that 
\begin{equation}
\norm{ [ A, B] }\leq ε \norm{A}  \norm{B}  \ , \ \ \forall B \in \mathfrak{U}_{ \mathbb{Z}^D \setminus Λ}
\end{equation}
then we can approximate $A$ by a strictly local $σ_Λ (A) \in \mathfrak{U}_Λ$:
\begin{equation}
\norm{σ_Λ (A) - A } \leq 2ε \norm{A}.
\end{equation} 
\end{lemma}

This Lemma states that if an observable $A \in \mathfrak{U}$ almost commutes with every $B\in \mathfrak{U}_{\Z^D \setminus Λ}$ supported outside a finite $Λ$, then it can be well approximated by a local $σ_Λ(A) \in \mathfrak{U}_Λ$. Combined with the Lieb-Robinson bound \Cref{eq:liebrobinsonbound} we can prove (this is also described in \cite[Chapter 4.3]{naaijkens_quantum_2017}):
\begin{propo} \label{propo:LRapproximation}
    Consider a dynamical system with exponentially decaying interactions, the time evolution $τ_t A$ of a local $A \in \mathfrak{U}_Λ$ and the finite sets $Λ_r = \cup_{\boldsymbol x \in Λ}B_{\boldsymbol x}(r)$, $r=1,2,3, \dots$, where $B_{\boldsymbol x}(r)$ is the ball of radius $r>0$ around $\boldsymbol x$, i.e.\ $Λ_r$ is $Λ$ extended by a distance $r$ around all of its points. Then, we can approximate $τ_t A$ by the local $σ_{Λ_r}(τ_t A) \in \mathfrak{U}_{Λ_r}$ :
    \begin{equation}
        \norm{ σ_{Λ_r} (τ_t(A)) - τ_t(A)} \leq 2ε_r \norm{A}
    \end{equation}
    with $ε_r = C |Λ|N^{2|Λ|} \exp{-λ(r- υ_{LR}|t|)}$, $r=1,2,3 \dots$.
\end{propo}
Using this result we can show the first condition of space-like clustering:
\begin{proof}[Proof of \ref{defn:spacelikeclustering}.\ref{spacelike1}]
The projection $σ$ allows us to define the sequences $σ_r(A) \coloneqq σ_{Λ_r}A$ required by condition 1 of  \Cref{defn:spacelikeclustering}. 
 Consider $\boldsymbol n \in \Z^D$ and $A,B  \in \mathfrak{U}_{loc}$ with supports
\begin{equation}
    \supp(A) \coloneqq Λ_A \ , \ \ \supp(B) \coloneqq Λ \text{ and } \supp(ι_{\boldsymbol n} A) \coloneqq Λ_A+\boldsymbol n
\end{equation}
Let $r \in \N$ and by \Cref{propo:LRapproximation} consider the local approximation of $τ_t B$ by $σ_{Λ_r}(τ_t B)$, supported in $Λ_r = \cup_{\boldsymbol x \in Λ} B_{\boldsymbol x}(r)$:
\begin{equation}
\norm{ σ_{Λ_r} τ_t B - τ_t B} \leq 2C |Λ| N^{2|Λ|} \norm{B} \exp\big(  -λ(r-υ_{LR} |t|) \big) \label{ineq:localapproximationB}
\end{equation}
Using this we can approximate the time evolution of observables by a local sequence $σ_{r} (τ_t A) \coloneqq σ_{Λ_r} (τ_t A)$ so that $\lim_{r} σ_{r} (τ_t A) = τ_t A$. 
\end{proof}

We show the uniformity condition 2 of the elements $σ_r (τ_t A)$ in the end of the appendix, we first proceed to show the third condition of \Cref{defn:spacelikeclustering}.

\begin{proof}[Proof of \ref{defn:spacelikeclustering}.\ref{spacelike2}]
Consider the quantity $I \coloneqq | ω( ι_{\boldsymbol n}(A) σ_{Λ_r}(τ_t B)) - ω( A) ω( σ_{Λ_r}(τ_t B)) |$, $r \in \N$ where we are interested in the large $r$ limit.  By linearity and the triangle inequality, it holds for any $r,l \in \N$ :
\begin{equation} \label{inequality1}
\begin{array}{*3{>{\displaystyle}lc}p{5cm}}
I  &=& \big| ω\bigg( ι_{\boldsymbol n}(A) \big(σ_{Λ_r}(τ_t B) + σ_{Λ_l}(τ_t B)  - σ_{Λ_l}(τ_t B) \big)\bigg)  - ω( A) ω\big ( σ_{Λ_r}(τ_t B)+σ_{Λ_l}(τ_t B)  - σ_{Λ_l}(τ_t B)\big ) \big| \\
&\leq& \big| ω\big( ι_{\boldsymbol n}(A) σ_{Λ_l}(τ_t B) \big) -ω( A)  ω( σ_{Λ_l}(τ_t B))\big|+  \\
&+&\big| ω\big( ι_{\boldsymbol n}(A) \big(σ_{Λ_r}(τ_t B)  - σ_{Λ_l}(τ_t B)\big) \big) -ω( A)  ω\big( σ_{Λ_r}(τ_t B)  - σ_{Λ_l}(τ_t B)\big) \big|
\end{array} 
\end{equation}

The idea is to control the first part using exponential clustering and the second one using the approximation of time evolution. We now estimate the quantities $T_1=\big| ω\big(ι_{\boldsymbol n}(A) σ_{Λ_l}(τ_t B) \big) - ω( A ) ω(σ_{Λ_l}(τ_t B))\big|$ and  $T_2=\big| ω\big( ι_{\boldsymbol n}(A) \big(σ_{Λ_r}(τ_t B)  - σ_{Λ_l}(τ_t B)\big) \big ) - ω(  A ) ω\big( σ_{Λ_r}(τ_t B)  - σ_{Λ_l}(τ_t B)\big ) \big|$. Consider $r$ large enough and 
\begin{equation}
    l= \floor{ε \dist (Λ_A+\boldsymbol n, Λ)} + \floor{ε \diam (Λ_A \cup Λ)} + 2 \ \text{ for some } \frac12<ε<1 \label{eq:l}
\end{equation}
We first estimate $T_1$, using exponential clustering 
\begin{equation}
    T_1 \leq k_{Λ_A+\boldsymbol n, Λ_l}e^{-λ \dist(Λ_A+\boldsymbol n, Λ_l)} \label{ineq:T_1}
\end{equation}
Where we have the following bound bound for $k(Λ_A+\boldsymbol n, Λ_l)$:
\begin{equation}
   k_{Λ_A+\boldsymbol n, Λ_l} \leq u \norm{A} \norm{ σ_{Λ_l}( τ_t B)}  |Λ_A|^r |Λ_l|^r
\end{equation}
Obviously $|Λ_A+\boldsymbol n|=|Λ_A|$ and $|Λ_l| \leq |B_{\boldsymbol 0}(l)||Λ|$. ,where  $|B_{\boldsymbol 0}(l)|$ is the number of lattice points in the D-ball of radius $l$, which is a polynomial in $l$ of degree $D$, hence it can be bounded by $K l^D$ for some constant $K>0$, $\forall l \geq 1$. Using this, the chosen value for $l$ (\Cref{eq:l}) and the triangle inequality for $\dist(Λ_A +\boldsymbol n, Λ_l)$:
\begin{equation}
    \begin{array}{*3{>{\displaystyle}lc}p{5cm}}
 
   |Λ_l| &\leq& |Λ|K( \floor{ε \dist ( Λ_A +\boldsymbol n , Λ)}+ \floor{ε \diam(Λ_A \cup Λ)}+2)^D \\
   &\leq& |Λ|K ε^D \big( \dist(Λ_A + \boldsymbol n, Λ) + \diam( Λ_A \cup Λ) +4\big)^D \\
    &\leq& |Λ|K ε^D \big( |\boldsymbol n| + \dist(Λ_A, Λ) + \diam( Λ_A \cup Λ) +4\big)^D
    \end{array}
\end{equation}
Since $ (|\boldsymbol n| + \dist(Λ_A, Λ) + \diam( Λ_A \cup Λ) +2)^D$ is a polynomial in $|\boldsymbol n|$ of degree $D$ it can be bounded by $L|\boldsymbol n|^D$ for some constant $L>0$ (depending on $\dist(Λ_A,Λ)$, $\diam(Λ_A \cup Λ)$), $\forall |\boldsymbol n| \geq 1$.  Hence:
\begin{equation}
    |Λ_l| \leq |Λ|KL ε^{D} |\boldsymbol n|^D
\end{equation}
By the explicit construction of the projection $σ$ in \cite{nachtergaele_quasi-locality_2019} it holds that $\norm{σ_{Λ_l}C} \leq \norm{C}$, $\forall C\in \mathfrak{U}$. Additionally, since $τ_t$ is a $^*$-automorphism for all $t \in \R$, it is norm preserving, hence
\begin{equation}
    \norm{σ_{Λ_l}(τ_tB)} \leq \norm{B}
\end{equation}
Next, we can estimate $\dist(Λ_A +\boldsymbol n, Λ_l)$ with respect to $|\boldsymbol n|$ by simple geometric arguements:
\begin{equation}
        \dist(Λ_A+\boldsymbol n, Λ_l) \geq \dist(Λ_A +\boldsymbol n, Λ)-l \geq (1-ε) \dist(Λ_A +\boldsymbol n, Λ) - ε \diam(Λ_A \cup Λ)-2
\end{equation}
and
\begin{equation}
 \begin{array}{*3{>{\displaystyle}lc}p{5cm}}
    \dist(Λ_A+\boldsymbol n, Λ) &=& \min \{ |\boldsymbol y-\boldsymbol z| : \boldsymbol y \in Λ_A+\boldsymbol n, \boldsymbol z \in Λ \} \\
        &=& \min \{ |\boldsymbol y+\boldsymbol n-\boldsymbol z| : \boldsymbol y \in Λ_A, \boldsymbol z \in Λ \} \\
        &\geq& \min \{ |\boldsymbol n| - |\boldsymbol y-\boldsymbol z| :\boldsymbol y \in Λ_A, \boldsymbol z \in Λ \} \\
        &=& |\boldsymbol n| - \max\{ |\boldsymbol y-\boldsymbol z| : \boldsymbol y \in Λ_A ,\boldsymbol z \in Λ \} \\
        &\geq& |\boldsymbol n| - \diam(Λ_A \cup Λ)
    \end{array} \label{eq:distΛvecn}
\end{equation}
Putting these together we get
\begin{equation}
    e^{-λ \dist(Λ_A +\boldsymbol n,Λ)} \leq e^{2\lambda}e^{λ \diam(Λ_A \cup Λ)} e^{-λ(1-ε)|\boldsymbol n|}
\end{equation}
and combining all the estimates for the terms in $T_1$, inequality \eqref{ineq:T_1} becomes
\begin{equation}
    T_1 \leq u e^{2\lambda}K^rL^r |Λ_A|^r|Λ|^r ε^{rD} \norm{A}\norm{B} e^{λ \diam(Λ_A \cup Λ)} |\boldsymbol n|^{rD} e^{-λ(1-ε)|\boldsymbol n|} \label{ineq:T1}
\end{equation}
To estimate $T_2$, we use the triangle inequality and note that $ω$ has norm $1$:
\begin{equation}
    T_2 \leq  \norm{ι_{\boldsymbol n}A} \norm{σ_{Λ_r}(τ_t B)  - σ_{Λ_l}(τ_t B)}  +  \norm{A} \norm{σ_{Λ_r}(τ_t B)  - σ_{Λ_l}(τ_t B)}  \label{ineq:T2}
\end{equation}
where $ \norm{ι_{\boldsymbol n}A}=\norm{A}$. Using the approximation of $τ_tB$, inequality \eqref{ineq:localapproximationB}, we get
\begin{equation}
    T_2 \leq 2\norm{A}  2C |Λ| N^{2|Λ|} \norm{B} \big( \exp \{ -λ(r-υ_{LR} |t|)\} +  \exp\{-λ(l-υ_{LR} |t|)\}\big)
\end{equation}

We consider a compact subset $T\coloneqq ε^{\prime} υ^{-1}_{LR} [-|\boldsymbol n|,|\boldsymbol n|]$ for some $0<ε^{\prime}<ε$, and let $t \in T$. Then, for the value \eqref{eq:l} of $l$, since $\floor{y} \geq y-1, \forall y>0$, we have:
\begin{equation}
\begin{array}{*3{>{\displaystyle}lc}p{5cm}}
 l-υ_{LR} |t|  &\geq& ε \dist(Λ_A +\boldsymbol n,Λ)-1+ε\diam(Λ_A \cup Λ)-1+2 - ε^{\prime}|\boldsymbol n| \\
 &\geq& ε |\boldsymbol n| -ε \diam(Λ_A \cup Λ) +ε\diam(Λ_A \cup Λ)- ε^{\prime}|\boldsymbol n|  \\
 &=& (ε-ε^{\prime})|\boldsymbol n| >0 \label{ineq:lvtestimate}
 \end{array}
\end{equation}
where in the second line we used relation \eqref{eq:distΛvecn}.  Hence:
\begin{equation}\label{ineq:lvtestimateprime}
    \exp\{-λ(l-υ_{LR}|t|)\}  \leq \exp\{-λ(ε-ε^{\prime})|\boldsymbol n|\} \ , \ \ \forall t \in T, 0<ε^{\prime}<ε<1
\end{equation}
and  also $\lim_r \exp\{-λ(r-υ_{LR}|t|)\} =0$, $\forall t \in T$. We can now take $\lim_r$ of inequality \ref{inequality1}, where the left-hand side becomes $\lim_r I = |ω(  ι_{\boldsymbol n}A τ_tB ) - ω(  A ) ω(  B )|$ by continuity. The right-hand side is bounded by the estimates \eqref{ineq:T1}, \eqref{ineq:T2}, \eqref{ineq:lvtestimateprime}, to get:
{
\begin{equation}
 \begin{array}{*3{>{\displaystyle}lc}p{5cm}}
  |ω(  ι_{\boldsymbol n}A τ_tB ) - ω( A ) ω(  B ) | \leq \\
  ue^{2\lambda}K^rL^r|Λ_A|^r|Λ|^r ε^{rD} \norm{A} \norm{B} e^{λ\diam(Λ_A \cup Λ)}|\boldsymbol n|^{rD} e^{-λ(1-ε)|\boldsymbol n|} \\
  + 4 C N^{2|Λ|} |Λ| \norm{A} \norm{B} e^{-λ(ε-ε^{\prime}) |\boldsymbol n|}
    \end{array} 
\end{equation}
}
Concluding the proof, note that for any $0<\tilde{λ}<λ(1-ε)$ there exists a large enough $G>0$ s.t. $|\boldsymbol n|^{rD} e^{-\lambda (1-ε)|\boldsymbol n|} \leq G e^{-\tilde{λ}|\boldsymbol n|}$. Hence we arrive at the result by defining $μ \coloneqq \min(\tilde{λ}, λ(ε-ε^{\prime})) >0$ and
\begin{equation}
    c= \norm{A} \norm{B} |Λ| (u e^{-2}K^rL^r |Λ_A|^r |Λ|^{r-1}ε^{rD}e^{λ\diam(Λ_A \cup Λ)}G + 4 CN^{2|Λ|})
\end{equation}
to get that there exist $c>0$ and a $μ>0$ s.t.
\begin{equation}
    | ω( ι_{\boldsymbol n}A τ_t B ) - ω(  A ) ω(  B ) | \leq c e^{-μ|\boldsymbol n|}
\end{equation}
for all $\boldsymbol n \in \Z^D$ and for any $t \in T = ε^{\prime} υ_{LR} [ - |\boldsymbol n|, |\boldsymbol n| ]$, with any choice $0<ε^{\prime}<ε$ and $1/2<ε<1$. Hence, for any $υ=υ_{LR}/ε^{\prime}>υ_{LR}$ and choosing $t= - υ^{-1}|\boldsymbol n|$, we have
\begin{equation}
    |ω(  ι_{\boldsymbol n}A τ_{-υ^{-1}|\boldsymbol n|} B )- ω(  A ) ω(  B) | \leq c e^{-μ|\boldsymbol n|} \ , \forall \boldsymbol n \in \Z^D \label{eq:spacelikeergodicproof}
\end{equation}
and taking advantage of the time invariance of $ω$, we have established exponential space-like clustering, and as a consequence space-like $p$-clustering for all $p$.
\end{proof}

It remains to prove the uniformity condition, i.e.\ that the set $\{ (σ_n τ_t a , σ_m τ_s b) : n,m \in \N \}$ is uniformly clustering, \Cref{eq:uniformclustering} in \Cref{defn:mixing}.

\begin{proof}[Proof of \ref{defn:spacelikeclustering}.\ref{spacelike3}]
We start with $I \coloneqq | ω( ι_{\boldsymbol n}(A) σ_{Λ_r}(τ_t B)) - ω( A) ω( σ_{Λ_r}(τ_t B)) |$ and choose $l= \dist(Λ_A+\boldsymbol n, Λ)$. If $r< l$ then we bound $I$ by the space clustering property, as in \Cref{ineq:T_1}:
\begin{equation}
     I_{r<l} \leq u |Λ_A|^r |Λ_l|^r \norm{A} \norm{B}  e^{-λ \dist(Λ_A+\boldsymbol n, Λ_l)}
\end{equation}
and $e^{-λ \dist(Λ_A+\boldsymbol n, Λ_r)} \leq e^{-λ|\boldsymbol n|} e^{λ \diam (Λ_A \cup Λ) - λ l}= e^{-2λ|\boldsymbol n|} e^{2λ \diam (Λ_A \cup Λ) } $. Thus, we get
\begin{equation}
    I_{r < l} \leq u |Λ_A|^r |Λ|^r \norm{A} \norm{B} e^{2λ \diam (Λ_A \cup Λ) }  e^{-2λ|\boldsymbol n|} 
\end{equation}

Now, if $r>l$ we proceed as in \Cref{inequality1} to get
\begin{equation}
 \begin{array}{*3{>{\displaystyle}lc}p{5cm}}
    I &\leq& I_{r<l} + 2\norm{A}  2C |Λ| N^{2|Λ|} \norm{B} \big( \exp\{ -λ(r-υ_{LR} |t|)\} +  \exp\{-λ(l-υ_{LR} |t|)\}\big) \\
    &\leq& I_{r<l} +4\norm{A}  2C |Λ| N^{2|Λ|} \norm{B} \exp\{-λ(l- υ_{LR}t)\} \\
    &\leq& I_{r<l} +4\norm{A}  2C |Λ| N^{2|Λ|} \norm{B}  e^{υ_{LR}t} e^{-λ \dist(Λ_A+ \boldsymbol n,Λ) } \\
    &\leq& I_{r<l} +4\norm{A}  2C |Λ| N^{2|Λ|} \norm{B}  e^{υ_{LR}t} 
    e^{λ \diam( Λ_A \cup Λ)}  e^{-λ |\boldsymbol n| }
    \end{array}
\end{equation}

Hence, we can overall obtain a uniform in $r$ exponential clustering. Similarly we can repeat the process for the replacement $ι_{\boldsymbol n} A \rightarrow σ_{r^{\prime}}  τ_{s} ι_{\boldsymbol n} A$, which will finally yield uniform exponential clustering for the set $\{( σ_{r^{\prime}}  τ_{s}  a , σ_{r}  τ_{t}  b)  : r,r^{\prime} \in \N \}$ for any $t,s \in \R$. This obviously implies uniform $p$-clustering for all $p$.
\end{proof}

\section{Sesquilinear forms properties} \label{forms}
Here we prove some basic properties for the sesquilinear forms used throughout the main text.
\begin{lemma} \label{lemma:positive}
For any $\boldsymbol k \in \mathbb{R}^D$ and $A \in \mathfrak{U}_{\rm loc}$ in a $p$-clustering with $p>D$ dynamical system, it follows $\langle A , A \rangle_{\boldsymbol k} \geq 0$.
\end{lemma}
\begin{proof}
Define $\displaystyle B= \sum_{{\boldsymbol x} \in [0,L]^D} e^{i{\boldsymbol k}\cdot{\boldsymbol x}} A(\bm x) \in \mathfrak{U}_{\rm loc}$, for $L \in \mathbb{N}$. Then
\begin{equation}
 \begin{array}{*3{>{\displaystyle}lc}p{5cm}}
0 \leq ( B, B ) &=& \sum_{{\boldsymbol x}, \boldsymbol y \in [0,L]^D} e^{i{\boldsymbol k}\cdot({\boldsymbol x}- \boldsymbol y)} ( A({\boldsymbol x}-\boldsymbol y), A )  \\ &=& \sum_{\boldsymbol z \in [-L,L]^D} \bigg(\prod_{j=1}^{D} (L+1-|z_j|) \bigg)  e^{i{\boldsymbol k}\cdot\boldsymbol z} (  A(\boldsymbol z), A ) \ , \ \ {\boldsymbol z}=(z_1, z_2,...,z_D)\\
&=&  (L+1)^D \sum_{\boldsymbol z \in [-L,L]^D} e^{i{\boldsymbol k}\cdot{\boldsymbol z}} (  A(\boldsymbol z) , A )   +  o( (L+1)^{D-1}).
\end{array}
\end{equation}
We divide both sides of the inequality by $(L+1)^D$ and the Lemma follows by taking the limit $L \to \infty$. That is because the first term becomes $\langle A,A\rangle_{\boldsymbol k},$ while the term $o( (L+1)^{D-1})$  tends to $0$ for $p>D$, after diving by $(L+1)^D$ and taking $L \to \infty$. This can be seen when writing $o( (L+1)^D)$  as a sum over all permutations $σ$ of $z_1,z_2,...,z_D$  and all $k=1,2,...,D$, of terms of the form:

\begin{equation}
(L+1)^{D-k} \sum_{ {\boldsymbol z} \in [-L,L]^D} \prod |σ(z_j)| e^{i{\boldsymbol k}\cdot{\boldsymbol z}}  ( A(\boldsymbol z), A ) \ , \ \ k=1,2,...,D
\end{equation}
where the product is taken over a choice of $k$ elements of $z_1,z_2,...,z_D$. We can bound the absolute value of any such term as follows, where we denote $S_D (d) = \big| \{\boldsymbol x \in \mathbb{Z}^D : |\boldsymbol x| = d \} \big|$ the number of lattice points on a circle of radius $d$:

\begin{equation}
\begin{array}{*3{>{\displaystyle}lc}p{5cm}}
(L+1)^{D-k} \sum_{{\boldsymbol z} \in [-L,L]^D} |{\boldsymbol z}|^k \frac{1}{(|{\boldsymbol z}|+1)^p} \leq (L+1)^{D-k} \sum_{{\boldsymbol z} \in [-L,L]^D} \frac{(|{\boldsymbol z}|+1 )^k}{ (|{\boldsymbol z}|+1)^p} \\
\leq  (L+1)^{D-k} \sum_{d=0}^{DL} \frac{1}{(d+1)^{p-k}} S_D (d) \\
\leq (L+1)^{D-k}S_D(0) + C(L+1)^{D-k} \sum_{d=1}^{DL} \frac{d^{D-1}}{(d +1)^{p-k}} \ \ \text{, for $C>0$} \\
\leq O(L^{D-k})  + C(L+1)^{D-k} \sum_{d=1}^{DL} \frac{(d+1)^{D-1}}{(d +1)^{p-k}} \\
\leq O(L^{D-k}) +  C(L+1)^{D-k} \int_0^{DL} dx(x +1)^{D-p+k-1}  \\
= O(L^{D-k}) + C(L+1)^{D-k} \bigg(  (DL+1)^{D-p+k} - 1 \bigg).
\end{array}
\end{equation}
Dividing by  $(L+1)^D$ and taking $L\to \infty$ we see that the right-hand side vanishes for all $k=1,2,...,D$ when $p>D$.
\end{proof}

A similar result can be obtained for the forms $( \cdot , \cdot )^{\widehat {\boldsymbol k},\perp}$ and $(\cdot,\cdot)^{\widehat {\boldsymbol k}}_{\boldsymbol k'}$:
\begin{lemma}
For any $\boldsymbol k \in \mathbb{R}^D_{\mathbb{Q}}, \boldsymbol k' \in\R^D$ and $A \in \mathfrak{U}_{loc}$ in a $p$-clustering dynamical system with $p>D$, it follows that $( A , A )^{\widehat {\boldsymbol k},\perp} \geq 0$ and $(A,A)^{\widehat {\boldsymbol k}}_{\boldsymbol k'} \geq 0$.
\end{lemma}

\begin{proof} \label{technical_forms}
The important point about the proof of Lemma \ref{lemma:positive} is that the form is obtained by a summation over all elements of a (sub-)group of $\Z^D$, of the action of these elements on one factor of the form; the proof works for any subgroup as clustering gives {\em a fortiori} a strong enough decay. Specifically, to see that $(A, A )^{\widehat {\boldsymbol k},\perp}$ is non-negative we write 
\begin{equation}
    B= \sum_{\bm x \in [0,L]^{D-1}}  A(x_i \bm h^{(i)})
\end{equation}
and proceed as in Lemma \ref{lemma:positive} from $\langle B, B \rangle \geq 0$. Similarly for $(A,A)^{\widehat {\boldsymbol k}}_{\boldsymbol k'}$.
\end{proof}

\begin{lemma}
For any $A,B \in \mathfrak{U}$ that are $p$-clustering for $p>D$, $\langle A, B \rangle_k$ is uniformly bounded for $k \in \mathbb{R}^D$, and \begin{equation}
\lim_{\boldsymbol k \to 0} \langle A , B \rangle_{\boldsymbol k} = \langle A, B \rangle_0.
\end{equation}
The same holds for $(a,a)^{\widehat {\boldsymbol k}}_{\boldsymbol k'}$.
\end{lemma}

\begin{proof}
The proof is an immediate generalisation of the proof of \cite[Lemma 5.6]{doyon_hydrodynamic_2022} for $p>D$.
\end{proof}

\section{Unitarity of space-time translations on various completions of $\mathfrak{U}_{loc}$} \label{appendix:sesquilinear}
Consider the sesquilinear forms  $( \cdot , \cdot )^{\widehat {\boldsymbol k},\perp}$ and $(\cdot,\cdot)^{\widehat {\boldsymbol k}}_{\boldsymbol k'}$  defined in \Cref{lemma_sesquiliniar}. We have to establish that  $ι_x \coloneqq \iota_{x \widehat{\boldsymbol k}}$ ($x\in\Z$) and $τ_t$ ($t\in\R$) act unitarily on $\mathbb{H}_{\widehat{\boldsymbol k}}^{D-1}$ as representations of the groups $\Z$ and $\R$ respectively, and that $τ_t$ ($t\in\R$) act unitarily on $\mathcal H^{\widehat {\boldsymbol k}}_{\boldsymbol k'}$.

We need to first show that these actions are indeed well defined on the equivalence classes. If $A \sim^{\bm{\hat k},\perp} A^{\prime}$ then also $ι_zA \sim^{\bm{\hat k},\perp} ι_z A^{\prime}$, $\forall z \in \Z$, since 
\begin{equation}
    (ι_zA - ι_zA^{\prime} , ι_zA - ι_zA^{\prime})^{\bm{\hat k},\perp} = \sum_{\bm x \in \mathbb H^{D-1}_{\widehat{\boldsymbol k}} } \big( ι_{z\bm{\hat{k}}} ι_{\bm x} ( A - A^{\prime} ), ι_{z \bm{\hat{k}}} ( A - A^{\prime}) \big)
\end{equation}
and $ι$ acts unitarily inside the innder product $( \cdot , \cdot)$ defined by the connected correlation, by space-time invariance of the state. Unitarity is also immediately established from this fact.   Finally space translations are extended to the full Hilbert space $\mathcal{H}^{\bm{\hat{k}},\perp}$  by continuous extension. 

For $τ_t$ the situation is more complicated, as it maps local elements to non-local ones. However, the treatment is the same  as is in $D=1$, see \cite[Theorem 5.11]{doyon_hydrodynamic_2022}.

\end{appendices}

\bibliographystyle{unsrt}
\bibliography{references}

\begin{thebibliography}{10}

\bibitem{eisert_quantum_2015}
J.~Eisert, M.~Friesdorf, and C.~Gogolin.
\newblock Quantum many-body systems out of equilibrium.
\newblock {\em Nature Physics}, 11(2):124--130, February 2015.

\bibitem{gogolin_equilibration_2016}
Christian Gogolin and Jens Eisert.
\newblock Equilibration, thermalisation, and the emergence of statistical
  mechanics in closed quantum systems.
\newblock {\em Reports on Progress in Physics}, 79(5):056001, April 2016.

\bibitem{dalessio_quantum_2016}
Luca D'Alessio, Yariv Kafri, Anatoli Polkovnikov, and Marcos Rigol.
\newblock From quantum chaos and eigenstate thermalization to statistical
  mechanics and thermodynamics.
\newblock {\em Advances in Physics}, 65(3):239--362, May 2016.

\bibitem{hilbert_mathematical_1902}
David~R. Hilbert.
\newblock Mathematical {Problems}.
\newblock Bull. Amer. Math. Soc. 8 (1902), 437-479, 1902.

\bibitem{neumann_beweis_1929}
J.~v. Neumann.
\newblock Beweis des {Ergodensatzes} und {desH}-{Theorems} in der neuen
  {Mechanik}.
\newblock {\em Zeitschrift für Physik}, 57(1):30--70, January 1929.

\bibitem{goldstein_long-time_2010}
S.~Goldstein, J.~L. Lebowitz, R.~Tumulka, and N.~Zanghì.
\newblock Long-time behavior of macroscopic quantum systems.
\newblock {\em The European Physical Journal H}, 35(2):173--200, November 2010.

\bibitem{goldstein_normal_2009}
Sheldon Goldstein, Joel Lebowitz, Christian Mastrodonato, Roderich Tumulka, and
  Nino Zanghi.
\newblock Normal {Typicality} and von {Neumann}'s {Quantum} {Ergodic}
  {Theorem}.
\newblock {\em Proceedings of The Royal Society A Mathematical Physical and
  Engineering Sciences}, 466, July 2009.

\bibitem{shenker_black_2014}
Stephen~H. Shenker and Douglas Stanford.
\newblock Black holes and the butterfly effect.
\newblock {\em Journal of High Energy Physics}, 2014(3):67, March 2014.

\bibitem{hosur_chaos_2016}
Pavan Hosur, Xiao-Liang Qi, Daniel~A. Roberts, and Beni Yoshida.
\newblock Chaos in quantum channels.
\newblock {\em Journal of High Energy Physics}, 2016(2):4, February 2016.

\bibitem{maldacena_bound_2016}
Juan Maldacena, Stephen~H. Shenker, and Douglas Stanford.
\newblock A bound on chaos.
\newblock {\em Journal of High Energy Physics}, 2016(8):106, August 2016.

\bibitem{hashimoto_out--time-order_2017}
Koji Hashimoto, Keiju Murata, and Ryosuke Yoshii.
\newblock Out-of-time-order correlators in quantum mechanics.
\newblock {\em Journal of High Energy Physics}, 2017(10):138, October 2017.

\bibitem{spohn_large_1991}
Herbert Spohn.
\newblock {\em Large scale dynamics of interacting particles}.
\newblock Springer-Verlag, 1991.

\bibitem{demasi_mathematical_2006}
Anna DeMasi and Errico Presutti.
\newblock {\em Mathematical methods for hydrodynamic limits}.
\newblock Springer, 2006.

\bibitem{kipnis_scaling_1998}
Claude Kipnis and Claudio Landim.
\newblock {\em Scaling limits of interacting particle systems}, volume 320.
\newblock Springer Science \& Business Media, 1998.

\bibitem{doyon_drude_2017}
Benjamin Doyon and Herbert Spohn.
\newblock Drude {Weight} for the {Lieb}-{Liniger} {Bose} {Gas}.
\newblock {\em SciPost Physics}, 3:039, 2017.

\bibitem{doyoncorrelations}
Benjamin Doyon.
\newblock Exact large-scale correlations in integrable systems out of
  equilibrium.
\newblock {\em SciPost Phys.}, 5(5):54, 2018.

\bibitem{nardis_correlation_2021}
Jacopo Nardis, Benjamin Doyon, Marko Medenjak, and Miłosz Panfil.
\newblock {\em Correlation functions and transport coefficients in generalised
  hydrodynamics}.
\newblock to appear in J. Stat. Mech, April 2021.

\bibitem{ampelogiannis_almost_2021}
Dimitrios Ampelogiannis and Benjamin Doyon.
\newblock Almost everywhere ergodicity in quantum lattice models, 2021.
\newblock arXiv: 2112.12730 [math-ph].

\bibitem{castro-alvaredo_emergent_2016}
Olalla~A. Castro-Alvaredo, Benjamin Doyon, and Takato Yoshimura.
\newblock Emergent {Hydrodynamics} in {Integrable} {Quantum} {Systems} {Out} of
  {Equilibrium}.
\newblock {\em Physical Review X}, 6(4):041065, December 2016.

\bibitem{bertini_transport_2016}
Bruno Bertini, Mario Collura, Jacopo De~Nardis, and Maurizio Fagotti.
\newblock Transport in {Out}-of-{Equilibrium} {XXZ} {Chains}: {Exact}
  {Profiles} of {Charges} and {Currents}.
\newblock {\em Physical Review Letters}, 117(20):207201, November 2016.

\bibitem{bulchandani_bethe-boltzmann_2018}
Vir~B. Bulchandani, Romain Vasseur, Christoph Karrasch, and Joel~E. Moore.
\newblock Bethe-{Boltzmann} hydrodynamics and spin transport in the {XXZ}
  chain.
\newblock {\em Physical Review B}, 97(4):045407, January 2018.

\bibitem{doyon_lecture_2020}
Benjamin Doyon.
\newblock Lecture notes on generalised hydrodynamics.
\newblock {\em SciPost Phys. Lect. Notes}, 2020.

\bibitem{chakraborti_entropy_2021}
Subhadip Chakraborti, Abhishek Dhar, Sheldon Goldstein, Anupam Kundu, and
  Joel~L. Lebowitz.
\newblock Entropy growth during free expansion of an ideal gas.
\newblock {\em arXiv:2109.07742 [cond-mat]}, December 2021.

\bibitem{buca2019nonstationary}
Berislav Buca, Joseph Tindall, and Dieter Jaksch.
\newblock Non-stationary coherent quantum many-body dynamics through
  dissipation.
\newblock {\em Nature Communications}, 10(1):1730, April 2019.

\bibitem{buca2020quantum}
Berislav Buca, Archak Purkayastha, Giacomo Guarnieri, Mark~T. Mitchison, Dieter
  Jaksch, and John Goold.
\newblock Quantum many-body attractors, 2020.
\newblock arXiv: 2008.11166 [quant-ph].

\bibitem{medenjak2020rigo}
Marko Medenjak, Tomaz Prosen, and Lenart Zadnik.
\newblock Rigorous bounds on dynamical response functions and time-translation
  symmetry breaking.
\newblock {\em SciPost Phys.}, 9(1):3, 2020.
\newblock Publisher: SciPost.

\bibitem{buca2021local}
Berislav Buca.
\newblock Local hilbert space fragmentation and out-of-time-ordered crystals,
  2021.
\newblock arXiv: 2108.13411 [cond-mat.stat-mech].

\bibitem{gunawardana2021dynamical}
Thivan Gunawardana and Berislav Buca.
\newblock Dynamical l-bits in {Stark} many-body localization, 2021.
\newblock arXiv: 2110.13135 [cond-mat.dis-nn].

\bibitem{doyon_hydrodynamic_2022}
Benjamin Doyon.
\newblock Hydrodynamic {Projections} and the {Emergence} of {Linearised}
  {Euler} {Equations} in {One}-{Dimensional} {Isolated} {Systems}.
\newblock {\em Communications in Mathematical Physics}, 391(1):293--356, April
  2022.

\bibitem{birkhoff1931ergodic}
G.~D. Birkhoff.
\newblock Proof of the {Ergodic} {Theorem}.
\newblock {\em Proceedings of the National Academy of Sciences of the United
  States of America}, 17(12):656--660, December 1931.

\bibitem{brandao2015equivalence}
Fernando G. S.~L. Brandao and Marcus Cramer.
\newblock Equivalence of statistical mechanical ensembles for non-critical
  quantum systems, 2015.
\newblock tex.copyright: arXiv.org perpetual, non-exclusive license.

\bibitem{Simon_Reed_Functional}
M.~Reed and B.~Simon.
\newblock {\em I: {Functional} {Analysis}}.
\newblock Methods of {Modern} {Mathematical} {Physics}. Elsevier Science, 1981.

\bibitem{ollahydro}
S.~Olla, S.~R.~S. Varadhan, and H.~T. Yau.
\newblock Hydrodynamical limit for a {Hamiltonian} system with weak noise.
\newblock {\em Communications in Mathematical Physics}, 155(3):523--560, August
  1993.

\bibitem{doyon_thermalization_2017}
Benjamin Doyon.
\newblock Thermalization and {Pseudolocality} in {Extended} {Quantum}
  {Systems}.
\newblock {\em Communications in Mathematical Physics}, 351(1):155--200, April
  2017.

\bibitem{bratteli_operator_1987}
O.~Bratteli and D.W. Robinson.
\newblock {\em Operator {Algebras} and {Quantum} {Statistical} {Mechanics} 2:
  {Equilibrium} {States} {Models} in {Quantum} {Statistical} {Mechanics}}.
\newblock Operator {Algebras} and {Quantum} {Statistical} {Mechanics}. Springer
  Science {\textbackslash}\& Business Media, 1997.

\bibitem{bratteli_operator_1997}
O.~Bratteli and D.W. Robinson.
\newblock {\em Operator {Algebras} and {Quantum} {Statistical} {Mechanics} 1:
  {C}*- and {W}*-{Algebras}. {Symmetry} {Groups}. {Decomposition} of {States}}.
\newblock Operator {Algebras} and {Quantum} {Statistical} {Mechanics}. Springer
  Science {\textbackslash}\& Business Media, 1987.

\bibitem{Araki1975uniqueness}
Huzihiro Araki.
\newblock On uniqueness of {KMS} states of one-dimensional quantum lattice
  systems.
\newblock {\em Communications in Mathematical Physics}, 44:1--7, 1975.

\bibitem{Lieb:1972wy}
Elliott~H. Lieb and Derek~W. Robinson.
\newblock The finite group velocity of quantum spin systems.
\newblock {\em Communications in Mathematical Physics}, 28(3):251--257,
  September 1972.

\bibitem{hille_functional_1996}
E.~Hille and R.S. Phillips.
\newblock {\em Functional {Analysis} and {Semi}-groups}.
\newblock American {Mathematical} {Society}: {Colloquium} publications.
  American Mathematical Society, 1996.
\newblock Issue: v. 31, pt. 1.

\bibitem{BergelsonDiscreteToContinuous}
V.~Bergelson, A.~Leibman, and C.~G. Moreira.
\newblock Form discrete- to continuous-time ergodic theorems.
\newblock {\em arXiv:1109.1800 [math]}, September 2011.
\newblock arXiv: 1109.1800.

\bibitem{fidaleo2014Nonconventional}
Francesco Fidaleo.
\newblock Nonconventional ergodic theorems for quantum dynamical systems.
\newblock {\em Infinite Dimensional Analysis, Quantum Probability and Related
  Topics}, 17(02):1450009, June 2014.
\newblock Publisher: World Scientific Publishing Co.

\bibitem{ruelle_statistical_1974}
David Ruelle.
\newblock {\em Statistical mechanics: {Rigorous} results}.
\newblock World Scientific, 1974.

\bibitem{duvenhage2009Bergelson}
Rocco Duvenhage.
\newblock Bergelson’s theorem for weakly mixing {C}*-dynamical systems.
\newblock {\em Studia Mathematica}, 192(3):235--257, 2009.
\newblock Publisher: Institute of Mathematics, Polish Academy of Sciences.

\bibitem{Fidaleo2009Ergodic}
Francesco Fidaleo.
\newblock An {Ergodic} {Theorem} for {Quantum} {Diagonal} {Measures}.
\newblock {\em Infinite Dimensional Analysis, Quantum Probability and Related
  Topics}, 12(02):307--320, June 2009.
\newblock Publisher: World Scientific Publishing Co.

\bibitem{rigol_relaxation_2007}
Marcos Rigol, Vanja Dunjko, Vladimir Yurovsky, and Maxim Olshanii.
\newblock Relaxation in a {Completely} {Integrable} {Many}-{Body} {Quantum}
  {System}: {An} {Ab} {Initio} {Study} of the {Dynamics} of the {Highly}
  {Excited} {States} of {1D} {Lattice} {Hard}-{Core} {Bosons}.
\newblock {\em Physical Review Letters}, 98(5):050405, February 2007.

\bibitem{ilievski_complete_2015}
E.~Ilievski, J.~De~Nardis, B.~Wouters, J.-S. Caux, F.~H.~L. Essler, and
  T.~Prosen.
\newblock Complete generalized gibbs ensembles in an interacting theory.
\newblock {\em Physical Review Letters}, 115(15):157201, October 2015.

\bibitem{essler_quench_2016}
Fabian H~L Essler and Maurizio Fagotti.
\newblock Quench dynamics and relaxation in isolated integrable quantum spin
  chains.
\newblock {\em Journal of Statistical Mechanics: Theory and Experiment},
  2016(6):064002, June 2016.

\bibitem{khemani_velocity-dependent_2018}
Vedika Khemani, David~A. Huse, and Adam Nahum.
\newblock Velocity-dependent {Lyapunov} exponents in many-body quantum,
  semiclassical, and classical chaos.
\newblock {\em Physical Review B}, 2018.

\bibitem{conwayFunctionalAnalysis2007}
John~B. Conway.
\newblock {\em A {Course} in {Functional} {Analysis}}, volume~96 of {\em
  Graduate {Texts} in {Mathematics}}.
\newblock Springer New York, New York, NY, 2007.

\bibitem{frohlich_properties_2015}
Jürg Fröhlich and Daniel Ueltschi.
\newblock Some properties of correlations of quantum lattice systems in thermal
  equilibrium.
\newblock {\em Journal of Mathematical Physics}, 56(5):053302, May 2015.
\newblock Publisher: American Institute of Physics.

\bibitem{Eisert_locality}
M.~Kliesch, C.~Gogolin, M.~J. Kastoryano, A.~Riera, and J.~Eisert.
\newblock Locality of temperature.
\newblock {\em Physical Review X}, 4(3):031019, July 2014.
\newblock Number of pages: 19 Publisher: American Physical Society.

\bibitem{delvecchio2021hydro}
Giuseppe Del Vecchio Del~Vecchio and Benjamin Doyon.
\newblock The hydrodynamic theory of dynamical correlation functions in the
  {XX} chain.
\newblock arXiv: 2111.08420 [math-ph], 2021.

\bibitem{Doyon:2019oaf}
Benjamin Doyon.
\newblock Diffusion and {Superdiffusion} from {Hydrodynamic} {Projections}.
\newblock {\em Journal of Statistical Physics}, 186(2):25, January 2022.

\bibitem{naaijkens_quantum_2017}
Pieter Naaijkens.
\newblock {\em Quantum {Spin} {Systems} on {Infinite} {Lattices}: {A} {Concise}
  {Introduction}}, volume 933 of {\em Lecture {Notes} in {Physics}}.
\newblock Springer International Publishing, Cham, 2017.

\bibitem{nachtergaele_quasi-locality_2019}
Bruno Nachtergaele, Robert Sims, and Amanda Young.
\newblock Quasi-locality bounds for quantum lattice systems. {I}.
  {Lieb}-{Robinson} bounds, quasi-local maps, and spectral flow automorphisms.
\newblock {\em Journal of Mathematical Physics}, 60(6):061101, June 2019.
\newblock Publisher: American Institute of Physics.

\end{thebibliography}

\end{document}